\documentclass[12pt,halfline,a4paper, bookmarksopen,bookmarksdepth=2,breaklinks=true,hyphens,obeyspaces]{pythia}

\begin{document}

\title{Rise of the Machines? \\ 
\footnotesize{Intraday High-Frequency Trading Patterns of Cryptocurrencies}}
\author{
\name\large{{Alla A. Petukhina}}
\address\footnotesize{{Humboldt-Universit\"at zu Berlin. \\
Firamis GmbH, Germany.}\\
\email{alla.petukhina[at]wiwi.hu-berlin.de}}\vspace{-0.5cm}
\and
\name\large{{Raphael C. G. Reule}}
\address\footnotesize{{Humboldt-Universit\"at zu Berlin.}\\
\email{irtg1792.wiwi[at]wiwi.hu-berlin.de}}
\and
\name{\large{Wolfgang Karl H\"ardle}}
\footnotesize{\address{Humboldt-Universit\"at zu Berlin, IRTG 1792, Dorotheenstr. 1, 10117 Berlin, Germany
School of Business, Singapore Management University, 50 Stamford Road, Singapore 178899
Faculty of Mathematics and Physics, Charles University, Ke Karlovu 3, 121 16 Prague, Czech Republic
Department of Information Management and Finance, National Chiao Tung University, Taiwan, ROC}
\email{haerdle[at]wiwi.hu-berlin.de}}
}

\abstract{This research analyses high-frequency data of the cryptocurrency market in regards to intraday trading patterns related to algorithmic trading and its impact on the European cryptocurrency market. We study trading quantitatives such as returns, traded volumes, volatility periodicity, and provide summary statistics of return correlations to CRIX (CRyptocurrency IndeX), as well as respective overall high-frequency based market statistics with respect to temporal aspects. Our results provide mandatory insight into a market, where the grand scale employment of automated trading algorithms and the extremely rapid execution of trades might seem to be a standard based on media reports. Our findings on intraday momentum of trading patterns lead to a new quantitative view on approaching the predictability of economic value in this new digital market.\\

\footnotesize{
\vspace{0.1cm} \textit{The financial support of the Czech Science Foundation under grant no. 19-28231X, the Firamis GmbH, Robert-Kempner-Ring 27, 61440 Oberursel (Taunus), the Yushan Scholar Program, and European Union’s Horizon 2020  ”FIN-TECH” Project, under the grant no. 825215 (Topic ICT-35-2018, Type of actions: CSA), as well as data support by dyos solutions GmbH, Oberwallstr. 8, 10117 Berlin, are greatly acknowledged.}

\bigskip
This is a post-peer-review, pre-copyedit version of an article published in The European Journal of Finance. The final authenticated version is available online at: 

\url{https://doi.org/10.1080/1351847X.2020.1789684}
}

}


\maketitle

\newpage


\normalsize

\section{Motivation}
\label{sec1}

High-frequency trading takes advantage of the incredible rise of computing power provided by the steady development of ever more capable structures. Algorithms are already major players in a variety of financial applications, and have proven to be more efficient than their human counterparts. By employing these so-called ``algos", positive effects can be exploited to their maximum and market inefficiencies can potentially be eliminated \citep{ucc:2020}. However, just like for every coin, there is a flipside, such as the negative impact on capital markets caused by technological inefficiencies \citep{Emem:2018}. One of the most noted events of an early point of attack for these algorithms was the Flash Crash of 2010.\\

No matter what, the machines are here to stay and their influence, possibly powered by ``learning" algorithms, will certainly increase even more with time - especially in regards to new emerging markets such as cryptocurrencies. The rising popularity and acceptance of this alternative asset, as it has yet to be understood as an alternative to fiat currency, requires for specialised strategies to maximise the potential return of investments \citep[][2019]{akhtaruzzaman2019influence, platanakis2019should, trimborn2018investing, PetTriHaeEle:2019}.

\vspace{-0.5cm}

\begin{figure}[H]
	\centering
	\includegraphics[keepaspectratio,width=16cm]{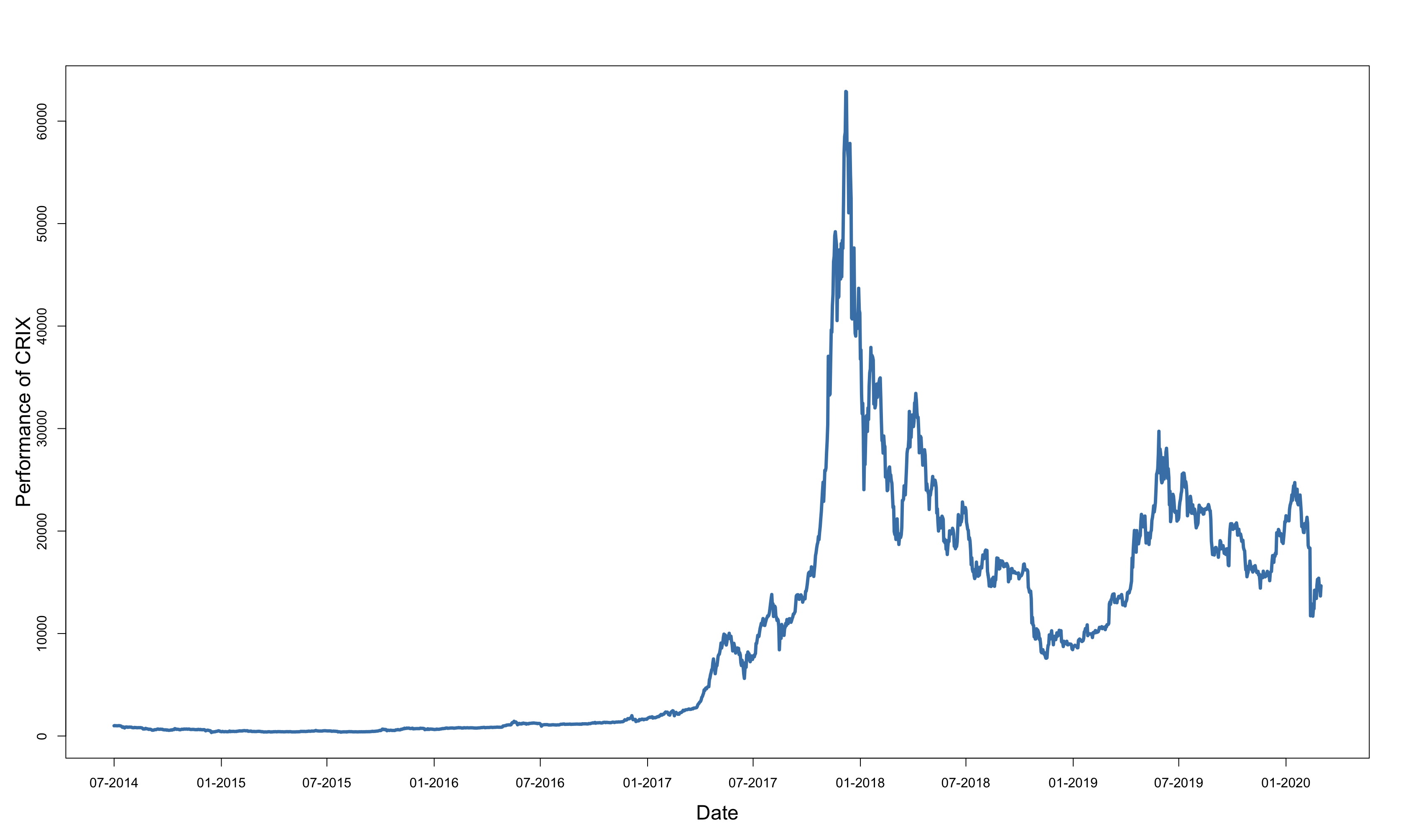}\href{https://github.com/QuantLet/CrixToDate/tree/master/CTD_CRIX_Full_24h/Full}{\includegraphics[keepaspectratio,width=0.4cm]{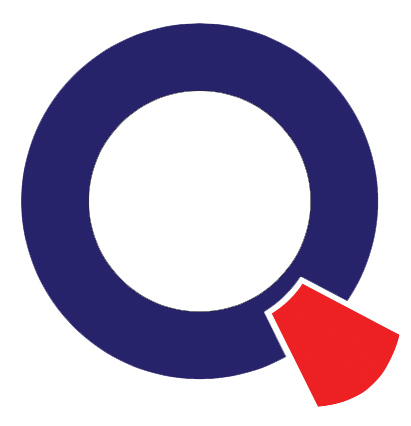}}
	\caption{CRIX Time Series.} 
\end{figure}

\newpage

Yet, did the quantlets or algorithms really venture in the realm of autonomous  machines, the digital world, or are they still with the world of the humans, the world of  manual labour oil and baby nutrition companies?\\

This is especially of interest since the cryptocurrency market has significantly matured in recent years and has attracted enormous investments, not only by major players but especially by individuals. Especially $FinTech$ $Startups$ are of high interest, as absurd amounts of financial backing was (and is still to some extent) being generated by just presenting a briefly written $Whitepaper$-PDF marketing outlet \citep{Zetsch:19}. The early cryptocurrency market kick-off starting in late 2017 is evidently presenting such happenings. The discrepancy between sentiment and tone generated by marketing versus the delivered performance is fascinating  \citep[with further references][]{ucc:2020,Romeo:2019,Qian:2019}.\\

In this research, we are analysing high-frequency data (5-minute intervals) gained from the cryptocurrency market and see, if there is really 24/7 algorithmic trading, or if there are still people sitting behind their computers creating and executing orders by hand after they have returned from their daily jobs.\\

Previous research outputs on this theme, such as \citet{ZhChChNa:2019}, have used time spans ranging from 1 hour to 12 hours. Their methods yielded results, which lead to different conclusions, yet opened up further thoughts towards factors such as trading patterns, variations in returns, volatility and trading volume. \citet{ZhWaLiSh:2018} are also looking at the same aspects as the previous research, with the additional finding of a power-law correlation
between price and volume. \cite{stylecc:2017} respectively build a uni- and multivariate analysis of quantitative facts to show off stylized facts of cryptocurrencies. \citet{SchReKra:2019} analyzed limit order data from cryptocurrency exchanges. Besides their recovery of common qualitative facts, they find that these data exhibit many of the properties found for classic limit order exchanges, such as a symmetric average limit order book, the autocorrelation of returns only at the tick level and the timing of large trades. Yet they find that cryptocurrency exchanges exhibit a relatively shallow limit order book with quickly rising liquidity costs for larger volumes, many small trades and an extended distribution of limit order volume far beyond the current mid-price.\\

Given the search for the most efficient trading strategies, \citet{Capo:19} provide a range of historic scientific works on the time of day effects to reap abnormal profits. In contrast to their work, we aim at identifying the market drivers, which are responsible for how this new emerging market, which is still full of conundrums for many, behaves - i.e. do market movements fit into human activity patterns or are these independent from time.\\

Preliminary research has therefore not touched the highly topical question of human impact in the wake of digital systems. There are many papers with interesting approaches and solutions, but only for problems that are already known and have been rebrewn for some time now. Yet, with the advent and popular discussion of the employment of Long Short Term Memory Neural Networks (LSTM) and hence deep learning for finance, AI advisory, essentially based on the human factor of sentiment in the realm of cryptocurrencies \citep{Romeo:2019}, will play a major role in especially this completely digital market. This, as a circular argument, brings us once again to the fundamental idea of enforcing the understanding of market behaviour based on the time of the day and the agents acting in these markets that are predestined to be ruled by the machines.\\

As a polemic term, we are using $Proof-Of-Human$ ($PoH$; derived from $Proof-Of-Work$, $Proof-Of-Stake$ et cetera consensus algorithms) to underline the hypotheses that not algorithms are the major players in this market, but humans. Humans don't act as programmed like algorithms - they act based on biological and psychological input, such as hunger or fatigue. The majority of humans will have certain times at which they are active, and at which they rest and are therefore inactive. Alternatively spoken, algorithms need humans to start and then exacerbate a price trend - the question is, therefore if the cryptocurrency market is dominated by human or algorithmic behaviour. Eventually, we can differentiate algorithmic and human trading patterns expressed within the market \citep[with further references][]{Capo:16}.\\

The paper is structured by giving a \hyperref[sec2]{brief general introduction and data source disclosure} and \hyperref[sec3.1]{methodology section}, followed by a respective \hyperref[sec3.2]{intraday data analysis}, which is concluded by a section on \hyperref[sec3.3]{Time-Of-Day effects and the Proof-Of-Human}.\\

All presented graphical and numerical examples shown are reproducible and can be found on \includegraphics[keepaspectratio,width=0.4cm]{media/qletlogo_tr.png} \url{www.quantlet.de} \citep{BH:2018} and are indicated as \href{http://github.com/QuantLet/CCID}{\includegraphics[keepaspectratio,width=0.4cm]{media/qletlogo_tr.png}}\href{http://github.com/QuantLet/CCID}{CCID}.


\section{High-Frequency Cryptocurrency Data}
\label{sec2}

To understand the dynamics of this new high-frequency market, it is mandatory to investigate the statistical properties
of various high-frequency variables, for example, trading volume or volatility, to find respective answers to questions like option pricing and forecasting. Preliminary research to visualize the cryptocurrency market was done by \citet{CRIX:2018} with the CRyptocurrency IndeX, \href{https://github.com/QuantLet/CRIX}{\includegraphics[keepaspectratio,width=0.4cm]{media/qletlogo_tr.png}}\href{https://github.com/QuantLet/CRIX}{ CRIX} (\href{http://www.crix.berlin}{crix.berlin}), in order to represent the performance of the cryptocurrency market with the help of the most mature and accepted cryptocurrencies, such as Bitcoin (BTC), Ethereum (ETH), or Ripple (XRP) - \hyperref[Appendix]{see appendix section 5.1} for further used abbreviations. As the CRIX index family covers a range of cryptocurrencies based on different liquidity rules and various model selection criteria, we have chosen this as the main data source. CRIX represents the cryptocurrency market, but by its very nature is dominated by a few main players with BTC being the absolute market driver over time. \\

Furthermore, we used data provided by dyos solutions GmbH (\href{https://dyos.io/}{dyos.io}) compiled from various exchanges' data, to ensure that our findings are coherent with other data available. It is important to keep in mind, that the 5-minute data analysed in this research is gained from sources located in the $European$ $markets$ (+1h GMT) and therefore the time-of-day effects may look different for markets from the Americas or Asia. We will make an exegesis on this important point in \hyperref[sec3.3]{subsection 3.3}.\\

In addition, the analysed data sample belongs to the time period after the cryptocurrency market heated up immensely around the end of 2017, followed by a sharp cooldown at the beginning of 2018. By that time a plethora of euphoric media outlets was praising the endless possibilities which the blockchain technology may provide - and what eventually also lead to quite a lot of ICO scams \citep{Zetsch:19}.   At that time, algorithmic trading in cryptocurrency markets was not seen as being a mere idea, but reality by more or less promising FinTech startups. These emerging enterprises are offering a wide variety of blockchain-related services, such as trading, asset management, or technical support. Especially FinTech startups related to the financial sector, in contrast to for example supply chain oriented ventures, are heavily interested in ArtificialIntelligence (”learning” algorithms) and are marketing their individual related products as groundbreaking and ready-to-use. Given the chosen typical vacation period, July and August, one should hence expect a less pronounced human, but algorithmic driven market behavior to contradict the hypotheses of the PoH concept - more on that as well in \hyperref[sec3.3]{subsection 3.3}.\\

Regarding data handling, we are coherent with previous research on high-frequency data based on traditional data sources, such as the NYSE, which has underlined data preparation issues and the specific statistical properties of various high-frequency variables \citep{Hautsch:2011}. As we are dealing with a subject, where individuals can act directly with the market without involving a middle-man, the characteristics of our data observed on transaction
level, therefore, are especially irregularly spaced in time and without interruption - see \hyperref[sec3]{section 3}.


\section{Intraday Data Analysis}
\label{sec3}

 In the following chapter, we provide an overview of the methods employed to analyze our high-frequency data at hand with further statistical intraday cryptocurrency market observations.

\subsection{Methodology}
\label{sec3.1}

This paper undertakes a fresh empirical investigation of key financial variables of cryptocurrency market, such as volatility, returns and trading volume. Following, for example, \citet{Hussein:2011}, intraday return volatility is calculated as absolute log-returns as defined in (\ref{logret}). As we are looking at high-frequency data, there is no need to use measures like, for example, the compounded annual growth rate (CAGR) instead of absolute returns, which is used to get the per-annum returns and does not support the analysis in this case.\\

The simple return $Ret_t$ is defined as
\begin{align}
		\label{ret}
Ret_t = \frac{P_t - P_{t-1}}{P_{t-1}},
\end{align}
where $P_t$ und $P_{t-1}$ are prices of coins at time points $t$ and $t-1$ respectively.
The log return $ret_t$ is defined as
\begin{align}
		\label{logret}
ret_t = \log \frac{P_t}{P_{t-1}} = \log ( 1+Ret_t).
\end{align}

In order to expressively visualize some features of our high dimensional and nonstationary time series gained from our large high-frequency dataset of the specifically chosen period of time, a $Generalized$ $Additive$ $Model$ (GAM) is best suited. A GAM is a generalized linear model (GLM), where the nonlinear predictor is given by a specified sum of smooth functions of the covariates, as well as a conventional parametric component of the linear predictor \citep{WKH:90}. The basic advantage of GAM is the possibility to model highly complex nonlinear relationships given a large number of potential predictors. In particular, recent computational developments in GAM fitting methods, such as \cite{wood2015generalized}, \cite{wand2017fast}, and \cite{wood2017generalized}, have made it possible to use these models to explore very large datasets. Moreover, in the last two decades, GAM methods have intensively developed in terms of the range of models that can be fitted. All these advantages make GAMs a feasible tool to investigate intraday seasonality patterns with high-frequency trading data. In general, the model has a structure something like:

\begin{equation}
g\left\{E\left(y_{i}\right)\right\}=\beta_{0}+f_{1}\left(x_{i 1}\right)+\cdots+f_{p}\left(x_{i p}\right)
\end{equation}

where $y=\left(y_{1}, \ldots, y_{n}\right)^{\top}$ observation of a response variable $Y$, $g$ is a link function (identical, logarithmic or inverse, etc.),  $x_1 \dots x_p$ are independent variables, $\beta_0$ is an intercept, $f_{1}\left(x_{i 1}\right) \dots f_{p}\left(x_{i p}\right)$ are unknown nonparametric smooth functions, and $\varepsilon_{i}$ is an i.i.d. random error.
In our application we use the identity link function, since the LHS of our equations are features/variables observed or measured on a continuous scale, to fit the following statistical model:

\begin{equation}
y_{i}=f_{1}\left(x_{1, i}\right)+f_{2}\left(x_{2, i}\right)+\ldots+f_{p}\left(x_{p, i}\right)+\varepsilon_{i}
\end{equation}

Here $y_i$ will be a trading volume, volatility, or returns as defined in (\ref{logret}), $x_{q,i}$  will be  the daily  and weekly effects. The nonlinear function $f_q$ is a smooth function, composed by sum of basis functions $b_{j}^{q}$ (for example B-splines, P-splines or cubic splines) and their corresponding regression coefficients $\beta_{q,j}$. Thus, each function $f_q$ is expressed as:

\begin{equation}
f_{q}(x)=\sum_{j=1}^{k_{q}} \beta_{q, j} b_{j}^{q}(x)
\end{equation}

where $k_q$ is the dimension of the spline basis. \\The smooth function  ${m(x_1, ... , x_p) = \sum_{q=1}^{p} f_q (x_q)}$  is estimated by penalized regression:

\begin{equation}
\sum_{i=1}^{n}\left(y_{i}-\sum_{q=1}^{p} f_{q}\left(x_{i}\right)\right)^{2}+\sum_{q=1}^{p} \lambda_{q} \int\left\|f_{q}^{\prime \prime}(x)\right\|^{2} d x
\end{equation}

where the penalty parameter $\Lambda=\left(\lambda_{1}, \ldots, \lambda_{p}\right)$  is a smoothing parameter controlling the fit–smoothness trade‐off for $f_q$ and can be selected by minimization of the Generalized Cross
Validation (GCV) score, see \citep{wood2004stable} and \citep{wood2011fast}. Denoting $B$ the matrix formed by concatenation of the $b^q_j$, one has to solve the following problem:

\begin{equation}
	\label{objf}
\widehat{\beta} = \arg \min _{\lambda,\beta}\left\{\|Y-B \beta\|^{2}+\sum_{q=1}^{p} \lambda_{\dot{q}} \beta^{\top} S_{q} \beta\right\}
\end{equation}

where $\beta=\left(\beta_{1}, \ldots, \beta_{p}\right)^{\top}$ is the vector of the unknown regression parameters, $S_q$ is a  matrix of known coefficients (a smoothing matrix) and depends on the spline basis. Thus, given $\lambda$, expression (\ref{objf}) may readily be minimized to yield the coefficient estimates $\hat{\beta_\lambda}$. The method of obtaining the estimate of the $\beta$  is called Penalized Iteratively Re-weighted Least Squares (P-IRLS) which is implemented in the mgcv R package,  see \citep{WoodR}.


\subsection{Summary Statistics}
\label{sec3.2}

As an introduction to the data analyzed in this brief research, we are providing  summary statistics regarding its statistical properties to form a basic understanding of the market at hand. Firstly, the trading data density of cryptocurrencies against the normal distribution of BTC is far from normally distributed, see figure 2. Hence the behaviour of agents in this market is far from what we would see in classic markets. This implies, that new rules are being employed, and therefore we have to rethink our common way on how to approach the quantitative analysis of markets in general. We will start our discussion on the specific research question by first providing a general overview of the cryptocurrency market with increasingly narrowed focus and attention to detail regarding specific timeframes and parameters for individual crypto-assets.\\

\vspace{-0.5cm}

\begin{figure}[H]
	\centering
	\includegraphics[keepaspectratio,width=11cm]{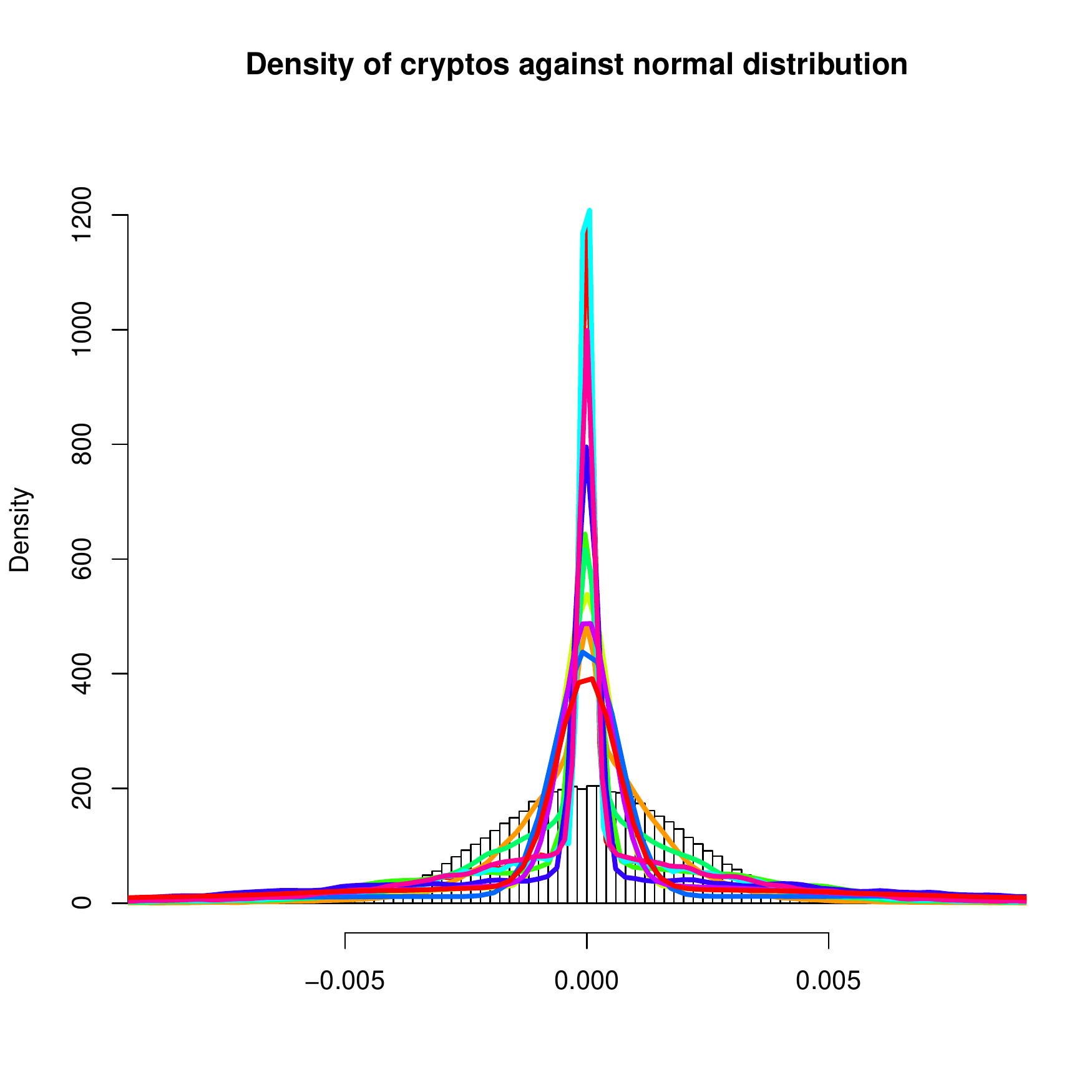}\href{https://github.com/QuantLet/CCID/tree/master/CCIDHistReturnsDensity}{\includegraphics[keepaspectratio,width=0.4cm]{media/qletlogo_tr.png}}
	\caption{Density of intraday CCs returns. 01. July 2018 - 31. August 2018. The probability density functions of the distributions of daily returns for the analized cryptocurrencies with the following colour code: \textcolor{BCH}{BCH}, \textcolor{BTC}{BTC},
\textcolor{DASH}{DASH}, \textcolor{ETC}{ETC}, \textcolor{ETH}{ETH}, \textcolor{LTC}{LTC}, \textcolor{REP}{REP}, \textcolor{STR}{STR}, \textcolor{XMR}{XMR}, \textcolor{XRP}{XRP}, \textcolor{ZEC}{ZEC}. A normal distribution with the same mean and standard deviation as the returns on BTC is displayed as a histogram in the background}
\end{figure}

Secondly, using GAM, we gain interesting insights into the trading activities in this 24/7 market. Cryptocurrencies are being traded without any forced break, as we know it from classic markets, for example, if the stock exchange closes for the night or especially for weekends. In addition  to this fact, we have to consider, that there is no centralized trading in the act, but a plethora of service providers, so-called cryptocurrency exchanges. As we disclose the origin of our data, we underline, that caused by this very decentralized nature of cryptocurrency genesis and their respective trading, partially greatly diverging price data is available for each individual cryptocurrency. Again, this is caused by the decentralized root of individual, unsupervised and unregulated, places for exchange. There is no fixed price for BTC contrary to, for example, for exchange rates of USD-EUR.

\begin{figure}[H]
	\centering
	\includegraphics[keepaspectratio,width=15cm]{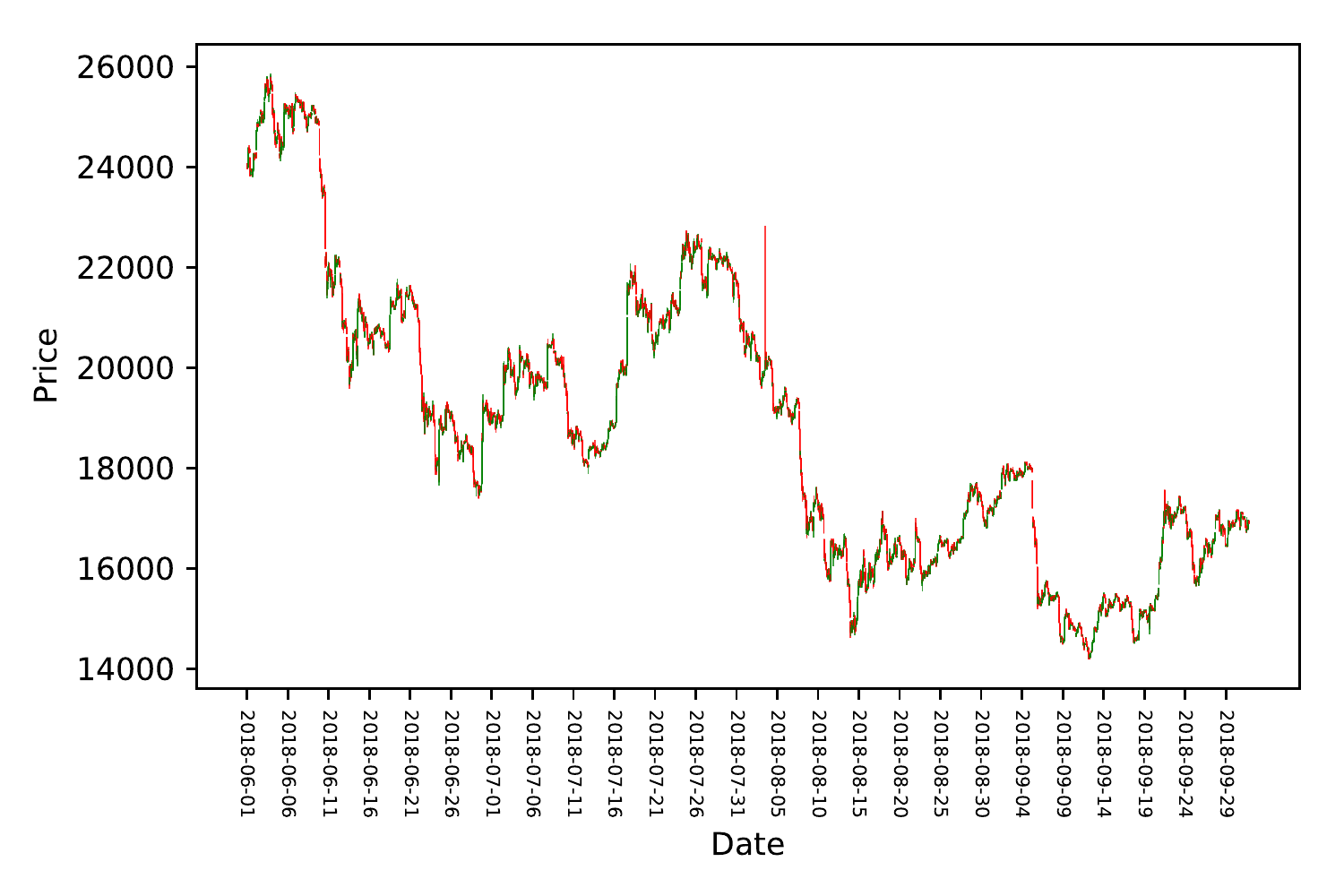}
	\href{https://github.com/QuantLet/CCID/tree/master/CCIDCandles}{\includegraphics[keepaspectratio,width=0.4cm]{media/qletlogo_tr.png}}
	\caption{Candlestick chart of CRIX. 01. July 2018 - 29. September 2018.}
\end{figure}

\begin{figure}[H]
\hfill
\subfigure[BTC]{\includegraphics[width=5cm]{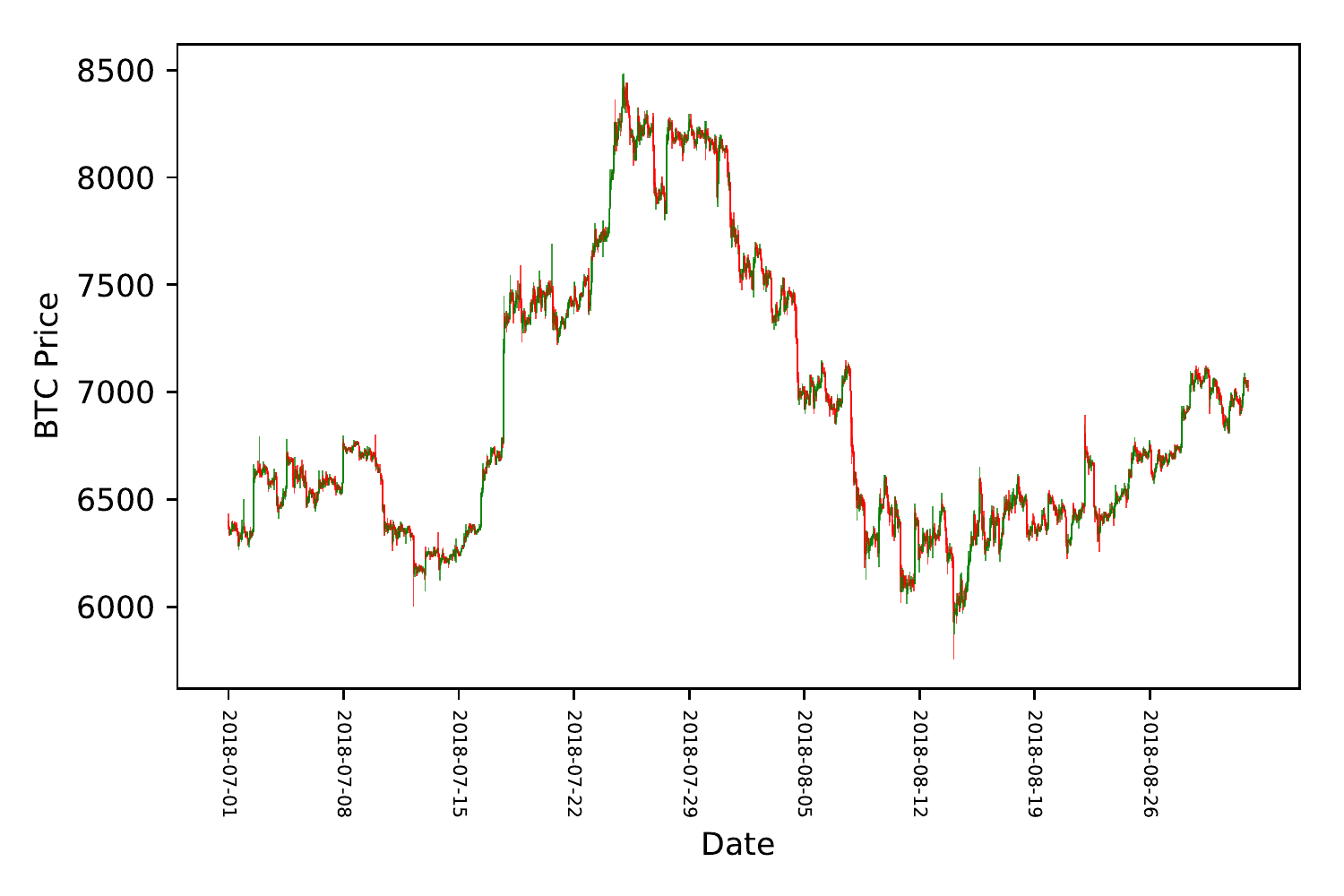}}
\hfill
\subfigure[ETH]{\includegraphics[width=5cm]{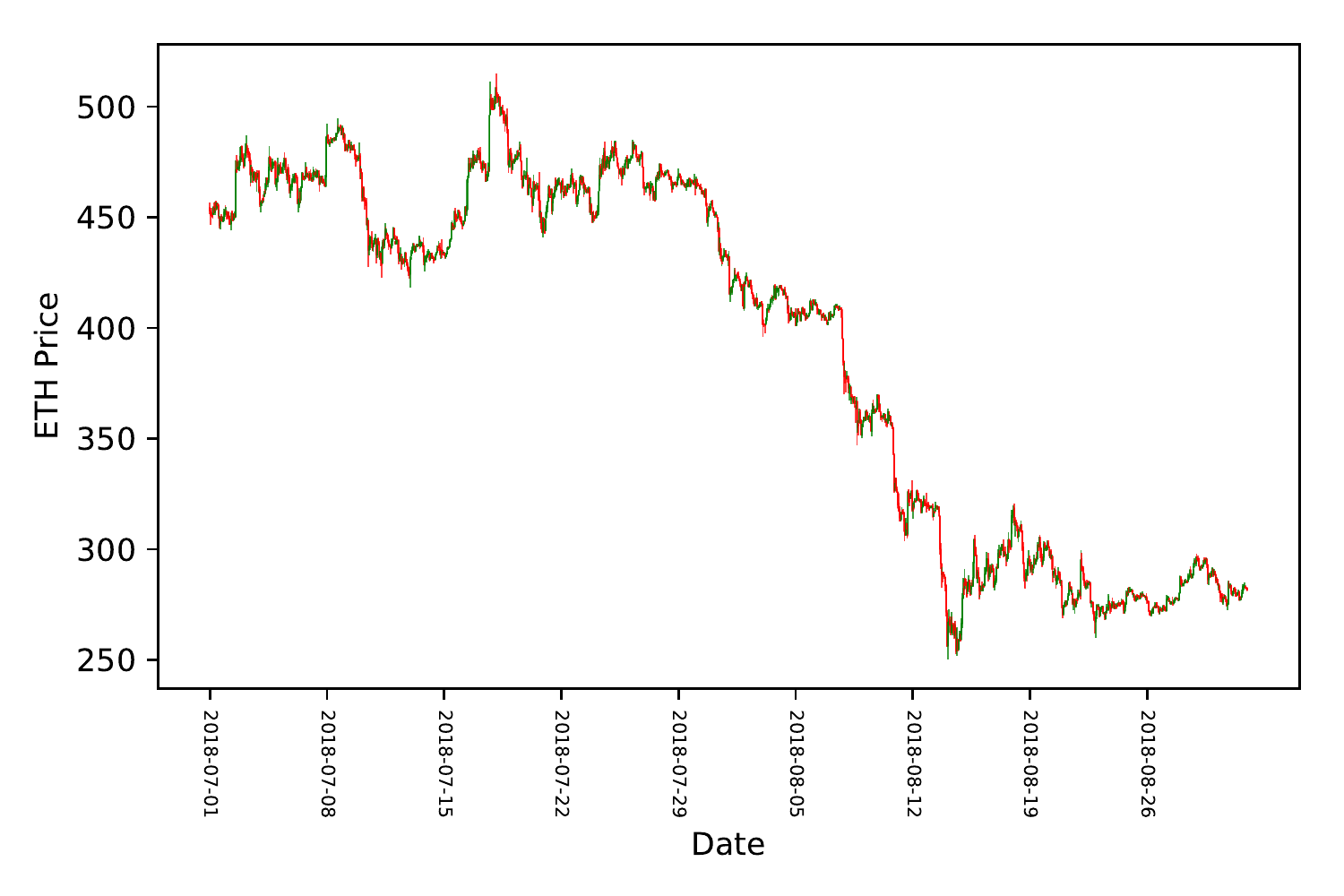}}
\hfill
\subfigure[XRP]{\includegraphics[width=5cm]{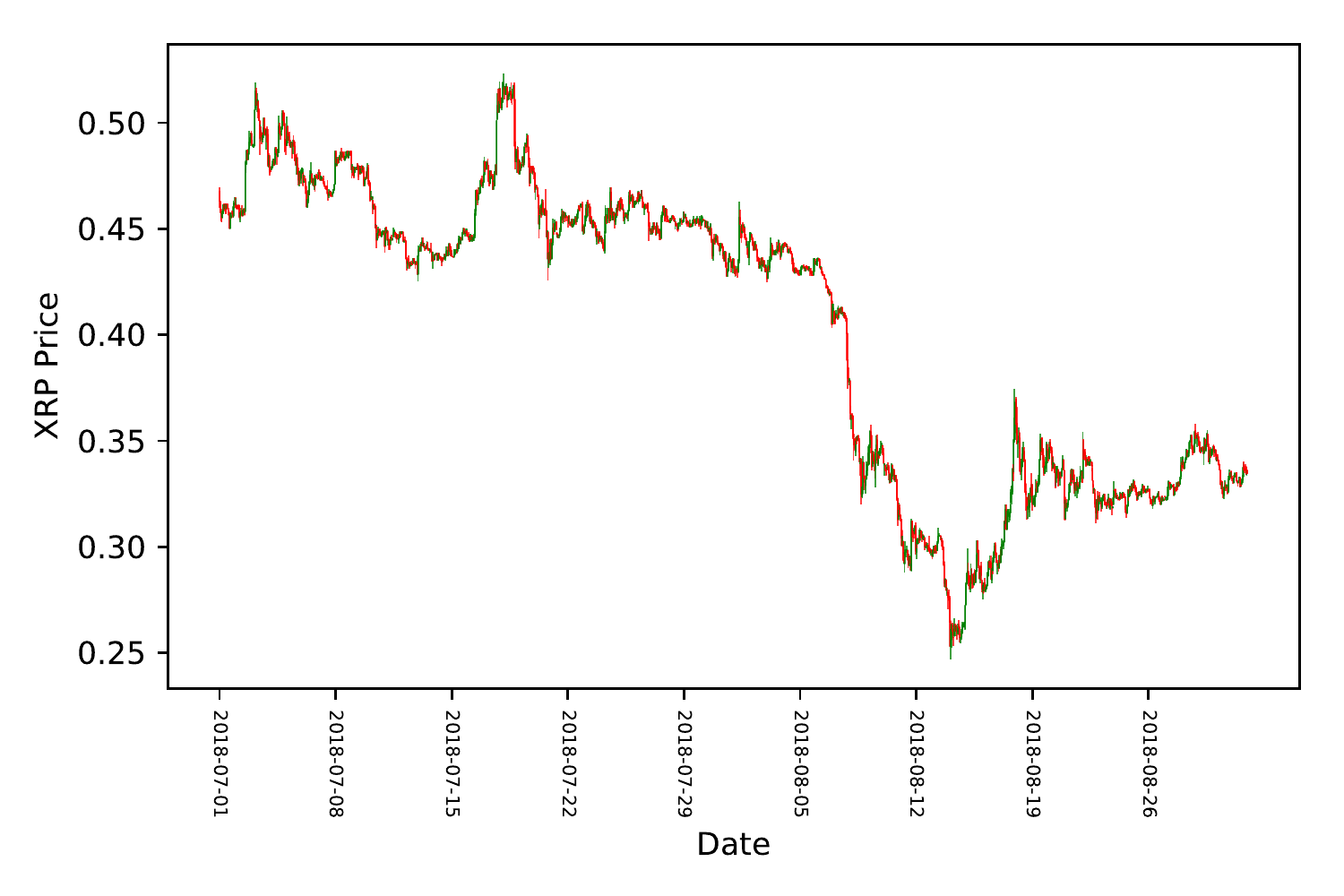}}
\hfill
\href{https://github.com/QuantLet/CCID/tree/master/CCIDCandles}{\includegraphics[keepaspectratio,width=0.4cm]{media/qletlogo_tr.png}}
\caption{Chandlestick charts for individual price movements. 01. July 2018 - 31. August 2018.}
\end{figure}

In contrast to the CRIX candlestick chart presented in figure 3, where five minute high-frequency data is aggregated to 60 minutes, we present respective individual plots for each examined cryptocurrency, as shown in figure 4 to give an easier entry to understand this volatile market. Consistency between Figure 1, 3, and 4 can be seen in the context of the findings in \citet{Romeo:2019}, where the impact of sentiment on cryptocurrency prices is evident \citep[with further references][]{Sonic:2019,Qian:2019}. Furthermore, when recurring to the observable price structure of the Flash Crash of 2010 as well, we can see quite many jumps in these figures - a phenomenon also described in \citet{Junjie:2019} and \citet{Qian:2019}.\\

Figure 5, shows the intraday 5-minutes returns for the period from the 01. July 2018 to the 31. August 2018. As indicated, overall returns across the board are very extreme - a phenomenon generally unknown to classic financial markets. In addition, we can observe an extreme activity cluster around the second half of August. We can link this activity to increased media outlets regarding cryptocurrencies: the more investors flooded into this market, the higher the trading activity, fueled by sentiment, became - leading to partially absurd returns; positive as well as negative.

\begin{figure}[H]
\hfill
\subfigure[BTC]{\includegraphics[width=5cm]{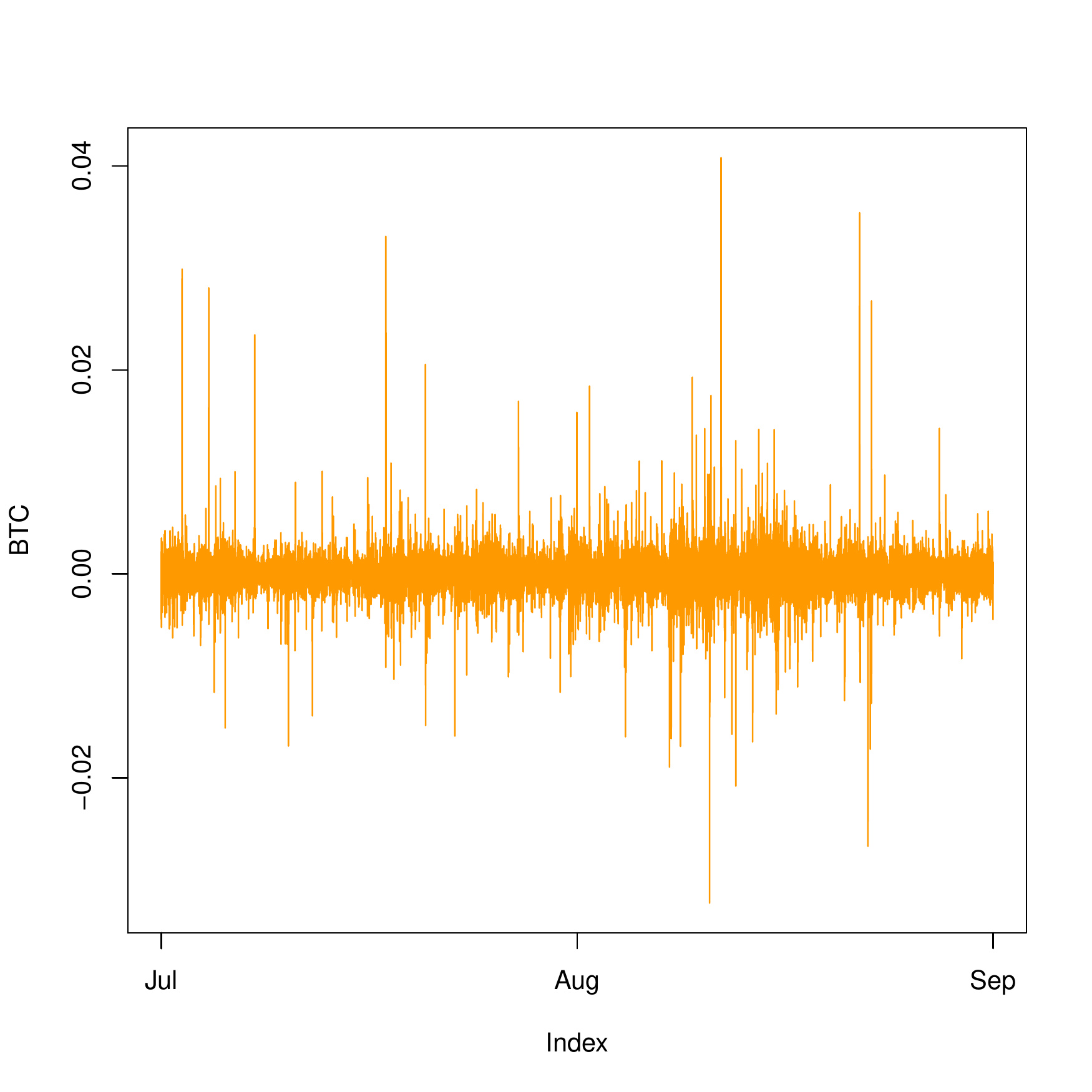}}
\hfill
\subfigure[ETH]{\includegraphics[width=5cm]{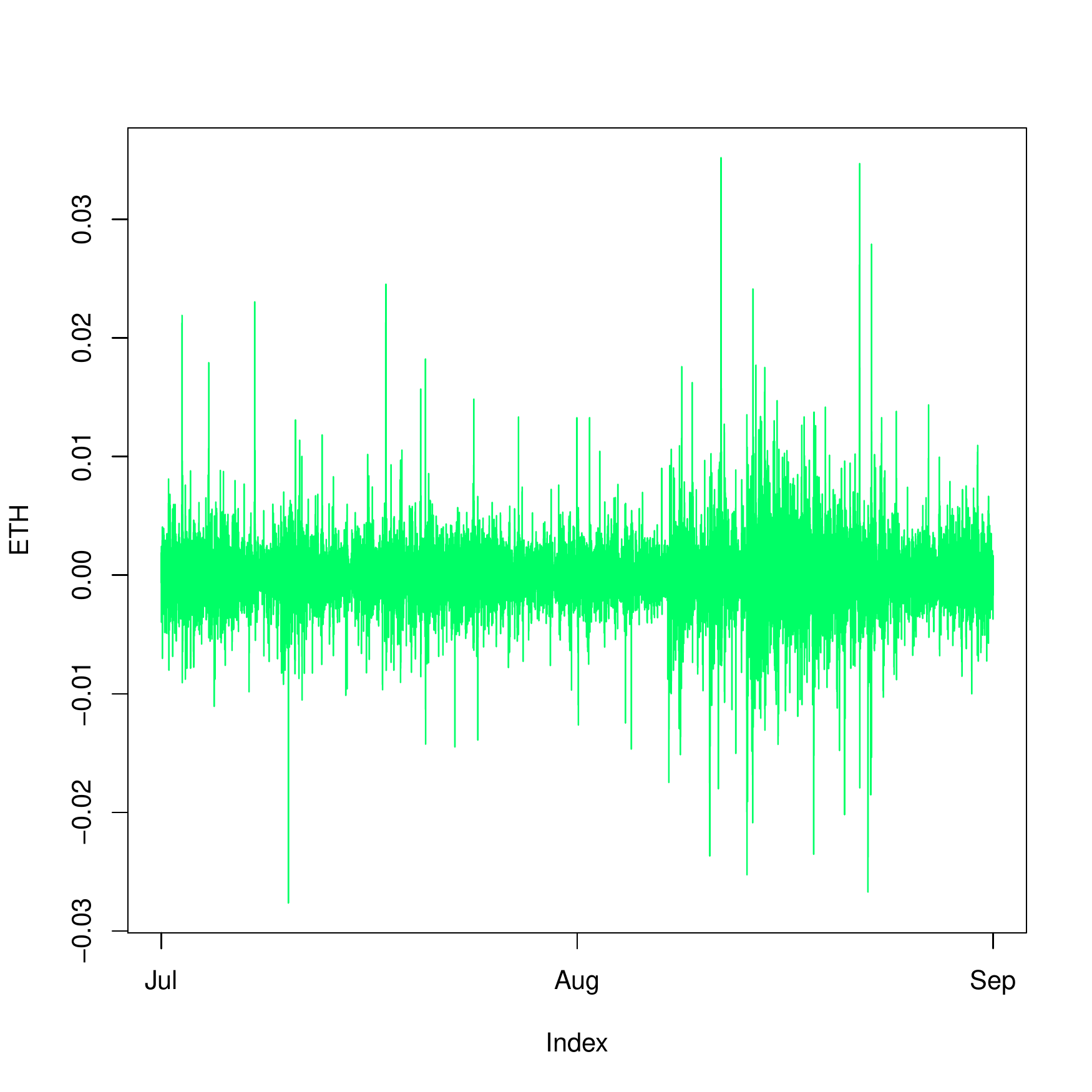}}
\hfill
\subfigure[XRP]{\includegraphics[width=5cm]{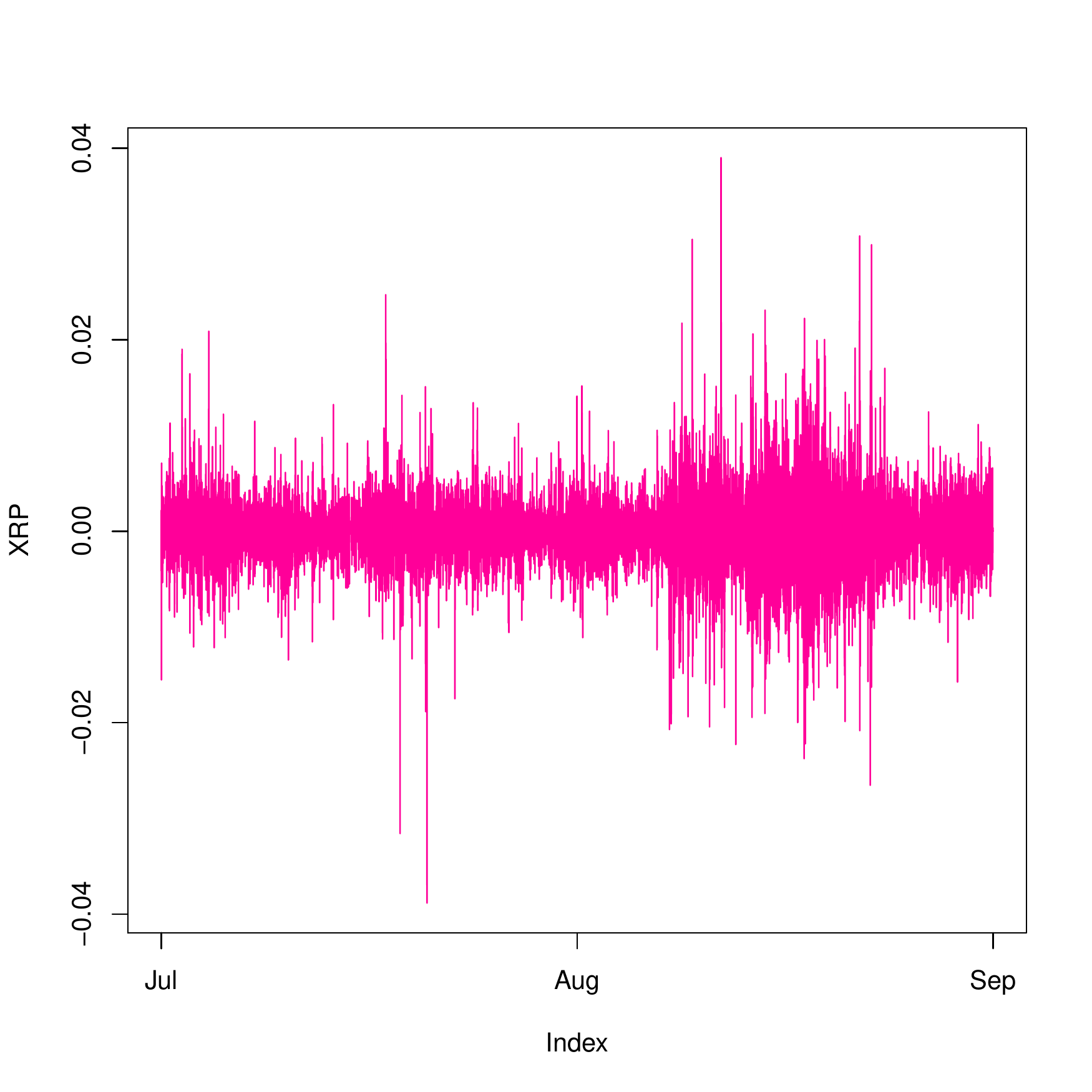}}
\hfill
\href{https://github.com/QuantLet/CCID/tree/master/CCIDHistRet}{\includegraphics[keepaspectratio,width=0.4cm]{media/qletlogo_tr.png}}
\caption{Intraday Returns (5 minutes). 01. July 2018 - 31. August 2018.}
\end{figure}

\vspace{-0.6cm}

\begin{figure}[H]
\hfill
\subfigure[BTC]{\includegraphics[width=5cm]{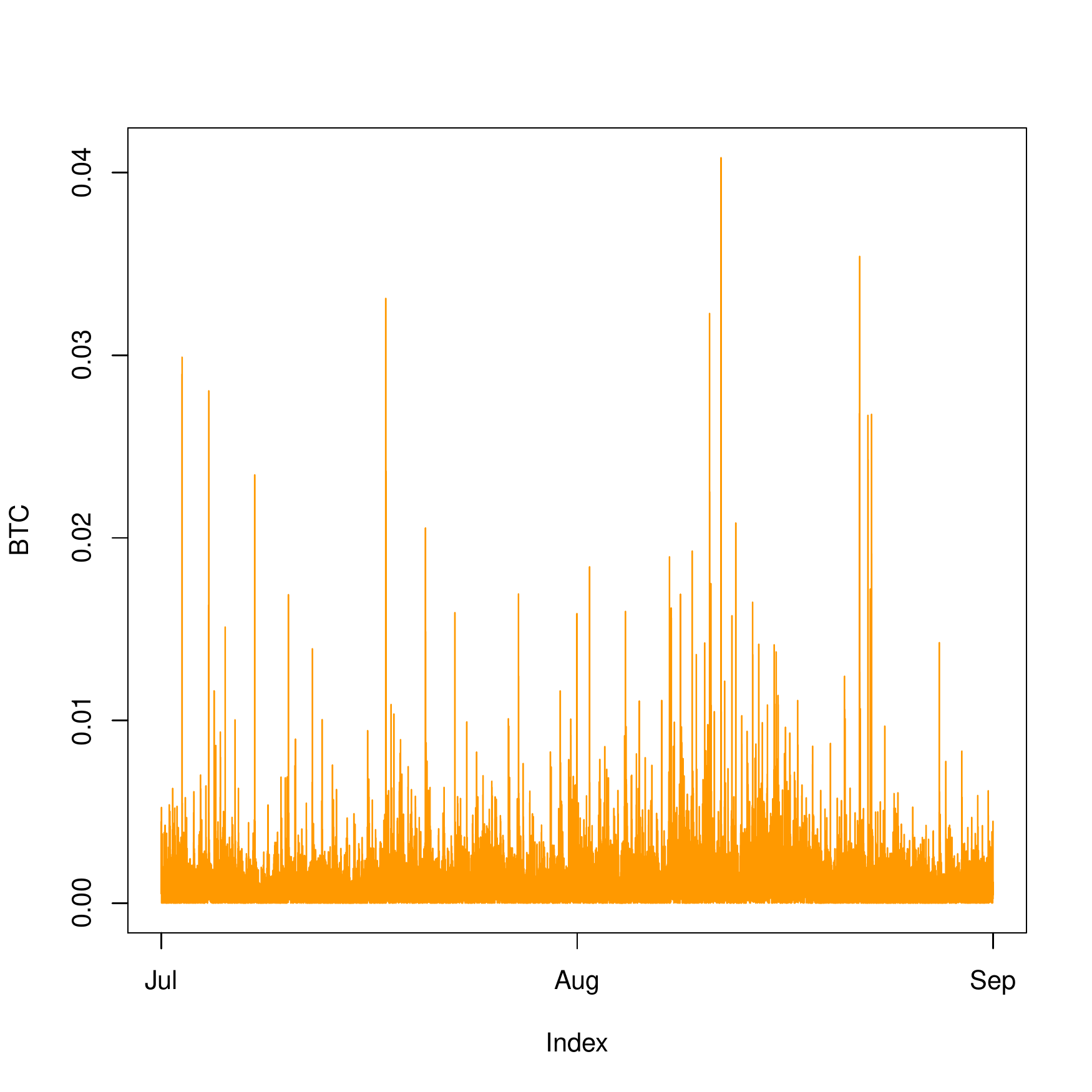}}
\hfill
\subfigure[ETH]{\includegraphics[width=5cm]{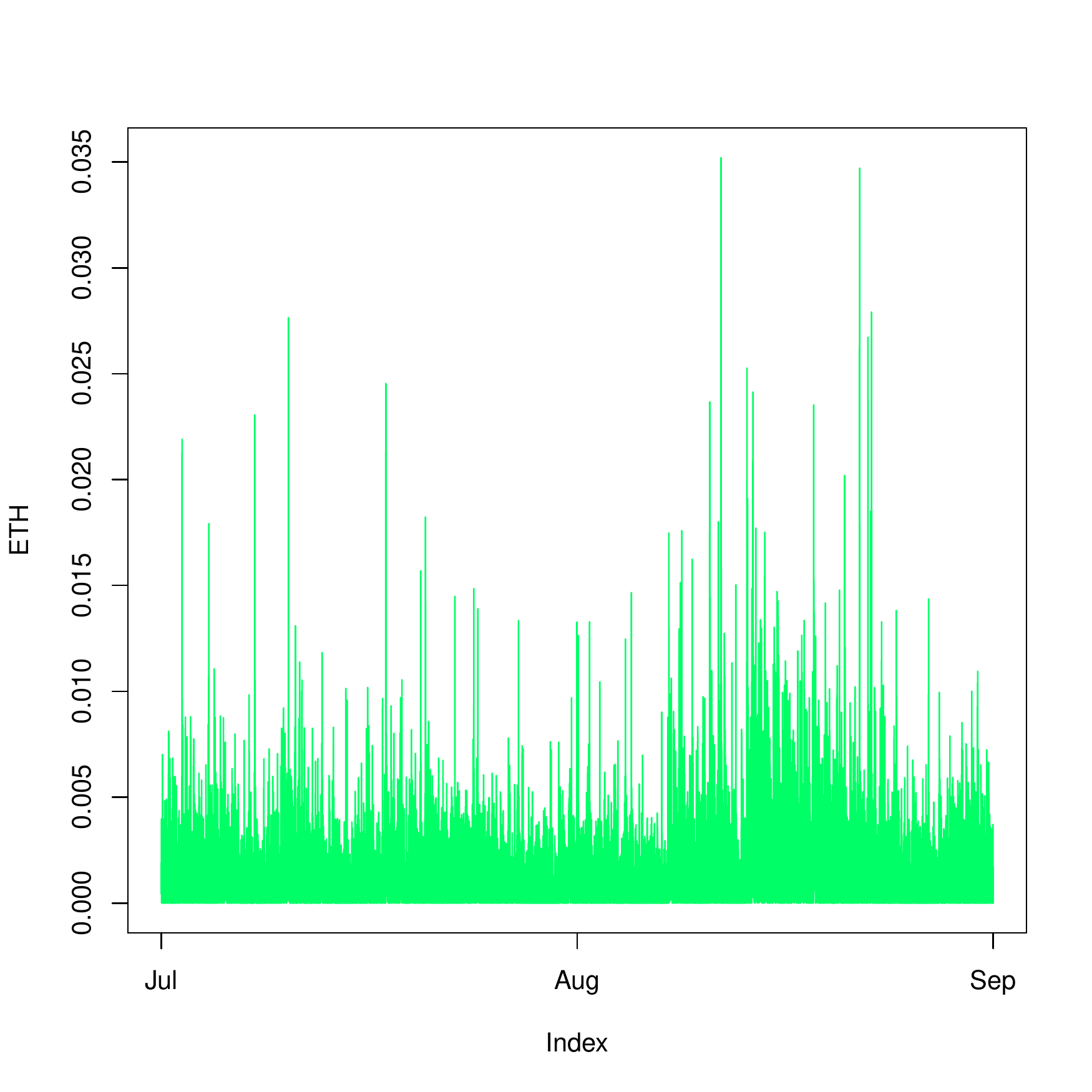}}
\hfill
\subfigure[XRP]{\includegraphics[width=5cm]{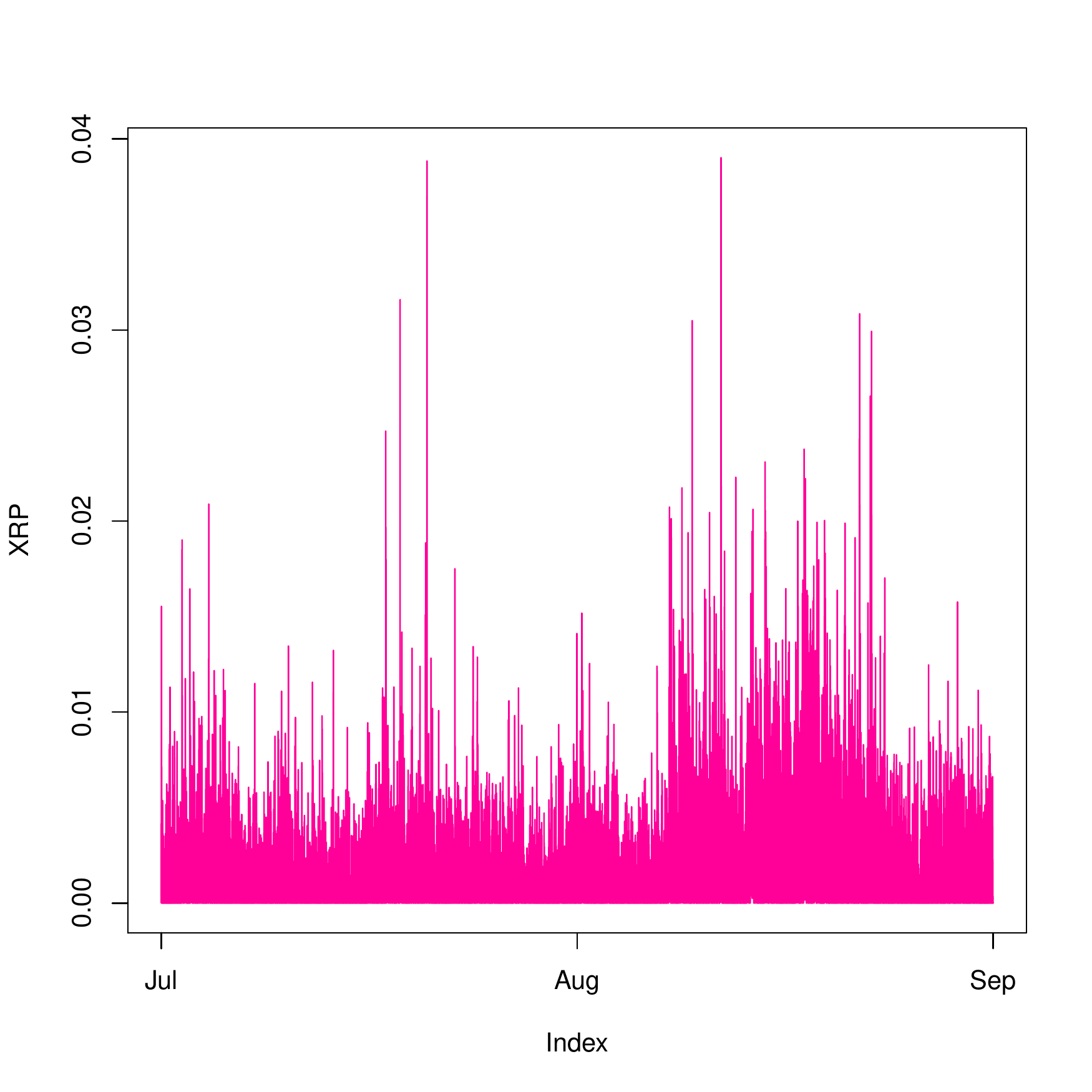}}
\hfill
\href{https://github.com/QuantLet/CCID/tree/master/CCIDHistVola}{\includegraphics[keepaspectratio,width=0.4cm]{media/qletlogo_tr.png}}
\caption{Intraday Volatility. 01. July 2018 - 31. August 2018.}
\end{figure}

Figure 6 adds to this finding, presenting the overall volatility from the beforehand stated period. As we can see, the return activity cluster in August from figure 5 is mirrored in the volatility activity cluster in figure 4. Hence, we proof the beforehand stated claim of cryptocurrency activity being fueled by media outlets as well as sentiment, as being attested. \\

Table 1 displays the estimated values of selected parameters for the cryptocurrency intraday trading for the given period of the 01. July 2018 to the 31. August 2018. The largest autocorrelation is for DASH (0.01), the smallest autocorrelation is for STR (-0.09).

\begin{table}[H]
\centering
\caption{Estimated first-order autocorrelation of the returns, $\widehat{\rho}\textsubscript{1}(ret\textsubscript{t})$, the squared returns, $\widehat{\rho}\textsubscript{1}(ret_{t}^{2})$, and the absolute returns, $\widehat{\rho}\textsubscript{1}(\vert ret\textsubscript{t} \vert)$, as well as the estimated skewness, $\widehat{S}$, the estimated excess kurtosis, $\widehat{e. Kurt}$, and the Jarque-Bera test statistic, JB, with the respective, obviously very small, p-value for the overall summed intraday high-frequency data from the 01. July 2018 to the 31. August 2018. }
\vspace{0.5cm}
\begin{tabular}{lrrrrrrrr}
\hline
\hline
  \toprule
& $\widehat{\rho}\textsubscript{1}(ret\textsubscript{t})$ & $\widehat{\rho}\textsubscript{1}(ret_{t}^{2})$ & $\widehat{\rho}\textsubscript{1}(\vert ret\textsubscript{t} \vert)$ & $\widehat{S} $& $\widehat{e. Kurt}$ & JB & JB p-value \\ 
  \midrule
  \hline
BCH & -0.01 & 0.12 & 0.20 & 0.49 & 13.69 & 140148.24 & 0.00 \\ 
BTC & -0.05 & 0.13 & 0.24 & 1.30 & 49.44 & 1823779.80 & 0.00 \\ 
DASH &  0.01 & 0.17 & 0.20 & 0.73 & 28.98 & 626596.64 & 0.00 \\ 
ETC &  -0.06 & 0.26 & 0.26 & 0.70 & 26.07 & 507374.39 & 0.00 \\ 
ETH &  -0.01 & 0.18 & 0.27 & 0.17 & 16.34 & 198777.58 & 0.00 \\ 
LTC &  -0.01 & 0.11 & 0.19 & 0.44 & 14.91 & 166121.81 & 0.00 \\ 
REP &  -0.08 & 0.22 & 0.19 & 0.35 & 21.89 & 356937.91 & 0.00 \\ 
STR &  -0.09 & 0.12 & 0.18 & 0.28 & 8.12 & 49354.96 & 0.00 \\ 
XMR &  -0.07 & 0.13 & 0.14 & 0.03 & 10.51 & 82241.48 & 0.00 \\ 
XRP &  -0.05 & 0.17 & 0.25 & 0.11 & 11.44 & 97390.58 & 0.00 \\ 
ZEC &  -0.07 & 0.25 & 0.22 & 1.30 & 26.66 & 534032.89 & 0.00 \\ 
   \bottomrule
\hline
\hline
&&&&&&&\href{https://github.com/QuantLet/CCID/tree/master/CCIDReturns}{\includegraphics[keepaspectratio,width=0.4cm]{media/qletlogo_tr.png}}
\end{tabular}
\end{table}

While the first-order autocorrelation of the returns of all cryptocurrencies is all close to zero and mostly negative, the autocorrelations of the squared and absolute returns of all cryptocurrencies are positive and significantly larger than zero. Obviously, there is a linear relationship in the absolute and squared values of the chronologically sequential returns. Since the autocorrelation is positive, it can be concluded, that small absolute returns are followed sequentially by small absolute returns and large absolute returns are followed by large ones again. This means, that there are quiet periods with small price changes and dynamic periods with large oscillations.\\

Furthermore, whereas the estimate for skewness is mostly close to zero, except for BTC and ZEC, the estimate for excess kurtosis is in every case significantly larger than 3. The smallest estimated excess kurtosis is by STR (yet with an expressive $\widehat{e. Kurt}$ of 8.12), and the largest by BTC ($\widehat{e. Kurt}$ = 49.44). These values show, that the tested constituents are far from normally distributed. Negative skewness signals about increasing the downside risk and is a consequence of asymmetric volatility models. Positively skewed distributions have a longer right tail, meaning for investors a greater chance of extremely positive outcomes.  A well-known stylized fact about returns distributions highlights their leptokurtic nature: they have more mass around the centre and in the tails than a normal distribution. For example, \citet{Hussein:2011} reports relatively high levels of kurtosis in stock data from the United States of America. This phenomenon is known as kurtosis risk.\\

The combined test of the normal distribution from Jarque and Bera (JB) can be derived as asymptotically $\chi^2$ distribution with two degrees of freedom. The last column in table 1 shows, that in all cases the normal distribution hypothesis is clearly rejected. This is above all caused by the value of kurtosis, which is significantly larger than 3, caused by a very frequent appearance of outliers in this new market. The higher kurtosis, compared to a normal distribution, proves that these extreme points result in $leptokurtic$ distributions and are evidence of fat tails relative to the normal distribution's tail. However, as this asymmetry is common to financial markets, it is especially strong in the cryptocurrency markets with potentially extreme returns and a very pronounced volatility.\\

The following tables respectively show the individual correlation to CRIX, if the market is acting positively, table 2, or negatively, table 3. Extensive care should be put on our main actors - BTC, ETH and XRP - when studying these. As these enjoy a large market acceptance and hence are long-term drivers of the cryptocurrency market, we can once again, underline our findings given beforehand.\\

On a side note, tables 2 and 3 show that among the top 11 cryptocurrencies, most pairs exhibit low return correlations, what suggest strong diversification benefits in a portfolio, especially outside the major cryptocurrencies presented, see also \citep{PetTriHaeEle:2019}.\\

We can observe, that the correlation to CRIX in both tables presents itself as clustered around well-known cryptocurrencies, namely BTC, ETH, XRP, as well as BCH, and ETC. Therefore, this activity can be interpreted in a way, which indicates these constituents as the market drivers. This finding also correlates with the long term trading activity registered on many online sources for these coins. We should note, without going into detail, that LTC and BCH are closely related to BTC, and that ETC is closely tied to the history of ETH. XRP itself was able to carve out its very specific niche early enough for certain applications, especially in the banking sector - in contrast, BTC can be seen as the genesis of digital currency without any intrinsic value, whereas the ETH system enables many different applications, majorly through so-called ``smart contracts".

\begin{table}[H]
\centering
\caption{Pairwise crypto-currency correlations of returns for positive market-movement days, as defined by returns on CRIX. 01. July 2018 - 31. August 2018.}
\vspace{0.5cm}
\begin{tabular}{lrrrrrrrrrrr}
\hline
\hline
\textbf{UP}   & BCH                          & BTC                          & DASH                         & ETC                          & ETH                          & LTC                          & REP                          & STR                          & XMR                          & XRP                          & ZEC                          \\ \hline
BCH  &                              & \cellcolor[HTML]{FD6864}0.50 & \cellcolor[HTML]{FFCCC9}0.23 & \cellcolor[HTML]{FD6864}0.33 & \cellcolor[HTML]{FD6864}0.47 & \cellcolor[HTML]{FD6864}0.46 & \cellcolor[HTML]{FFCCC9}0.13 & \cellcolor[HTML]{FFCCC9}0.29 & \cellcolor[HTML]{FFCCC9}0.25 & \cellcolor[HTML]{FD6864}0.37 & \cellcolor[HTML]{FFCCC9}0.23 \\
BTC  & \cellcolor[HTML]{FD6864}0.50 &                              & \cellcolor[HTML]{FFCCC9}0.27 & \cellcolor[HTML]{FD6864}0.36 & \cellcolor[HTML]{FD6864}0.55 & \cellcolor[HTML]{FD6864}0.49 & \cellcolor[HTML]{FFCCC9}0.18 & \cellcolor[HTML]{FD6864}0.34 & \cellcolor[HTML]{FD6864}0.30 & \cellcolor[HTML]{FD6864}0.40 & \cellcolor[HTML]{FFCCC9}0.27 \\
DASH & \cellcolor[HTML]{FFCCC9}0.23 & \cellcolor[HTML]{FFCCC9}0.27 &                              & \cellcolor[HTML]{FFCCC9}0.17 & \cellcolor[HTML]{FFCCC9}0.22 & \cellcolor[HTML]{FFCCC9}0.22 & \cellcolor[HTML]{FFCCC9}0.10 & \cellcolor[HTML]{FFCCC9}0.17 & \cellcolor[HTML]{FFCCC9}0.17 & \cellcolor[HTML]{FFCCC9}0.22 & \cellcolor[HTML]{FFCCC9}0.14 \\
ETC  & \cellcolor[HTML]{FD6864}0.33 & \cellcolor[HTML]{FD6864}0.36 & \cellcolor[HTML]{FFCCC9}0.17 &                              & \cellcolor[HTML]{FD6864}0.37 & \cellcolor[HTML]{FD6864}0.31 & \cellcolor[HTML]{FFCCC9}0.11 & \cellcolor[HTML]{FFCCC9}0.21 & \cellcolor[HTML]{FFCCC9}0.17 & \cellcolor[HTML]{FFCCC9}0.28 & \cellcolor[HTML]{FFCCC9}0.14 \\
ETH  & \cellcolor[HTML]{FD6864}0.47 & \cellcolor[HTML]{FD6864}0.55 & \cellcolor[HTML]{FFCCC9}0.22 & \cellcolor[HTML]{FD6864}0.37 &                              & \cellcolor[HTML]{FD6864}0.47 & \cellcolor[HTML]{FFCCC9}0.16 & \cellcolor[HTML]{FD6864}0.30 & \cellcolor[HTML]{FFCCC9}0.27 & \cellcolor[HTML]{FD6864}0.42 & \cellcolor[HTML]{FFCCC9}0.22 \\
LTC  & \cellcolor[HTML]{FD6864}0.46 & \cellcolor[HTML]{FD6864}0.49 & \cellcolor[HTML]{FFCCC9}0.22 & \cellcolor[HTML]{FD6864}0.31 & \cellcolor[HTML]{FD6864}0.47 &                              & \cellcolor[HTML]{FFCCC9}0.17 & \cellcolor[HTML]{FFCCC9}0.26 & \cellcolor[HTML]{FFCCC9}0.25 & \cellcolor[HTML]{FD6864}0.39 & \cellcolor[HTML]{FFCCC9}0.23 \\
REP  & \cellcolor[HTML]{FFCCC9}0.13 & \cellcolor[HTML]{FFCCC9}0.18 & \cellcolor[HTML]{FFCCC9}0.10 & \cellcolor[HTML]{FFCCC9}0.11 & \cellcolor[HTML]{FFCCC9}0.16 & \cellcolor[HTML]{FFCCC9}0.17 &                              & \cellcolor[HTML]{FFCCC9}0.12 & \cellcolor[HTML]{FFCCC9}0.11 & \cellcolor[HTML]{FFCCC9}0.11 & \cellcolor[HTML]{FFCCC9}0.11 \\
STR  & \cellcolor[HTML]{FFCCC9}0.29 & \cellcolor[HTML]{FD6864}0.34 & \cellcolor[HTML]{FFCCC9}0.17 & \cellcolor[HTML]{FFCCC9}0.21 & \cellcolor[HTML]{FD6864}0.30 & \cellcolor[HTML]{FFCCC9}0.26 & \cellcolor[HTML]{FFCCC9}0.12 &                              & \cellcolor[HTML]{FFCCC9}0.18 & \cellcolor[HTML]{FFCCC9}0.27 & \cellcolor[HTML]{FFCCC9}0.19 \\
XMR  & \cellcolor[HTML]{FFCCC9}0.25 & \cellcolor[HTML]{FD6864}0.30 & \cellcolor[HTML]{FFCCC9}0.17 & \cellcolor[HTML]{FFCCC9}0.17 & \cellcolor[HTML]{FFCCC9}0.27 & \cellcolor[HTML]{FFCCC9}0.25 & \cellcolor[HTML]{FFCCC9}0.11 & \cellcolor[HTML]{FFCCC9}0.18 &                              & \cellcolor[HTML]{FFCCC9}0.20 & \cellcolor[HTML]{FFCCC9}0.15 \\
XRP  & \cellcolor[HTML]{FD6864}0.37 & \cellcolor[HTML]{FD6864}0.40 & \cellcolor[HTML]{FFCCC9}0.22 & \cellcolor[HTML]{FFCCC9}0.28 & \cellcolor[HTML]{FD6864}0.42 & \cellcolor[HTML]{FD6864}0.39 & \cellcolor[HTML]{FFCCC9}0.11 & \cellcolor[HTML]{FFCCC9}0.27 & \cellcolor[HTML]{FFCCC9}0.20 &                              & \cellcolor[HTML]{FFCCC9}0.19 \\
ZEC  & \cellcolor[HTML]{FFCCC9}0.23 & \cellcolor[HTML]{FFCCC9}0.27 & \cellcolor[HTML]{FFCCC9}0.14 & \cellcolor[HTML]{FFCCC9}0.14 & \cellcolor[HTML]{FFCCC9}0.22 & \cellcolor[HTML]{FFCCC9}0.23 & \cellcolor[HTML]{FFCCC9}0.11 & \cellcolor[HTML]{FFCCC9}0.19 & \cellcolor[HTML]{FFCCC9}0.15 & \cellcolor[HTML]{FFCCC9}0.19 &                              \\ 
\hline
\hline
&&&&&&&&&&&\href{https://github.com/QuantLet/CCID/tree/master/CCIDcorr}{\includegraphics[keepaspectratio,width=0.4cm]{media/qletlogo_tr.png}}
\end{tabular}
\end{table}

\vspace{-0.5cm}

\begin{table}[H]
\centering
\caption{Pairwise crypto-currency correlations of returns for negative market-movement days, as defined by returns on CRIX. 01. July 2018 - 31. August 2018.}
\vspace{0.5cm}
\begin{tabular}{lrrrrrrrrrrr}
\hline
\hline
\textbf{DOWN} & BCH                          & BTC                          & DASH                         & ETC                          & ETH                          & LTC                          & REP                          & STR                          & XMR                          & XRP                          & ZEC                          \\ \hline
BCH  &                              & \cellcolor[HTML]{FD6864}0.48 & \cellcolor[HTML]{FFCCC9}0.21 & \cellcolor[HTML]{FD6864}0.32 & \cellcolor[HTML]{FD6864}0.47 & \cellcolor[HTML]{FD6864}0.43 & \cellcolor[HTML]{FFCCC9}0.15 & \cellcolor[HTML]{FFCCC9}0.27 & \cellcolor[HTML]{FFCCC9}0.23 & \cellcolor[HTML]{FD6864}0.37 & \cellcolor[HTML]{FFCCC9}0.22 \\
BTC  & \cellcolor[HTML]{FD6864}0.48 &                              & \cellcolor[HTML]{FFCCC9}0.26 & \cellcolor[HTML]{FD6864}0.36 & \cellcolor[HTML]{FD6864}0.52 & \cellcolor[HTML]{FD6864}0.45 & \cellcolor[HTML]{FFCCC9}0.19 & \cellcolor[HTML]{FD6864}0.33 & \cellcolor[HTML]{FD6864}0.30 & \cellcolor[HTML]{FD6864}0.41 & \cellcolor[HTML]{FFCCC9}0.24 \\
DASH & \cellcolor[HTML]{FFCCC9}0.21 & \cellcolor[HTML]{FFCCC9}0.26 &                              & \cellcolor[HTML]{FFCCC9}0.15 & \cellcolor[HTML]{FFCCC9}0.22 & \cellcolor[HTML]{FFCCC9}0.21 & \cellcolor[HTML]{FFCCC9}0.11 & \cellcolor[HTML]{FFCCC9}0.16 & \cellcolor[HTML]{FFCCC9}0.18 & \cellcolor[HTML]{FFCCC9}0.18 & \cellcolor[HTML]{FFCCC9}0.14 \\
ETC  & \cellcolor[HTML]{FD6864}0.32 & \cellcolor[HTML]{FD6864}0.36 & \cellcolor[HTML]{FFCCC9}0.15 &                              & \cellcolor[HTML]{FD6864}0.36 & \cellcolor[HTML]{FD6864}0.30 & \cellcolor[HTML]{FFCCC9}0.14 & \cellcolor[HTML]{FFCCC9}0.21 & \cellcolor[HTML]{FFCCC9}0.18 & \cellcolor[HTML]{FD6864}0.30 & \cellcolor[HTML]{FFCCC9}0.16 \\
ETH  & \cellcolor[HTML]{FD6864}0.47 & \cellcolor[HTML]{FD6864}0.52 & \cellcolor[HTML]{FFCCC9}0.22 & \cellcolor[HTML]{FD6864}0.36 &                              & \cellcolor[HTML]{FD6864}0.42 & \cellcolor[HTML]{FFCCC9}0.16 & \cellcolor[HTML]{FFCCC9}0.29 & \cellcolor[HTML]{FFCCC9}0.23 & \cellcolor[HTML]{FD6864}0.40 & \cellcolor[HTML]{FFCCC9}0.21 \\
LTC  & \cellcolor[HTML]{FD6864}0.43 & \cellcolor[HTML]{FD6864}0.45 & \cellcolor[HTML]{FFCCC9}0.21 & \cellcolor[HTML]{FD6864}0.30 & \cellcolor[HTML]{FD6864}0.42 &                              & \cellcolor[HTML]{FFCCC9}0.16 & \cellcolor[HTML]{FFCCC9}0.26 & \cellcolor[HTML]{FFCCC9}0.24 & \cellcolor[HTML]{FD6864}0.35 & \cellcolor[HTML]{FFCCC9}0.19 \\
REP  & \cellcolor[HTML]{FFCCC9}0.15 & \cellcolor[HTML]{FFCCC9}0.19 & \cellcolor[HTML]{FFCCC9}0.11 & \cellcolor[HTML]{FFCCC9}0.14 & \cellcolor[HTML]{FFCCC9}0.16 & \cellcolor[HTML]{FFCCC9}0.16 &                              & \cellcolor[HTML]{FFCCC9}0.11 & \cellcolor[HTML]{FFCCC9}0.12 & \cellcolor[HTML]{FFCCC9}0.13 & \cellcolor[HTML]{FFCCC9}0.08 \\
STR  & \cellcolor[HTML]{FFCCC9}0.27 & \cellcolor[HTML]{FD6864}0.33 & \cellcolor[HTML]{FFCCC9}0.16 & \cellcolor[HTML]{FFCCC9}0.21 & \cellcolor[HTML]{FFCCC9}0.29 & \cellcolor[HTML]{FFCCC9}0.26 & \cellcolor[HTML]{FFCCC9}0.11 &                              & \cellcolor[HTML]{FFCCC9}0.16 & \cellcolor[HTML]{FFCCC9}0.26 & \cellcolor[HTML]{FFCCC9}0.16 \\
XMR  & \cellcolor[HTML]{FFCCC9}0.23 & \cellcolor[HTML]{FD6864}0.30 & \cellcolor[HTML]{FFCCC9}0.18 & \cellcolor[HTML]{FFCCC9}0.18 & \cellcolor[HTML]{FFCCC9}0.23 & \cellcolor[HTML]{FFCCC9}0.24 & \cellcolor[HTML]{FFCCC9}0.12 & \cellcolor[HTML]{FFCCC9}0.16 &                              & \cellcolor[HTML]{FFCCC9}0.20 & \cellcolor[HTML]{FFCCC9}0.15 \\
XRP  & \cellcolor[HTML]{FD6864}0.37 & \cellcolor[HTML]{FD6864}0.41 & \cellcolor[HTML]{FFCCC9}0.18 & \cellcolor[HTML]{FD6864}0.30 & \cellcolor[HTML]{FD6864}0.40 & \cellcolor[HTML]{FD6864}0.35 & \cellcolor[HTML]{FFCCC9}0.13 & \cellcolor[HTML]{FFCCC9}0.26 & \cellcolor[HTML]{FFCCC9}0.20 &                              & \cellcolor[HTML]{FFCCC9}0.17 \\
ZEC  & \cellcolor[HTML]{FFCCC9}0.22 & \cellcolor[HTML]{FFCCC9}0.24 & \cellcolor[HTML]{FFCCC9}0.14 & \cellcolor[HTML]{FFCCC9}0.16 & \cellcolor[HTML]{FFCCC9}0.21 & \cellcolor[HTML]{FFCCC9}0.19 & \cellcolor[HTML]{FFCCC9}0.08 & \cellcolor[HTML]{FFCCC9}0.16 & \cellcolor[HTML]{FFCCC9}0.15 & \cellcolor[HTML]{FFCCC9}0.17 &                              \\
\hline
\hline
&&&&&&&&&&&\href{https://github.com/QuantLet/CCID/tree/master/CCIDcorr}{\includegraphics[keepaspectratio,width=0.4cm]{media/qletlogo_tr.png}}
\end{tabular}
\end{table}

\newpage

\subsection{Time-Of-Day Effects and Proof-Of-Human}
\label{sec3.3}

To support our hypothesis of mostly dealing with human agent initiated trades, which we coin as PoH, we present our findings regarding the time-of-day trading in this section.  Additional material on information arrival, news sentiment, volatilities and jumps of intraday returns can also be taken from \citet{Qian:2019}.\\

Cryptocurrency exchanges, as introduced in \hyperref[sec2]{section 2}, are often designed to serve a certain target group, for example by emphasizing compliance with national regulatory frameworks. By plotting the trade volume against the timestamps, we can also observe certain properties of market activity and draw coherent conclusions to the origin of the market participants: are these mostly human, who are doing trades by hand, or are we looking at a well oiled automatic machinery full of algorithms - just as commonly portrayed. Keep in mind, as mentioned in \hyperref[sec2]{section 2}, that our data is gained from Europe-based sources, and taken from periods that are overwhelmingly identifiable by corporate staff vacations. One should hence expect a less pronounced human, but algorithmic driven market behaviour to contradict our hypotheses.\\

To underline this argument, it is useful to imagine a transitional system, whereas human interference is completely removed or not relevant to a market system \citep[e.g.][]{Capo:16}, and where the trading pattern will, therefore, be independent of the time-of-day effects:\vspace{0.02cm}

\begin{flushleft}
$human + human + human \: \widehat{=} \: human \: driven \: network$\newline
$human + algorithm + human \: \widehat{=} \: predominantly \: human \: driven \: network$\newline
$human + algorithm + algorithm \: \widehat{=} \: predominantly \: machine \: driven \: network$\newline
$algorithm + algorithm + algorithm \: \widehat{=} \: algorithmic \: driven \: network$\newline
\end{flushleft}

\vspace{-0.5cm}

With increasing market participation of algorithms, we expect, for example, nighttime to have a negligible impact on the market activity. In contrast, we expect nighttime to have an impact on market activity if the market is dominated by human interaction.\\

The following figures employ GAM to observe daily and weekly patterns for intraday volatility and trading volume.  For daily seasonality cubic regression splines, for weekly seasonality $P$-splines are used, and a number of knots are logically set to the number of unique values, i.e 62 for daily patterns and 7 for weekly. The summary statistics of GAM for all cryptocurrencies demonstrate a high significance of  smooth terms combined with a quite low explanatory power (coefficients of determination are around 1\%). Nevertheless, we can observe distinct intraday seasonality patterns.

\begin{figure}[H]
\hfill
\subfigure[BTC]{\includegraphics[width=5cm]{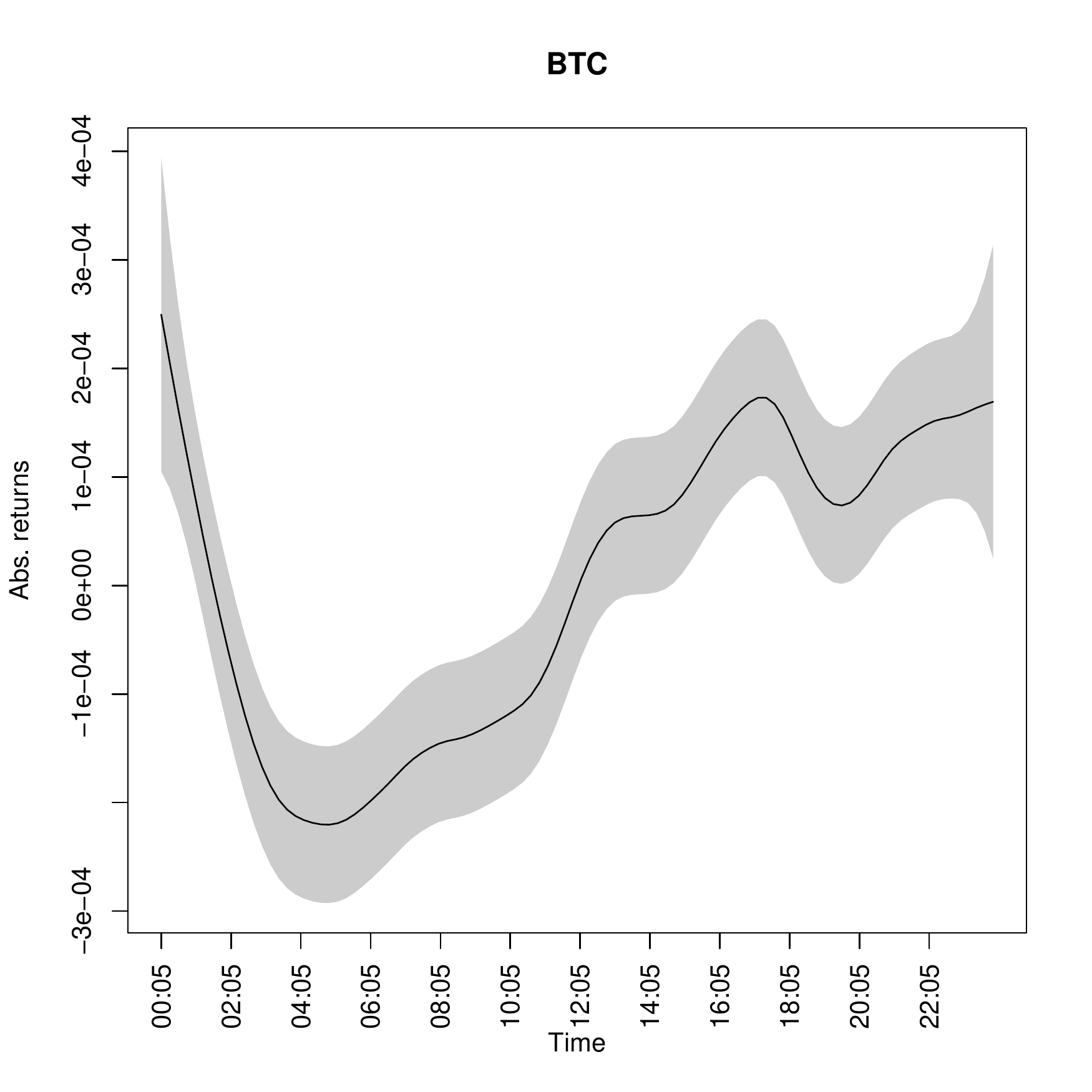}}
\hfill
\subfigure[ETH]{\includegraphics[width=5cm]{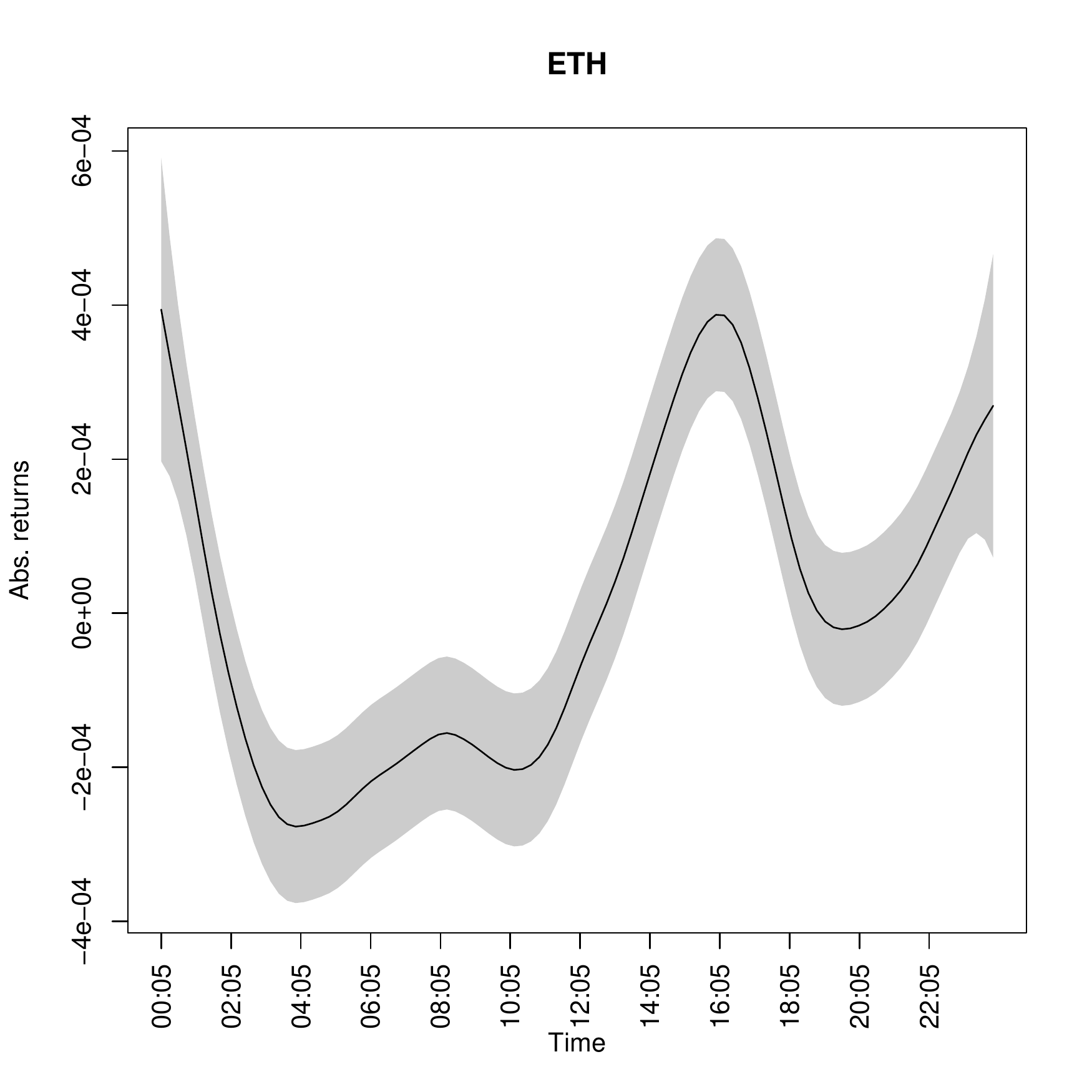}}
\hfill
\subfigure[XRP]{\includegraphics[width=5cm]{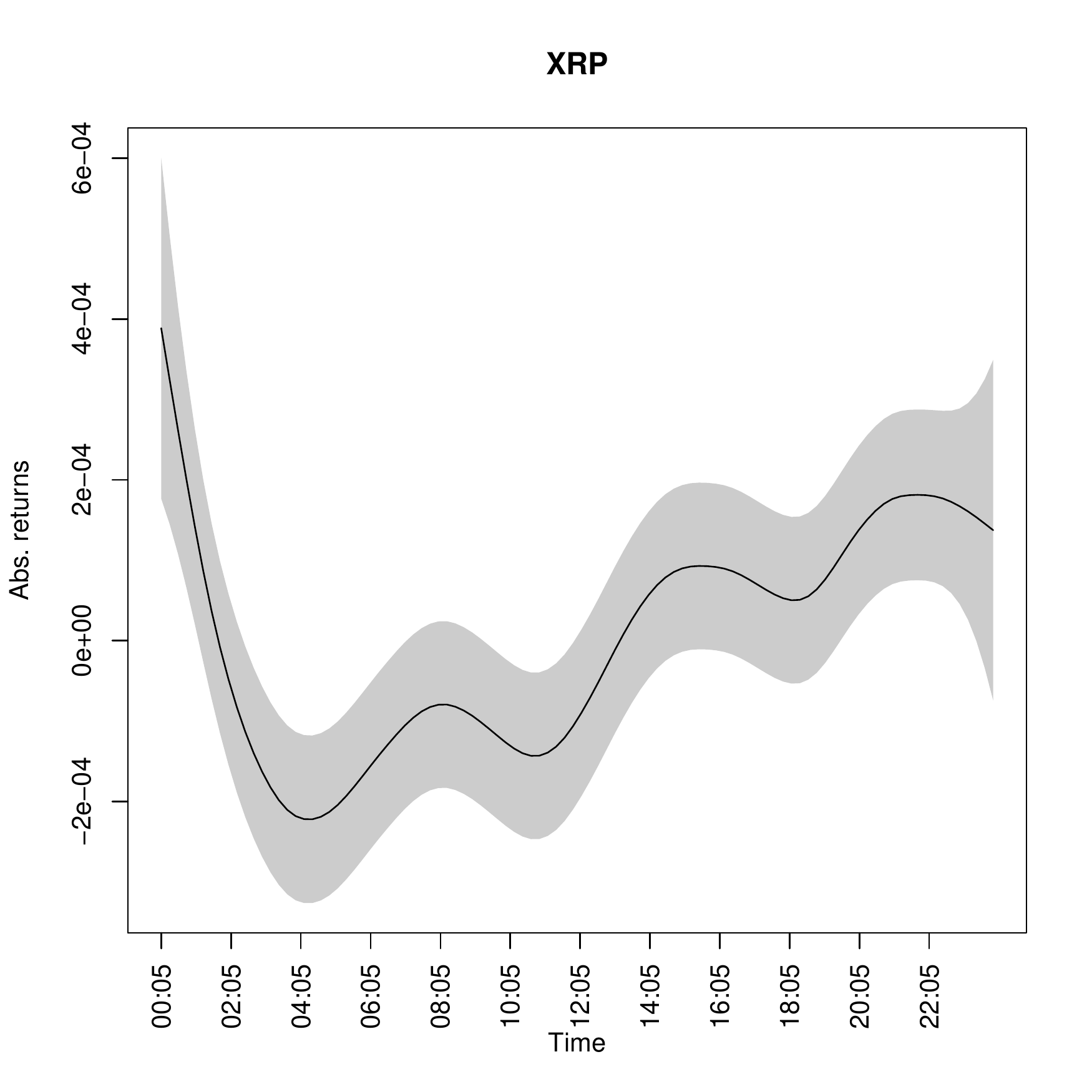}}
\hfill
\href{https://github.com/QuantLet/CCID/tree/master/CCIDvolaGAM}{\includegraphics[keepaspectratio,width=0.4cm]{media/qletlogo_tr.png}}
\caption{Daily seasonality: fit of Generalized Additive Model (5 min nodes)  with cubic regression splines for absolute returns of cryptocurrencies (shaded regions represent confidence bands for smooths), 01. July 2018 - 31. August 2018.}
\end{figure}

\vspace{-0.6cm}

\begin{figure}[H]
\hfill
\subfigure[BTC]{\includegraphics[width=5cm]{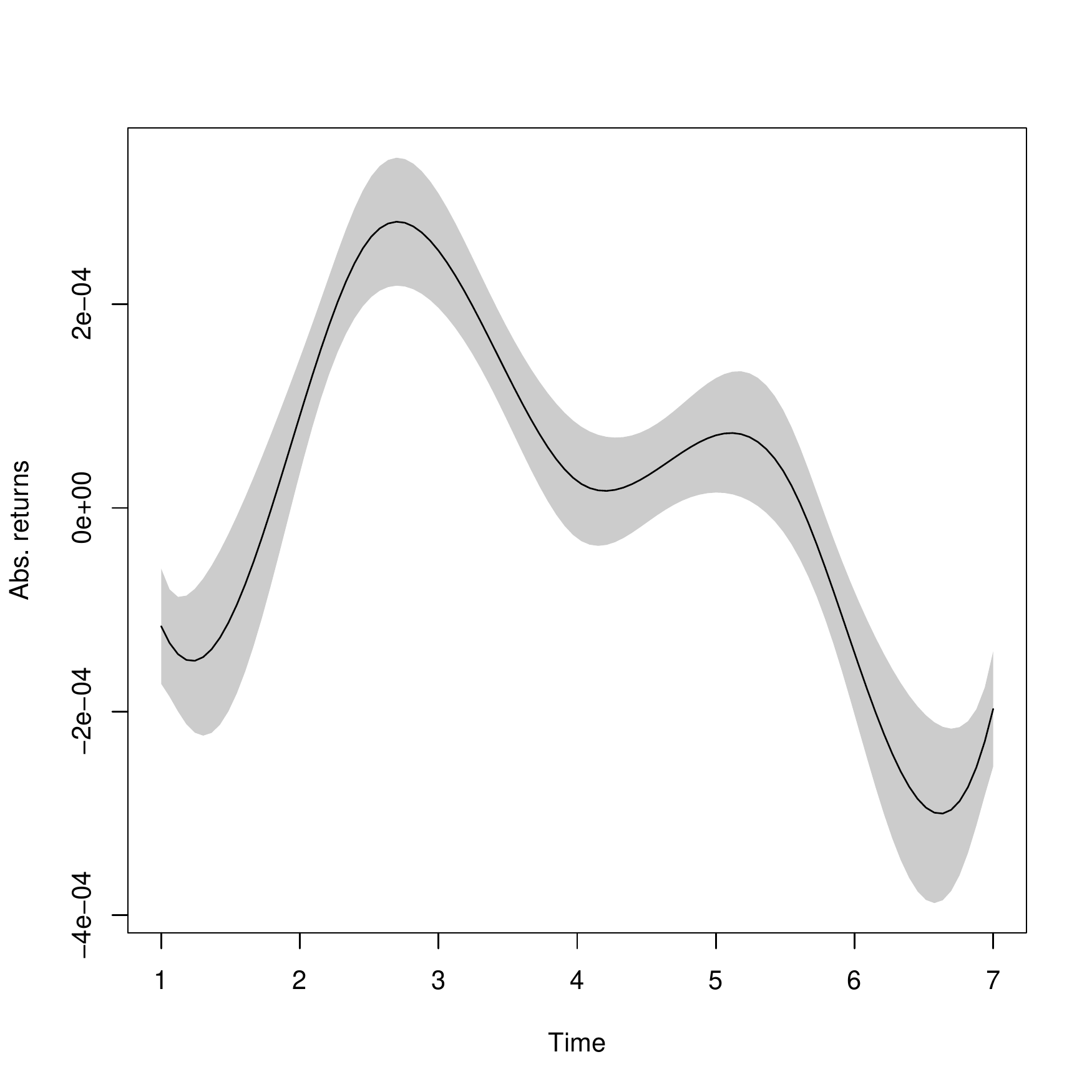}}
\hfill
\subfigure[ETH]{\includegraphics[width=5cm]{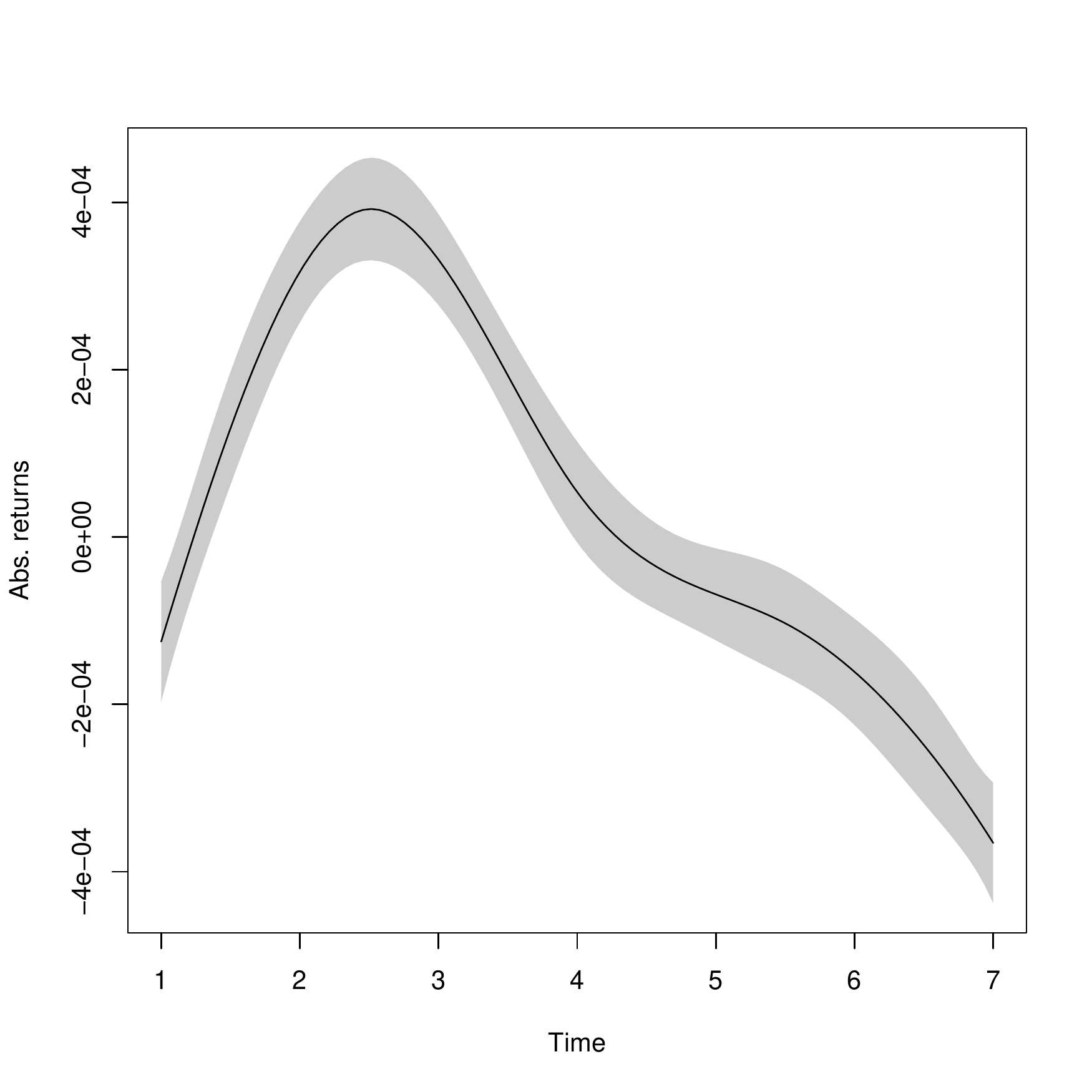}}
\hfill
\subfigure[XRP]{\includegraphics[width=5cm]{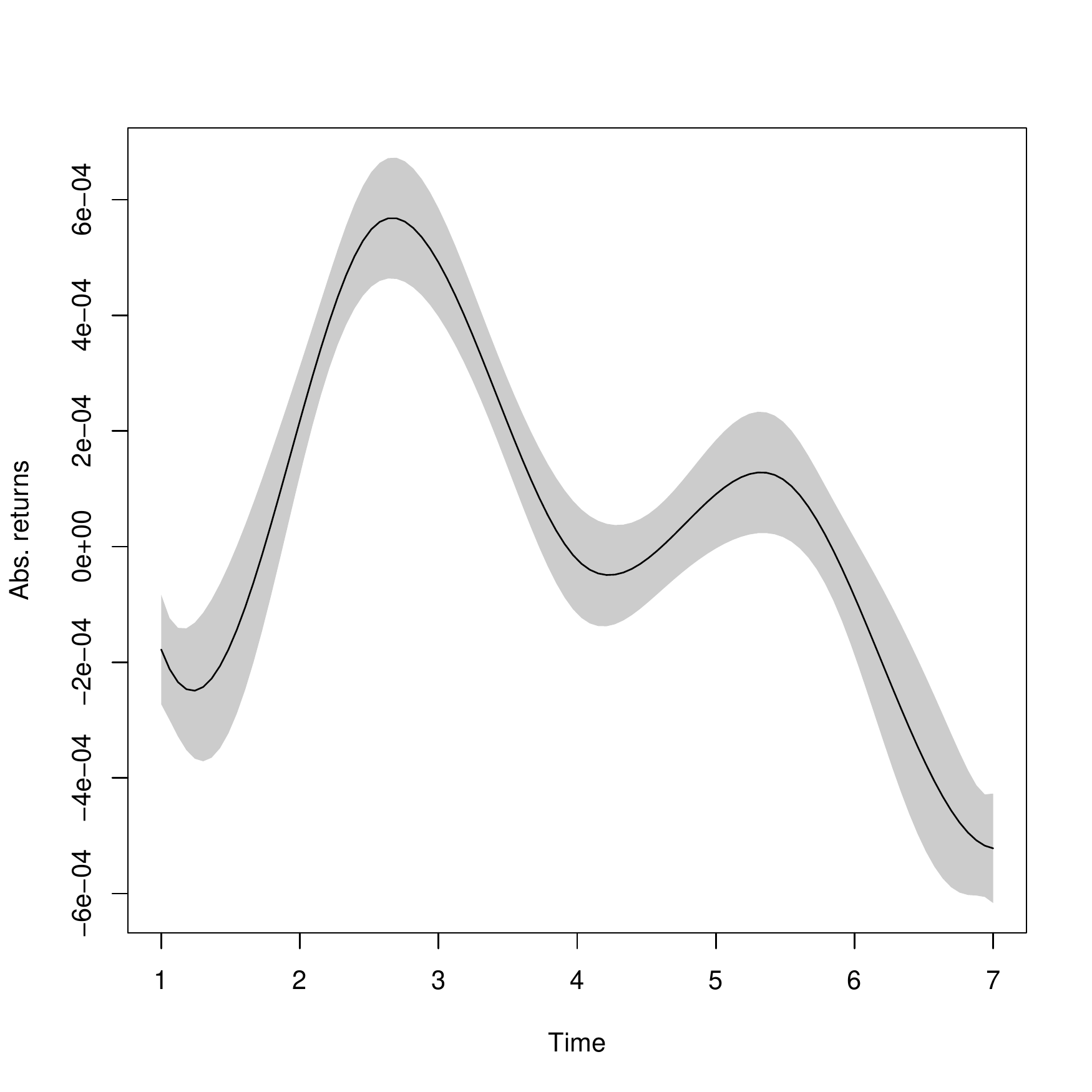}}
\hfill
\href{https://github.com/QuantLet/CCID/tree/master/CCIDvolaGAM}{\includegraphics[keepaspectratio,width=0.4cm]{media/qletlogo_tr.png}}
\caption{Weekly seasonality: fit of Generalized Additive Model   with p-splines for absolute returns of cryptocurrencies (shaded regions represent confidence bands for smooths), 01. July 2018 - 31. August 2018.}
\end{figure}


Assuming that the majority of employed persons do work from 09:00 to 17:00 o'clock in Europe, figures 7 and 8 (data time is +1 GMT) present us with a very clear picture of returns and volume. Characteristic $human$ $activity$ curves are presented by figure 7 showing the daily seasonality - a curve driven by algorithms as the main actor, or $Artificial$ $Intelligence$ in a FinTech startup buzzword context, should not present such a comparatively extreme low around a typical time for the majority of humans to be asleep. Following that point, the curves expresse a significant growth, only to flat out again around lunch break time. Most figures present a peak between 17:00 and 20:00 o'clock, just when most people finish their daily routine jobs, followed by an expressive decline of the curves. This is surprising, as media outlets and startup marketing generally praise the non-stop availability and easy access to cryptocurrency exchanges, and hence we would presume to see a curve different to that of a ``routine"-job. Further adding to this argument of trading being mostly done by humans organized in cooperations (regarding figure 8 with the seven numbers indicating the days of the week), is research on anomalies such as the ``Monday Effect" applied to our findings \citep[e.g.][]{Cross:73,Bash:06}. By applying both parametric and nonparametric methods, \citet{Capo:19} find abnormal returns for no other cryptocurrency than BTC, and that only on Mondays - yet, in figure 8 we can observe that weekly absolute returns across cryptocurrencies reach their peak only in the period from Tuesdays to around Thursdays, with a steep decline in activity during the weekends.

\begin{figure}[H]
\hfill
\subfigure[BTC]{\includegraphics[width=5cm]{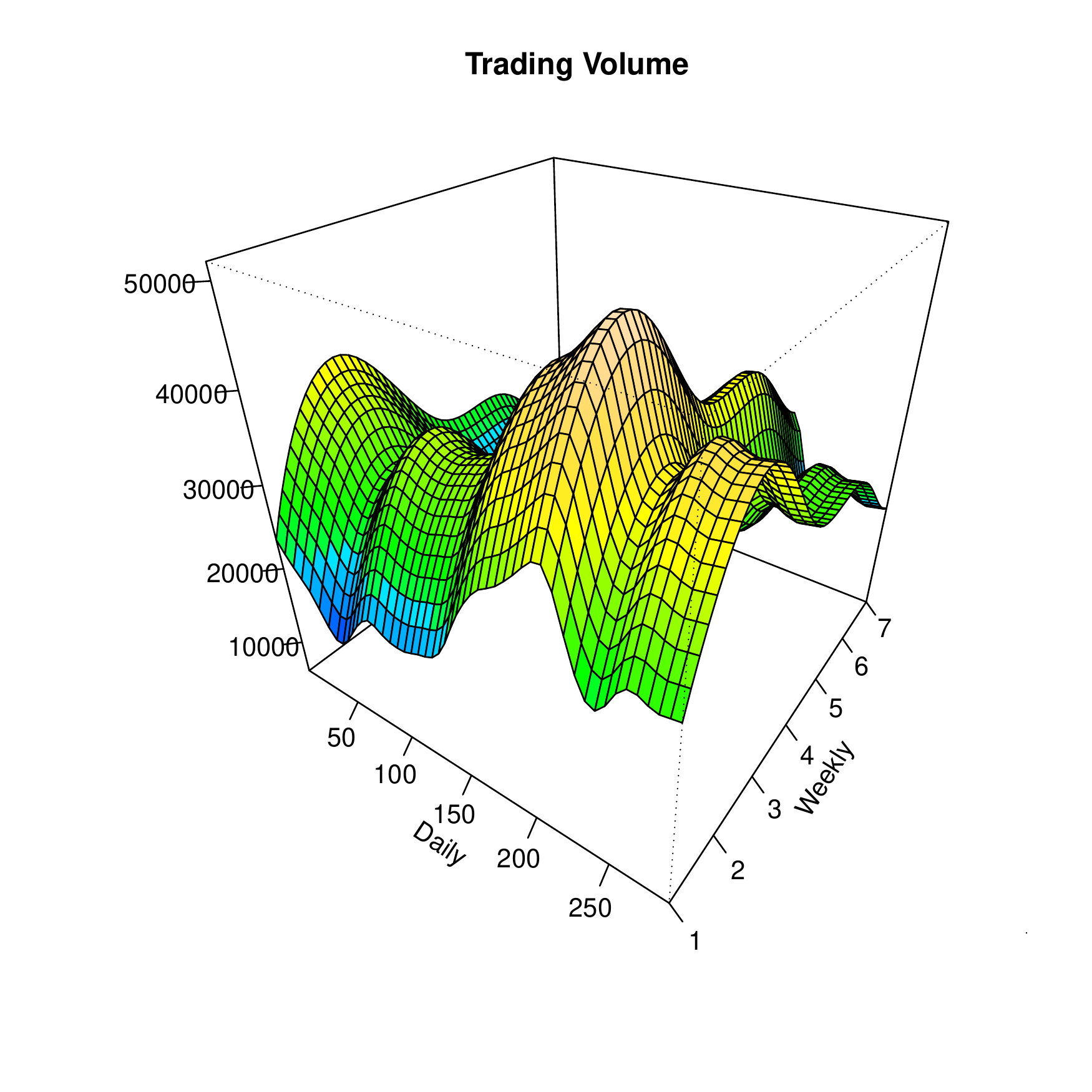}}
\hfill
\subfigure[ETH]{\includegraphics[width=5cm]{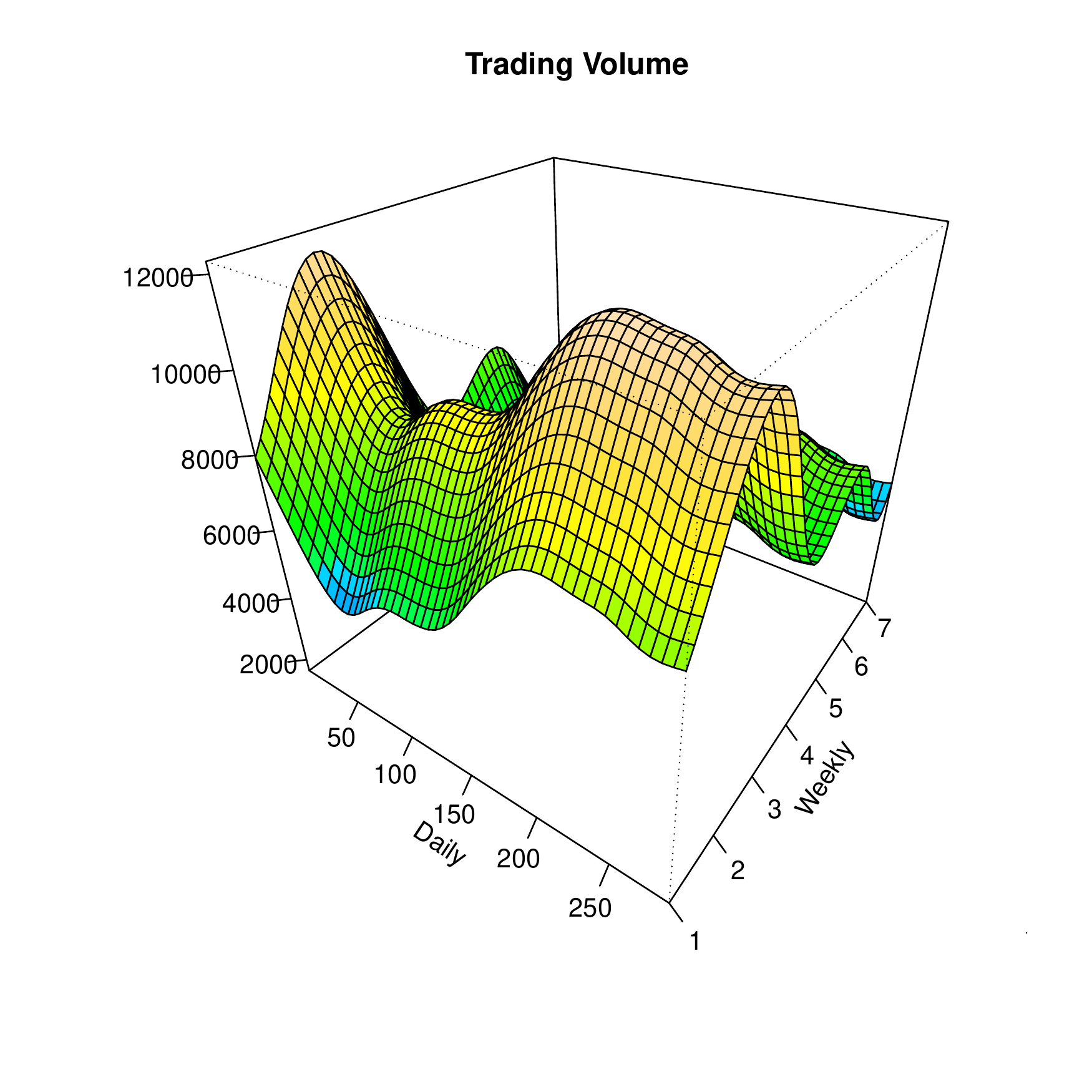}}
\hfill
\subfigure[XRP]{\includegraphics[width=5cm]{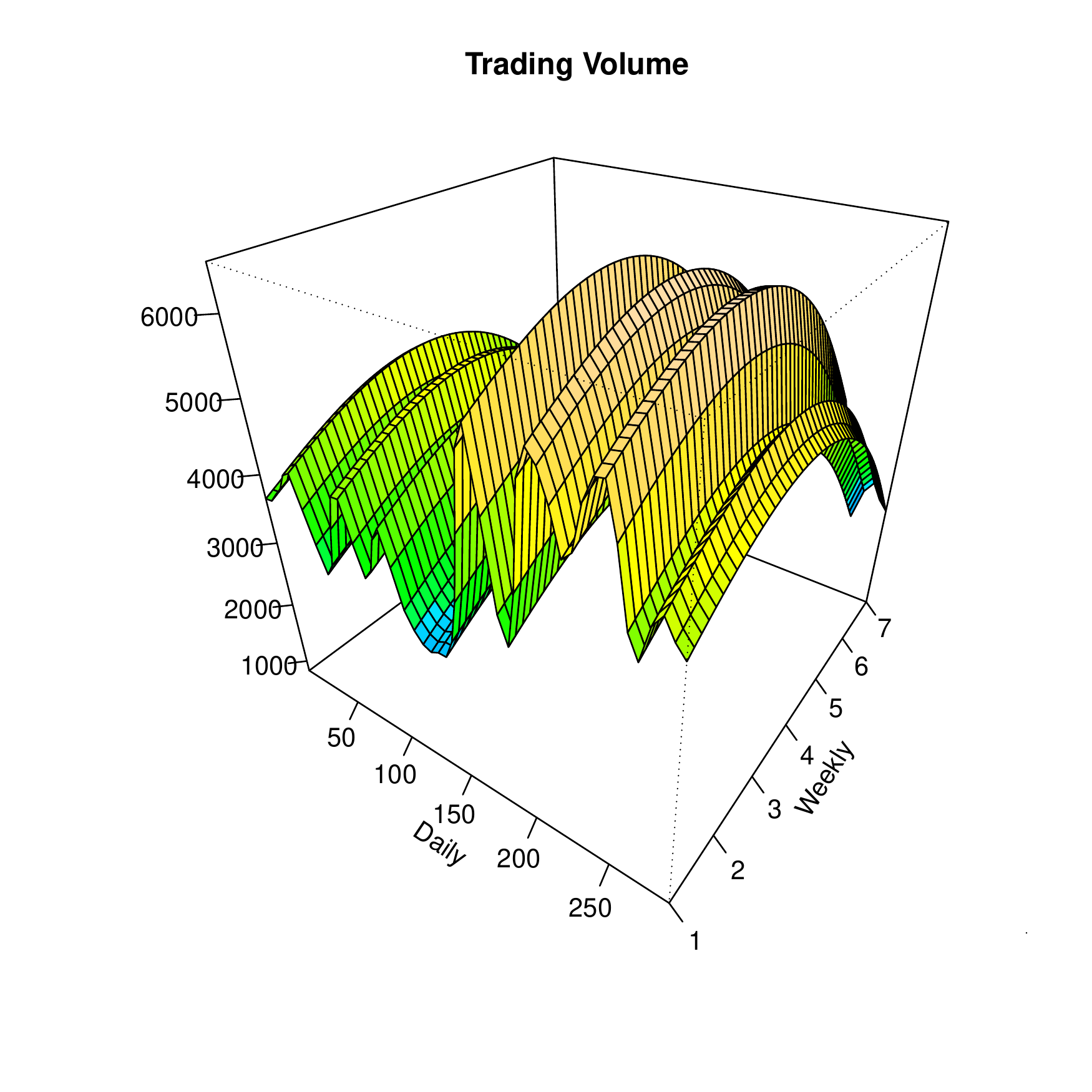}}
\hfill
\href{https://github.com/QuantLet/CCID/tree/master/CCIDvolumeGAM}{\includegraphics[keepaspectratio,width=0.4cm]{media/qletlogo_tr.png}}
\caption{Daily and weekly seasonality: fit of Generalized Additive Model   with cubic and p-splines for trading volume of cryptocurrencies (5 min nodes), 01. July 2018 - 31. August 2018. 01. July 2018 - 31. August 2018.}
\end{figure}

 Figure 9 presents us a respective lower trading volume during the weekends, compared to for example Thursdays or especially Fridays. Similar results can be seen in figure 10, presenting us with low volatility on the cryptocurrency market at said times - one assumption from this could be taken from the immense influx of financially potent startups organized as cooperations in this emerging market \citep[c.f.][]{Benedetti}. Yet, we can see that human interaction is shaping how the market behaves during the given time frames. Trade limited to regular working hours and days in Europe leads to the conclusion, that the majority of trades are not done by algorithms, which are active 24/7, but by human agents themselves making transactions and orders individually and by hand. This is especially obvious through figure 8, which is presenting a much lower activity pattern observable during the weekends. Should algorithms really be the drivers in this, technically predestined, fully digitized market, then this curve should not drop off as observable on Saturdays and Sundays. These findings are similar across the board \hyperref[Appendix]{(see appendix sections 5.2 - 5.4)}. While there is a plethora of well working, open-source trading bots available for these markets, for example via \href{https://github.com/topics/trading-bot}{Github} \citep{Nevskii}, as well as an abundance of commercially available trading bots \citep{Norry}, the trust in these - or the knowledge of how to employ them in this emerging market - is certainly low. This is especially surprising, as the possibility for arbitrage or mean reversion is obvious with multiple exchanges trading the same assets each with individually different prices, see \hyperref[sec3]{section 3}. The inherent possibility to take advantage of this inefficiency of the distributed trading, with near-simultaneous transactions, leads to great opportunities for traders unseen in most traditional markets for most assets. Hence we can assume, as algorithms need humans to get deployed and take action, like reacting to price changes, that the overall impact of these is not significant, if not negligible at all.

\begin{figure}[H]
\hfill
\subfigure[BTC]{\includegraphics[width=5cm]{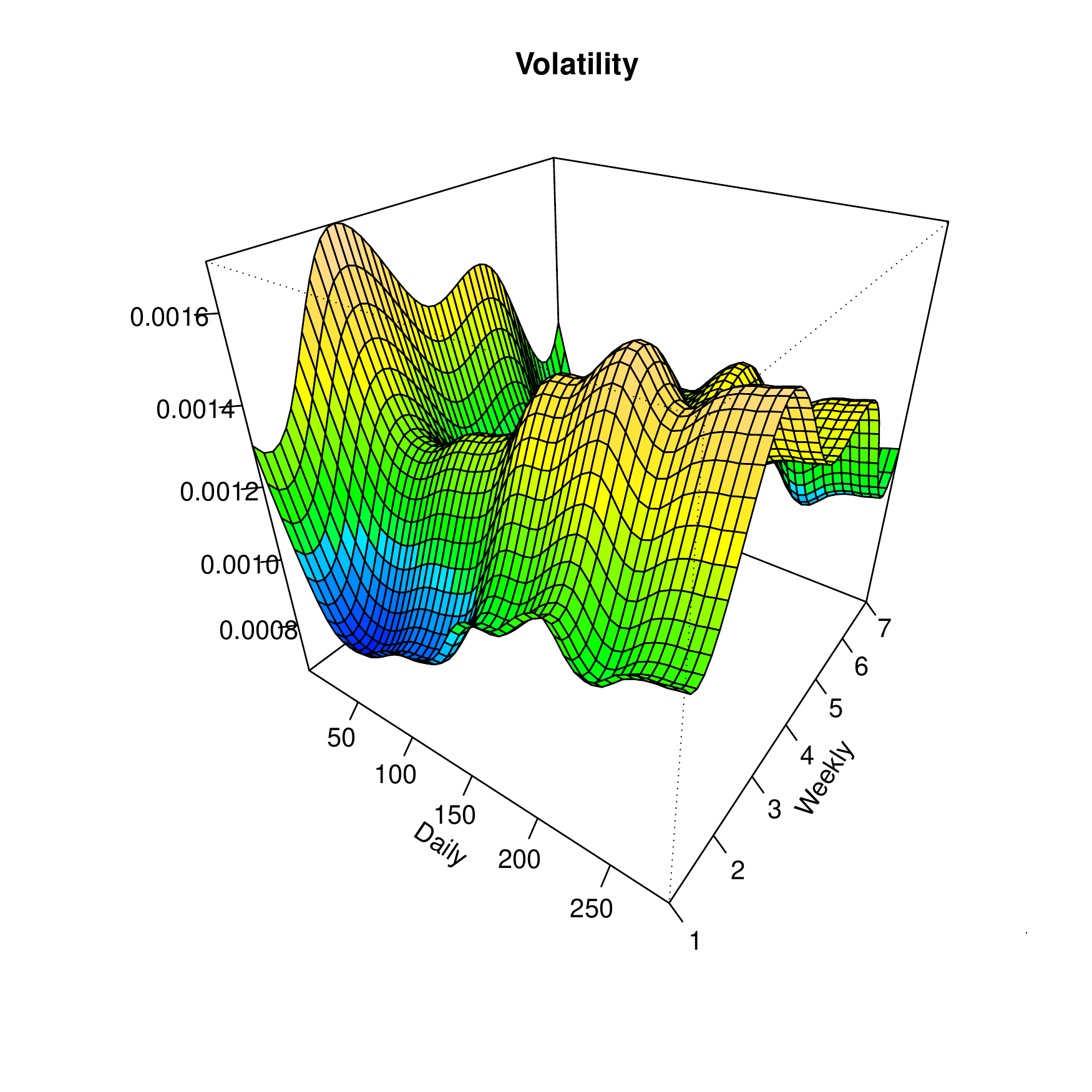}}
\hfill
\subfigure[ETH]{\includegraphics[width=5cm]{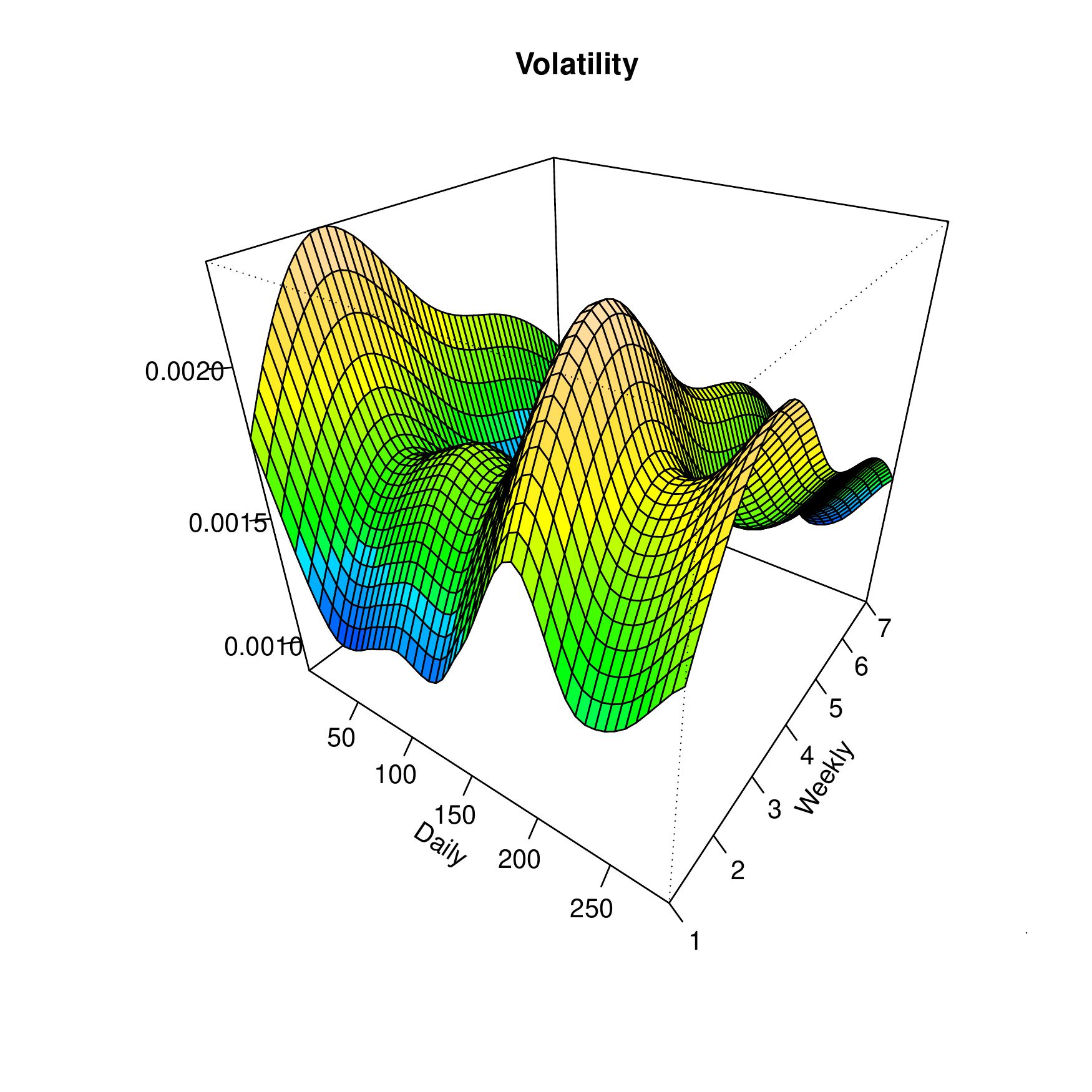}}
\hfill
\subfigure[XRP]{\includegraphics[width=5cm]{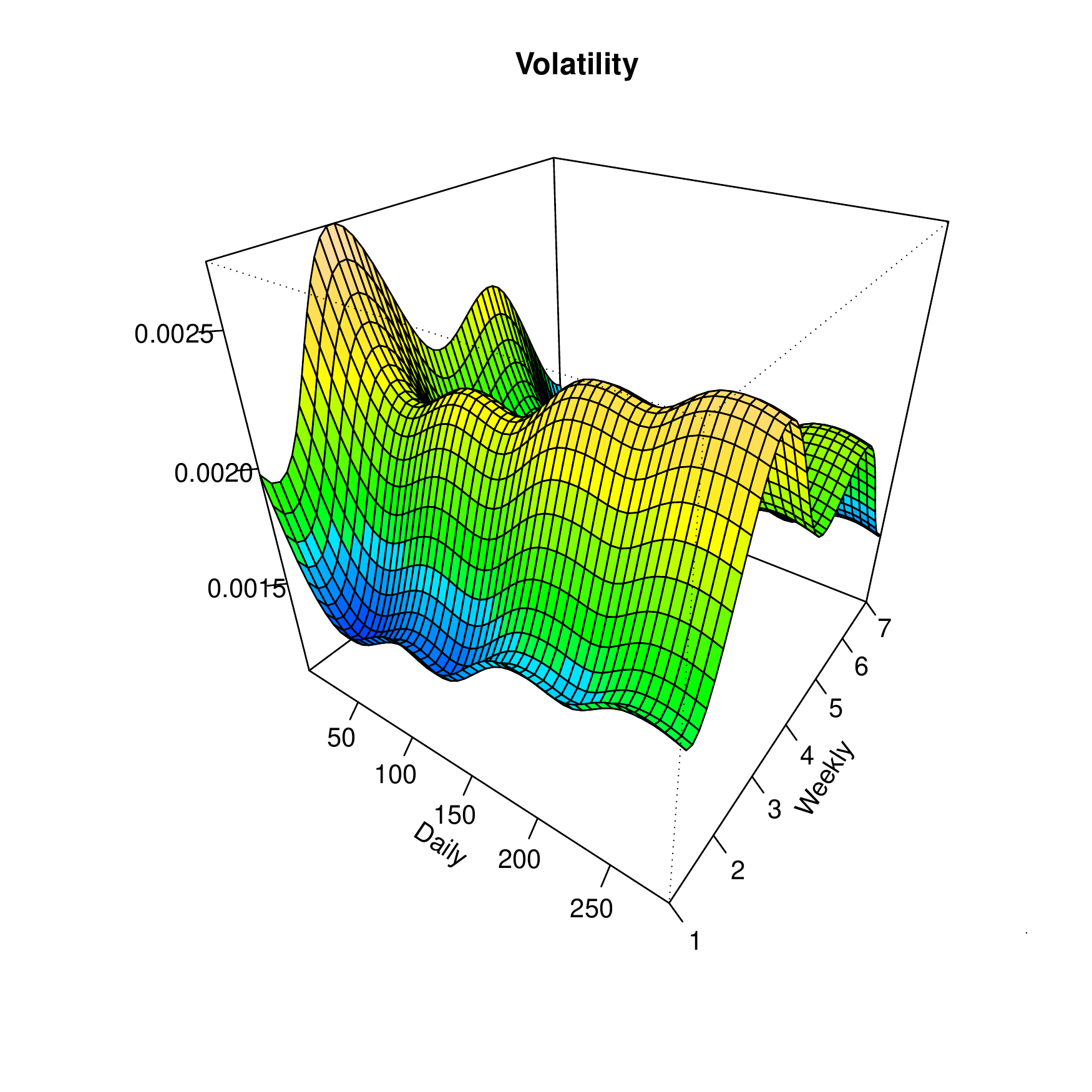}}
\hfill
\href{https://github.com/QuantLet/CCID/tree/master/CCIDvolaGAM}{\includegraphics[keepaspectratio,width=0.4cm]{media/qletlogo_tr.png}}
\caption{Daily and weekly seasonality: fit of Generalized Additive Model   with cubic and p-splines for volatility of cryptocurrencies (5 min nodes), 01. July 2018 - 31. August 2018. 01. July 2018 - 31. August 2018.}
\end{figure}

In total we can observe, that the activity patterns displayed in this market not only tend to express human interaction but also corporate structures as well, as most trading is done Mondays to Fridays, with the weekends expressing a low intensity of trades taking place. The previously mentioned immense increase of financially potent FinTech entities have attracted absurd amounts of financial backing compared to the output delivered via initial coin offerings, ICOs for short  \citep[c.f.][]{Benedetti,Zetsch:19}. To enable new industries using the blockchain technology, startups and commercial companies have been launching ICOs, similar to the initial public offerings (IPOs) of companies, to sell tokens in a transparent and decentralized manner and therefore creating a new method of raising funds without intermediaries, like traditional financial institutes. Some of these tokens are pegged to other (monetary) systems or even cryptocurrency constructions directly, as these have already gained a high market acceptance - especially the Ethereum ecosystem is facilitating this by providing excessive tools and documentaries, paired with a focused and growing community of developers, to create what they coined as ``coloured coins" in order to expand the utility of the existing blockchain \citep{Walters}. Besides the fact, that the legality of ICOs is disputed and potential responses from regulatory agencies are growing to be imminent, ICOs enable anyone within the community to participate in the investment, providing opportunities for small-scale investors. Hence the assumption would be, that especially these specialized corporate startups are working on their backend and maintain their ecosystem, whilst being active drivers of trading in this market - yet predominantly human ones.\\

Coming back to the 2010 Flash Crash mentioned in the introduction of this paper, one could argue, that such a flash crash is not possible due to the delay that is inherent to blockchains - the so-called blocktime \citep{ucc:2020}. However, as research has shown, it is easily possible to derive sentiment and therefore market reactions from Twitter, Facebook, Stocktwits, or similar public forums. As most activity can be seen on the respective cryptocurrency exchanges, where the order books are not handled on-chain, but necessarily off-chain to quickly process the exchange users trading requests \citep{Sonic:2019,Romeo:2019,Qian:2019}. Therefore a crash related to certain cryptocurrency prices may be seen only after the respective information has been seeded into the network and accepted as new information to the individual blockchain. This is creating an inherent risk, as market behaviour can not be seen by only relying on on-chain data to predict certain price movements. The previously mentioned ```learning" algorithms could therefore be, if they are employed in this manner, be dangerous if a ``false-postive" is identified and results in a respective process leading to dumping a certain asset, which in turn could then generate a waterfall when other algorithms, that respond to blockchain price data, reply to this movement \citep{ASOB:2019}. Hence, a grand scale application of algorithms needs to be finely tuned in order to avoid any humanely unforeseeable, but technically feasible, consequences.\\

With the cryptocurrency market being easy to join and to actively participate in, financial traders are becoming redundant - unless they provide specialized services. Making many transactions doesn't cost time to interact with a trader and money to pay this person, as one can do that by hand at home with very low transaction costs. This said, there is a big competition going on between the exchanges, who themselves may act as traders or brokers. The future has to tell if through this competition the rise of the machines and the respective mass employment of algorithmic trading in this digital realm will become reality.


\section{Closing remarks}
\label{sec6}

We have shown, that meanwhile there are certainly grand-scale employers of algorithmic trading around in this new emerging market of cryptocurrencies, yet, based on the time-of-day effects and the evidence gained, we can conclude, that the impact of 24/7 algorithmic trading is rather negligible given the empirical facts we have at hand. This leads us to the conclusion, that even though this new digital market appears predestined to be ruled by algorithms and specialised AI advisors, the digital realm of cryptocurrencies has yet to be conquered by the machines and is still firmly in the hands of humans or generally driven by respective startup's.\\

Further research should certainly step into this breach, that we have proven to be existent, and create means on how to best exploit this open ground on a market-oriented basis, as well as on an individual level, say in regards to the exchanges. Necessarily, such research not only needs to be of quantitative or technical origin, but also needs to include a regulatory point of view, as especially this field on blockchain research is more and more characterized by its evident interdisciplinary nature.



\begin{footnotesize}

\bibliography{literature}

\raggedright

\end{footnotesize}

\newpage

\section{Appendix}
\label{Appendix}

\vspace{1cm}

\subsection{List of cryptocurrencies in this research}
\label{sec7.1}

\begin{table}[ht]
\centering
\label{my-label}
\begin{tabular}{lll}
\hline
\hline
Abbrev.   & CC               & Website                    \\ \hline
BCH       & Bitcoin Cash     & \url{bitcoincash.org}           \\
BTC (XBT) & Bitcoin          & \url{bitcoin.com}, \url{bitcoin.org}   \\
DASH      & Dash             & \url{dash.org}                   \\
ETC       & Ethereum Classic & \url{ethereumclassic.github.io}  \\
ETH       & Ethereum         & \url{ethereum.org}               \\
LTC       & Litecoin         & \url{litecoin.com}, \url{litecoin.org} \\
REP       & Augur           & \url{augur.net}              \\
STR       & Stalker           & \url{staker.network}              \\
XMR       & Monero           & \url{getmonero.org}              \\
XRP       & Ripple           & \url{ripple.com}                 \\
ZEC       & Zcash           & \url{z.cash}              \\
\hline
\hline
\end{tabular}
\end{table}

\vspace{0.5cm}

\subsection{Appendix-Statistics for BCH, ETC and LTC}
\label{sec7.2}

\begin{figure}[H]
\hfill
\subfigure[BCH]{\includegraphics[width=5cm]{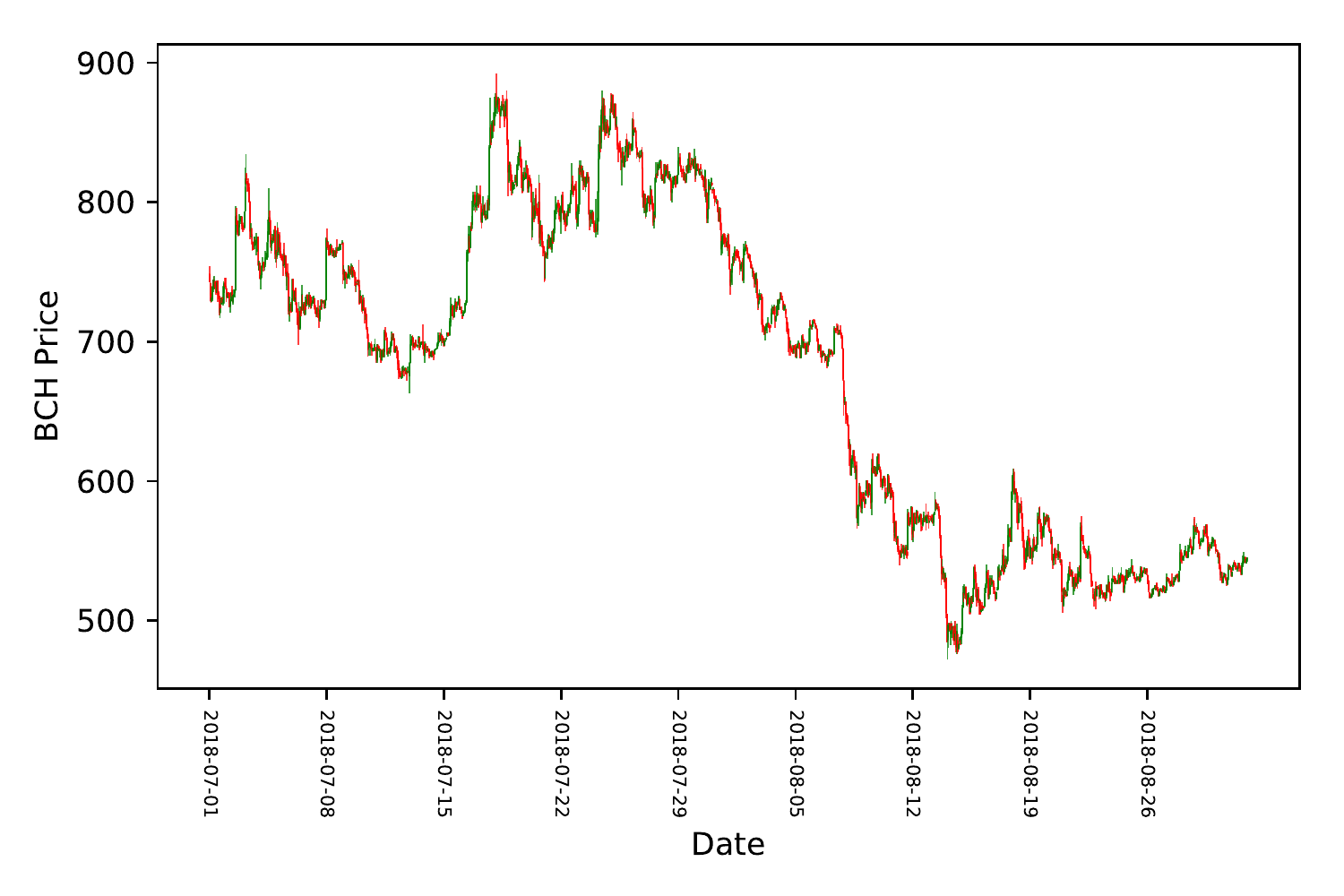}}
\hfill
\subfigure[ETC]{\includegraphics[width=5cm]{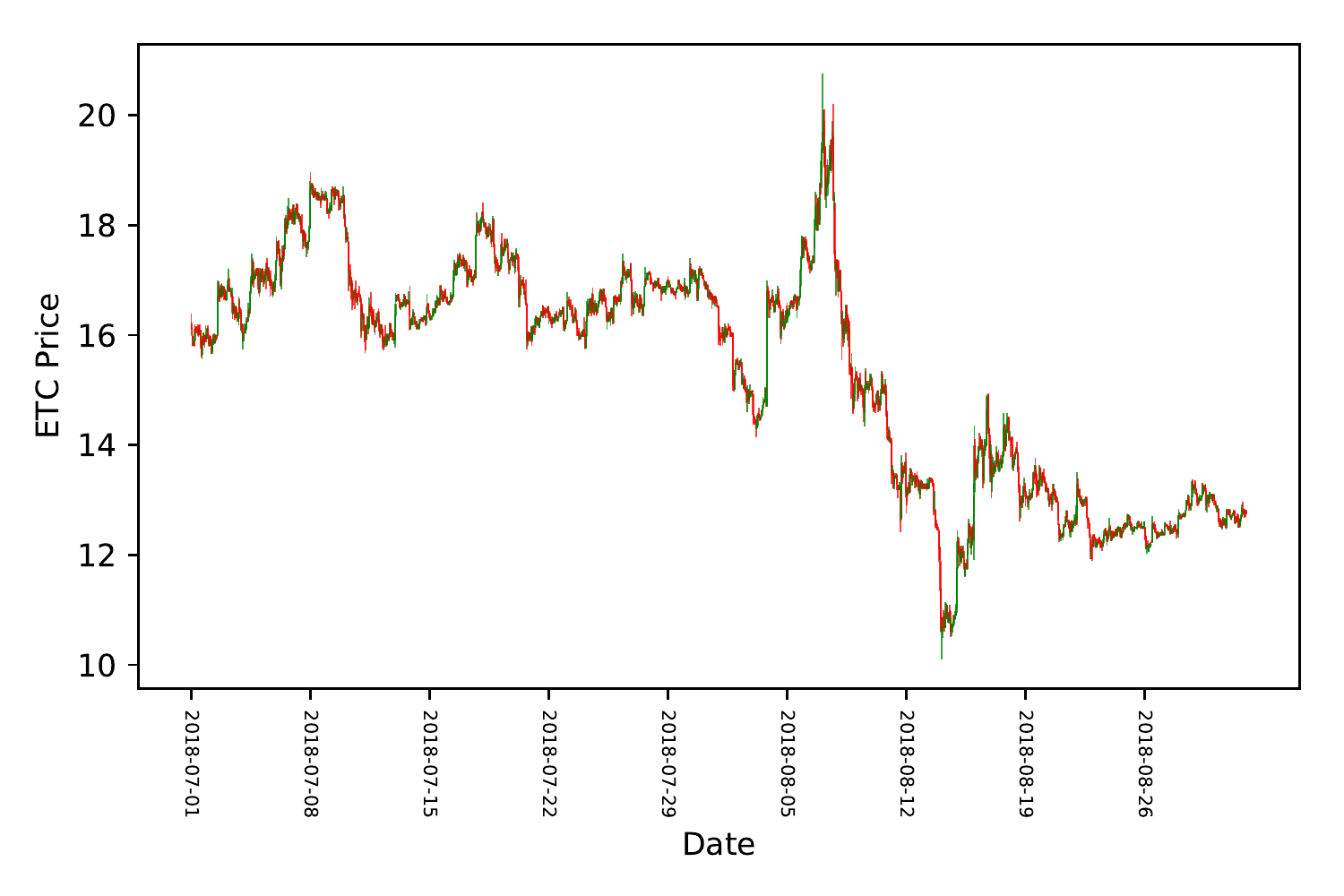}}
\hfill
\subfigure[LTC]{\includegraphics[width=5cm]{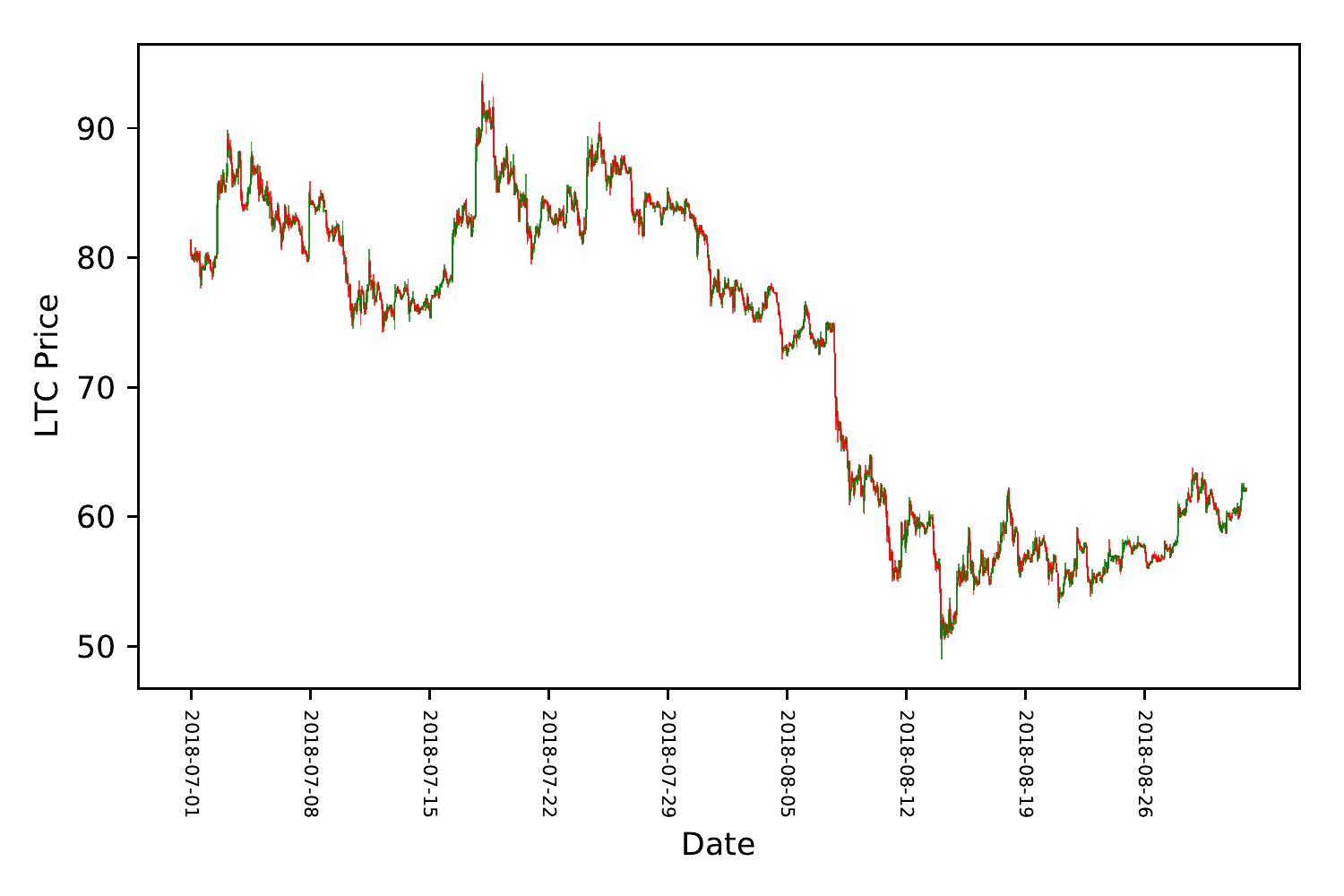}}
\hfill
\href{https://github.com/QuantLet/CCID/tree/master/CCIDCandles}{\includegraphics[keepaspectratio,width=0.4cm]{media/qletlogo_tr.png}}
\caption{Candlestick charts for individual price movements. 01. July 2018 - 31. August 2018.}
\end{figure}

\begin{figure}[H]
\hfill
\subfigure[BCH]{\includegraphics[width=5cm]{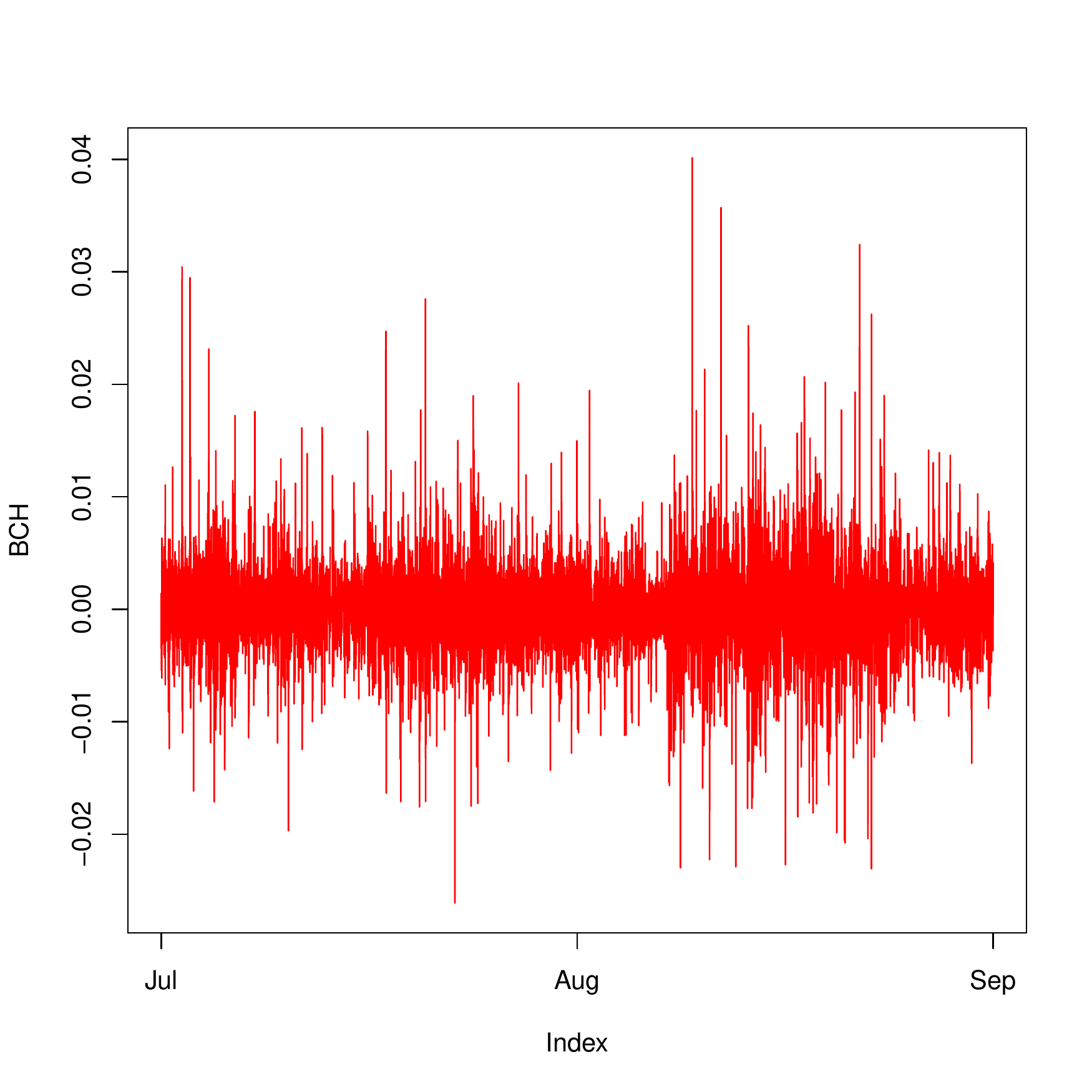}}
\hfill
\subfigure[ETC]{\includegraphics[width=5cm]{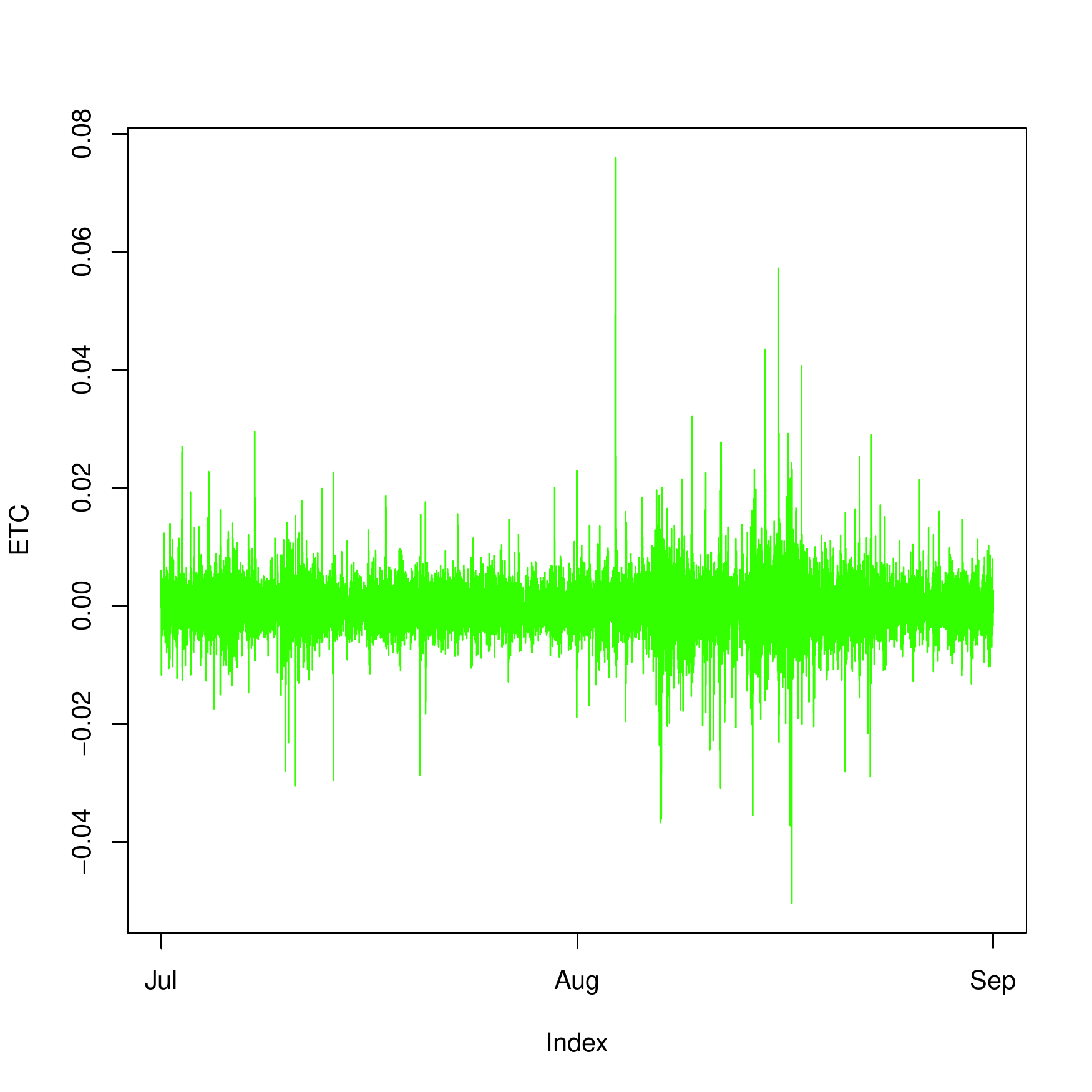}}
\hfill
\subfigure[LTC]{\includegraphics[width=5cm]{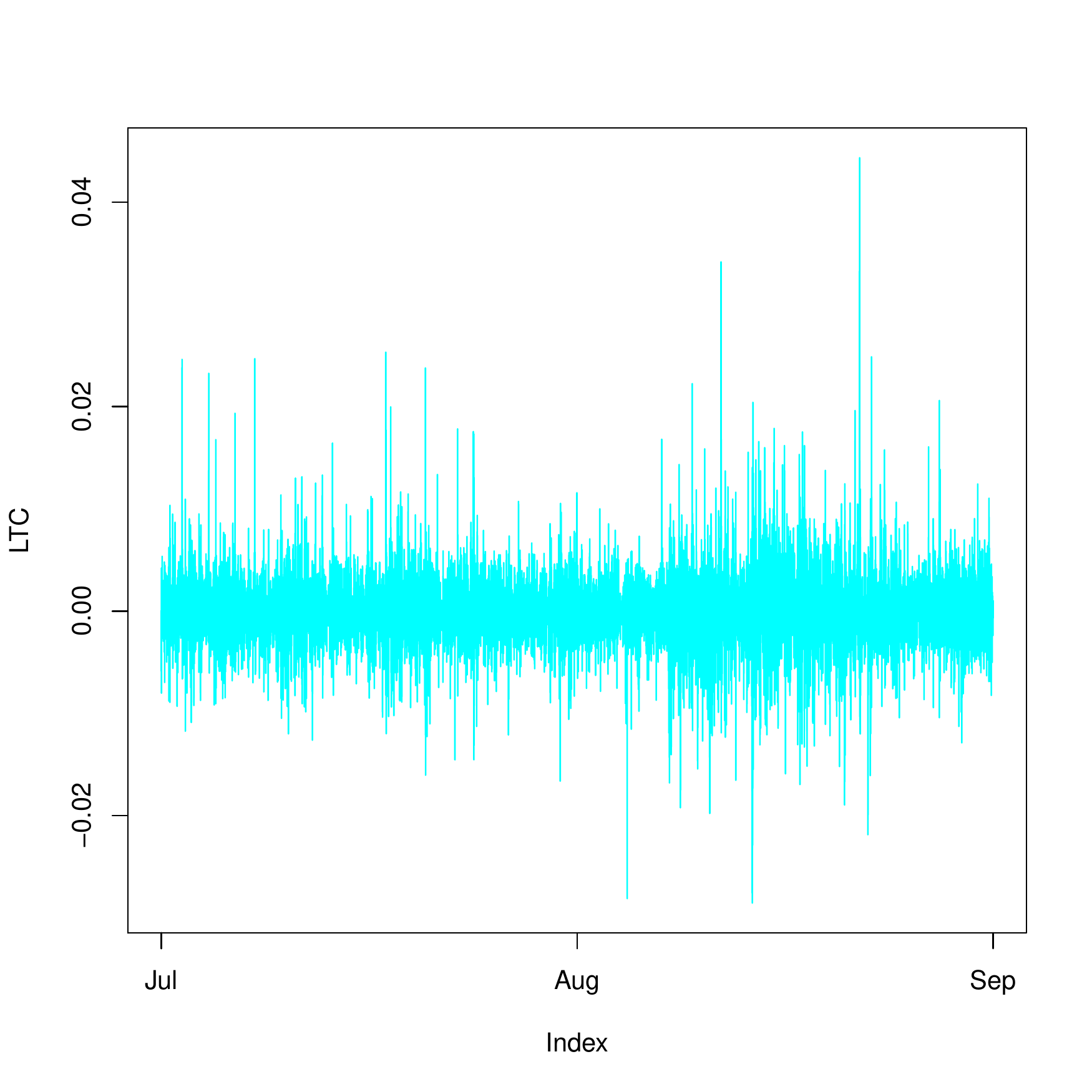}}
\hfill
\href{http://github.com/QuantLet/CCID}{\includegraphics[keepaspectratio,width=0.4cm]{media/qletlogo_tr.png}}
\caption{Intraday 5-minutes log-returns. 01. July 2018 - 31. August 2018.}
\end{figure}

\begin{figure}[H]
\hfill
\subfigure[BCH]{\includegraphics[width=5cm]{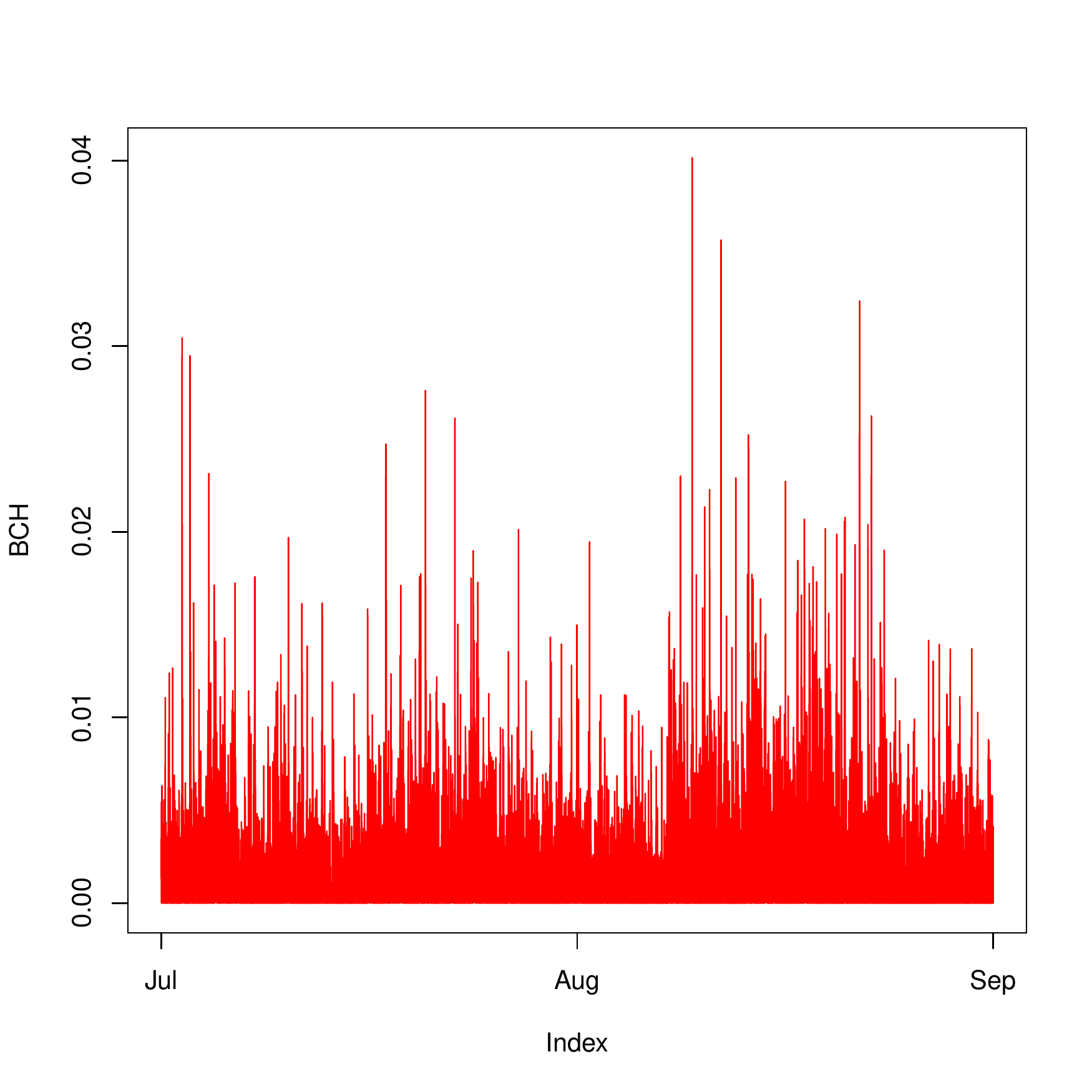}}
\hfill
\subfigure[ETC]{\includegraphics[width=5cm]{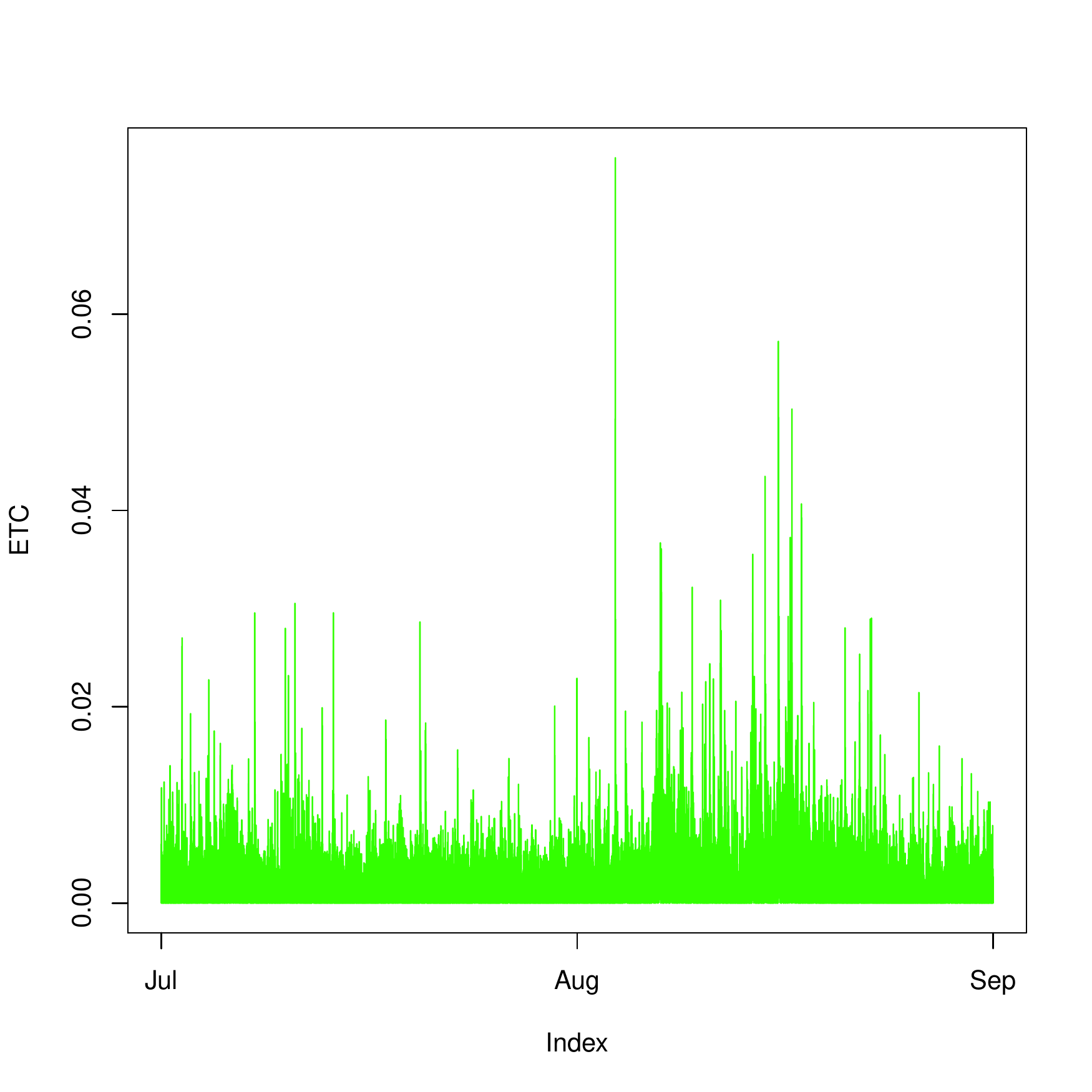}}
\hfill
\subfigure[LTC]{\includegraphics[width=5cm]{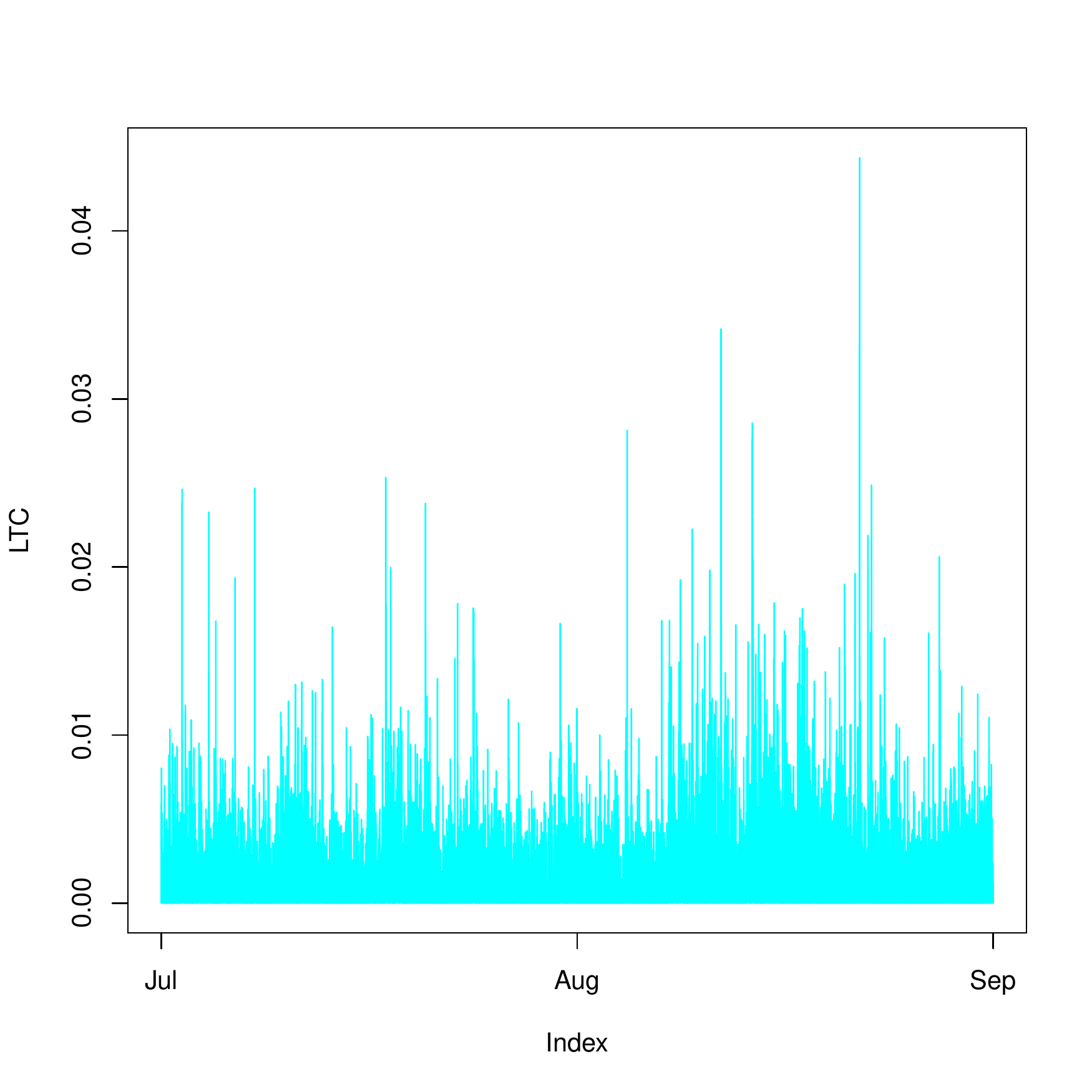}}
\hfill
\href{https://github.com/QuantLet/CCID/tree/master/CCIDHistRet}{\includegraphics[keepaspectratio,width=0.4cm]{media/qletlogo_tr.png}}
\caption{Intraday volatility (absolute values of 5-minutes log-returns) . 01. July 2018 - 31. August 2018.}
\end{figure}

\begin{figure}[H]
\hfill
\subfigure[BCH]{\includegraphics[width=5cm]{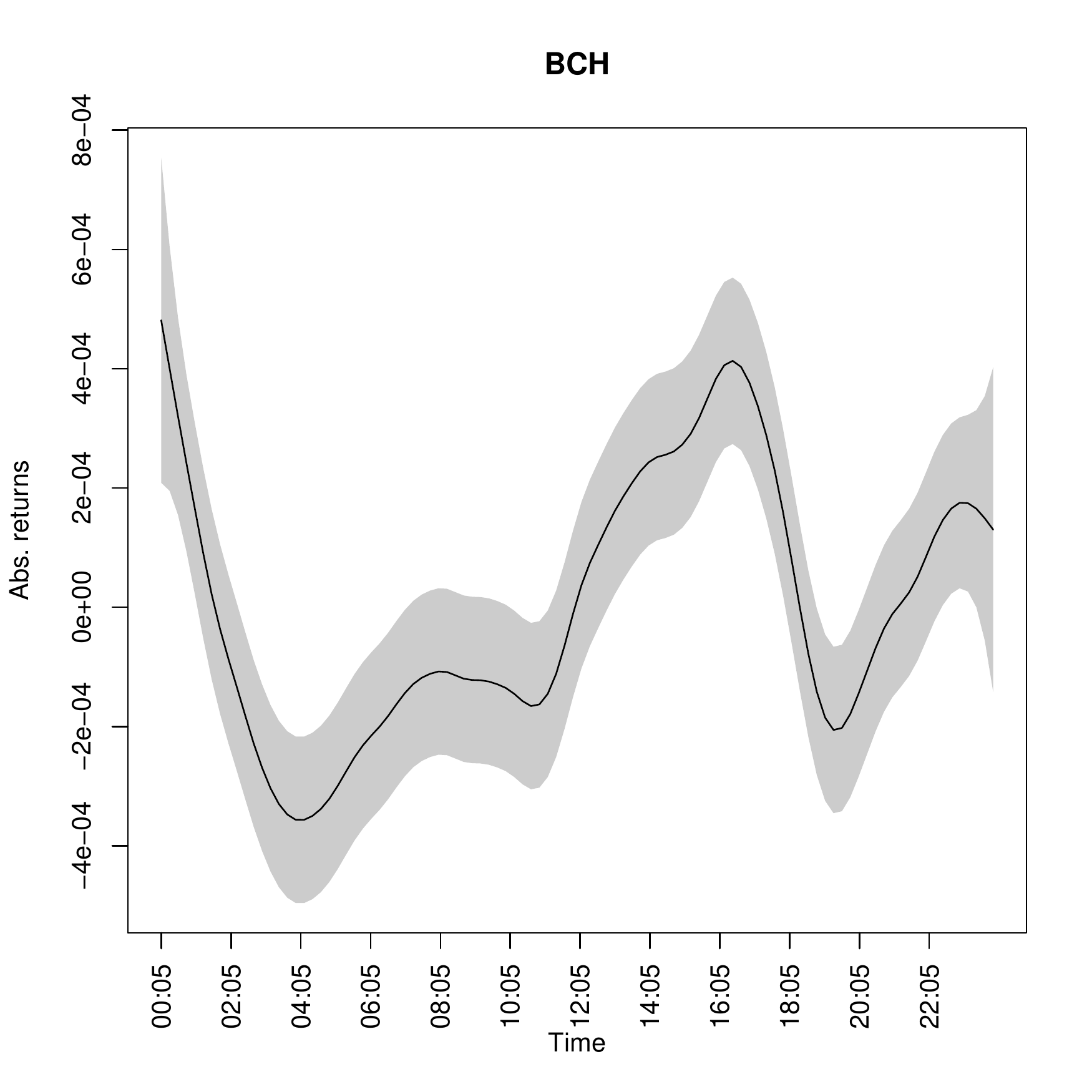}}
\hfill
\subfigure[ETC]{\includegraphics[width=5cm]{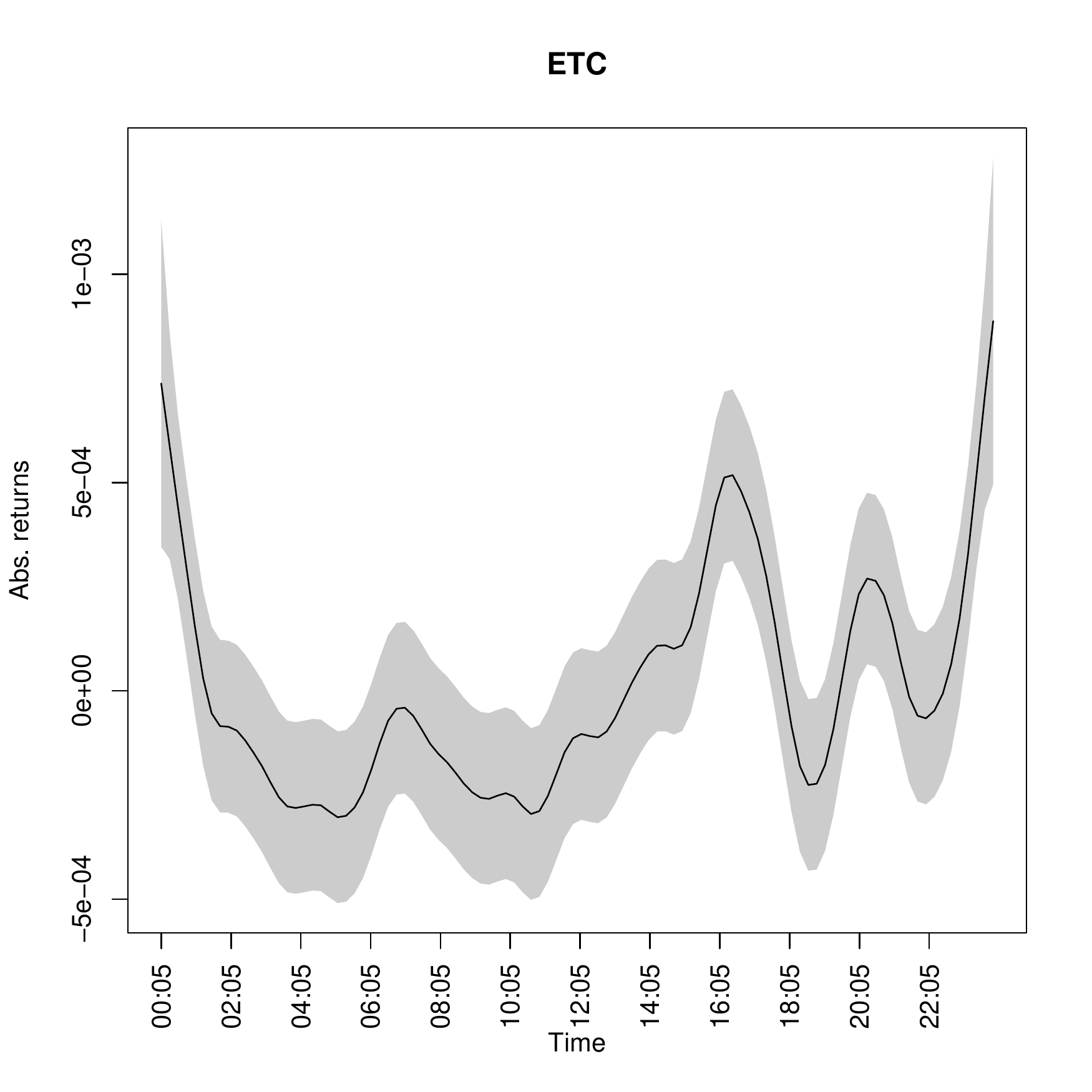}}
\hfill
\subfigure[LTC]{\includegraphics[width=5cm]{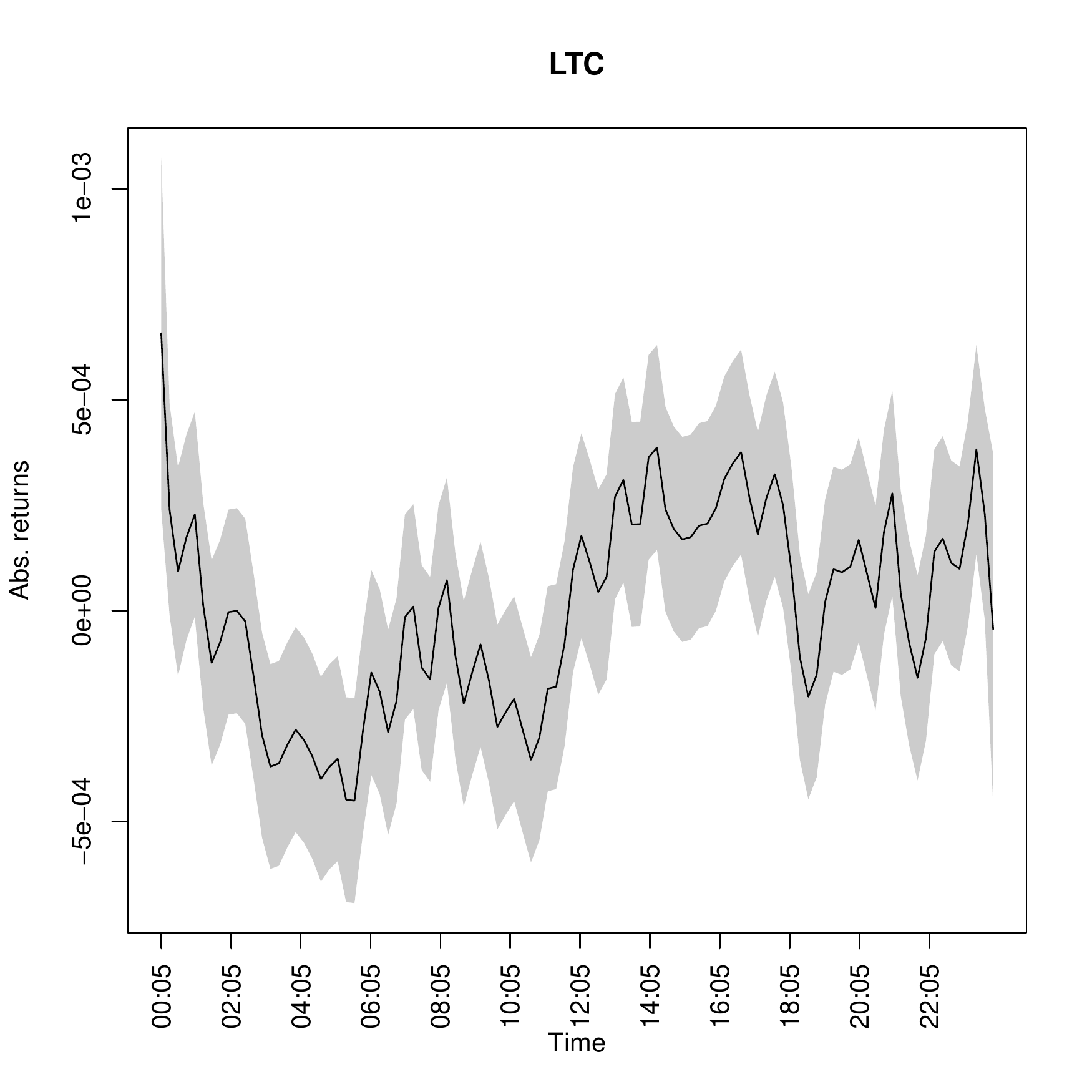}}
\hfill
\href{https://github.com/QuantLet/CCID/tree/master/CCIDvolaGAM}{\includegraphics[keepaspectratio,width=0.4cm]{media/qletlogo_tr.png}}
\caption{Generalized additive model of volatility. 01. July 2018 - 31. August 2018.}
\end{figure}

\begin{figure}[H]
\hfill
\subfigure[BCH]{\includegraphics[width=5cm]{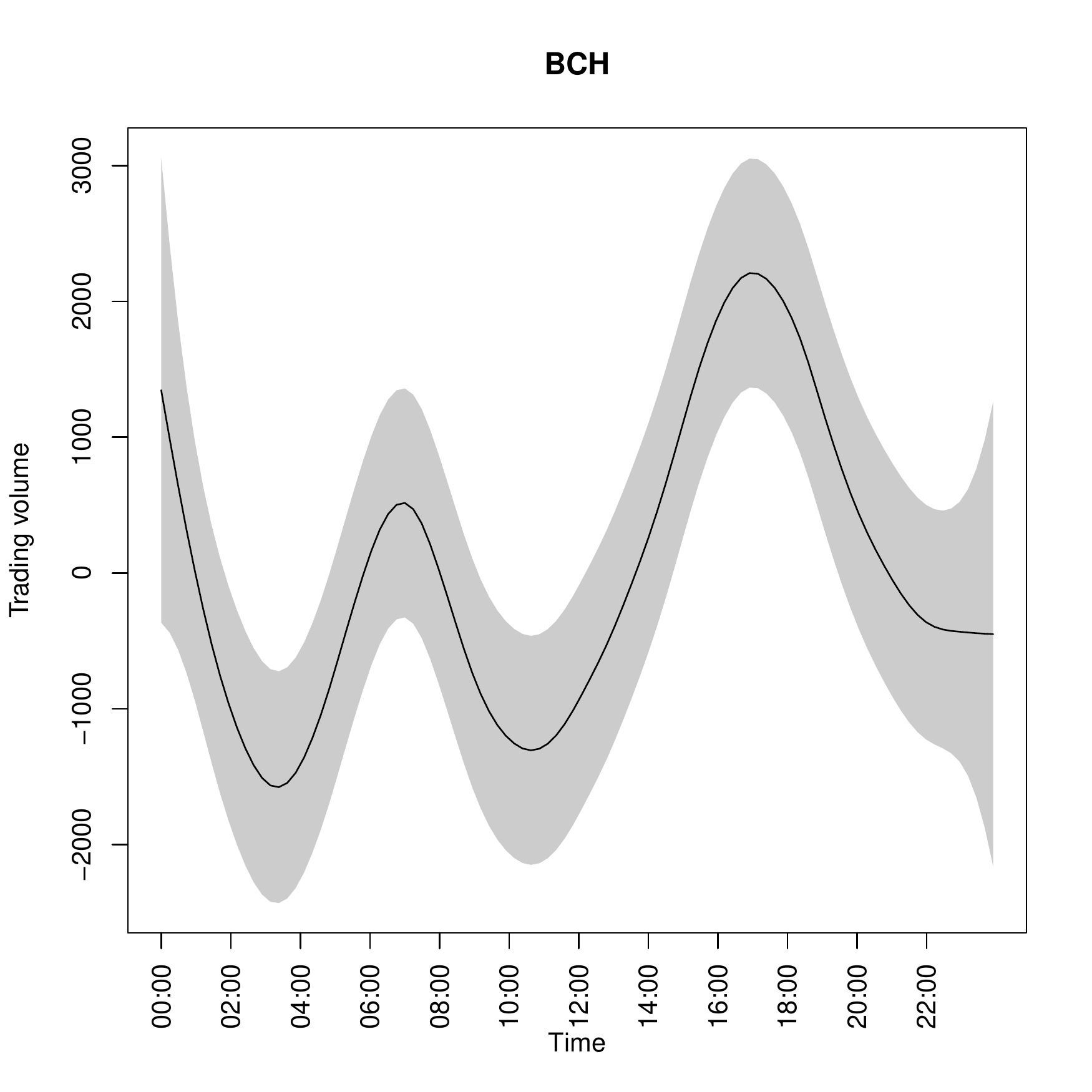}}
\hfill
\subfigure[ETC]{\includegraphics[width=5cm]{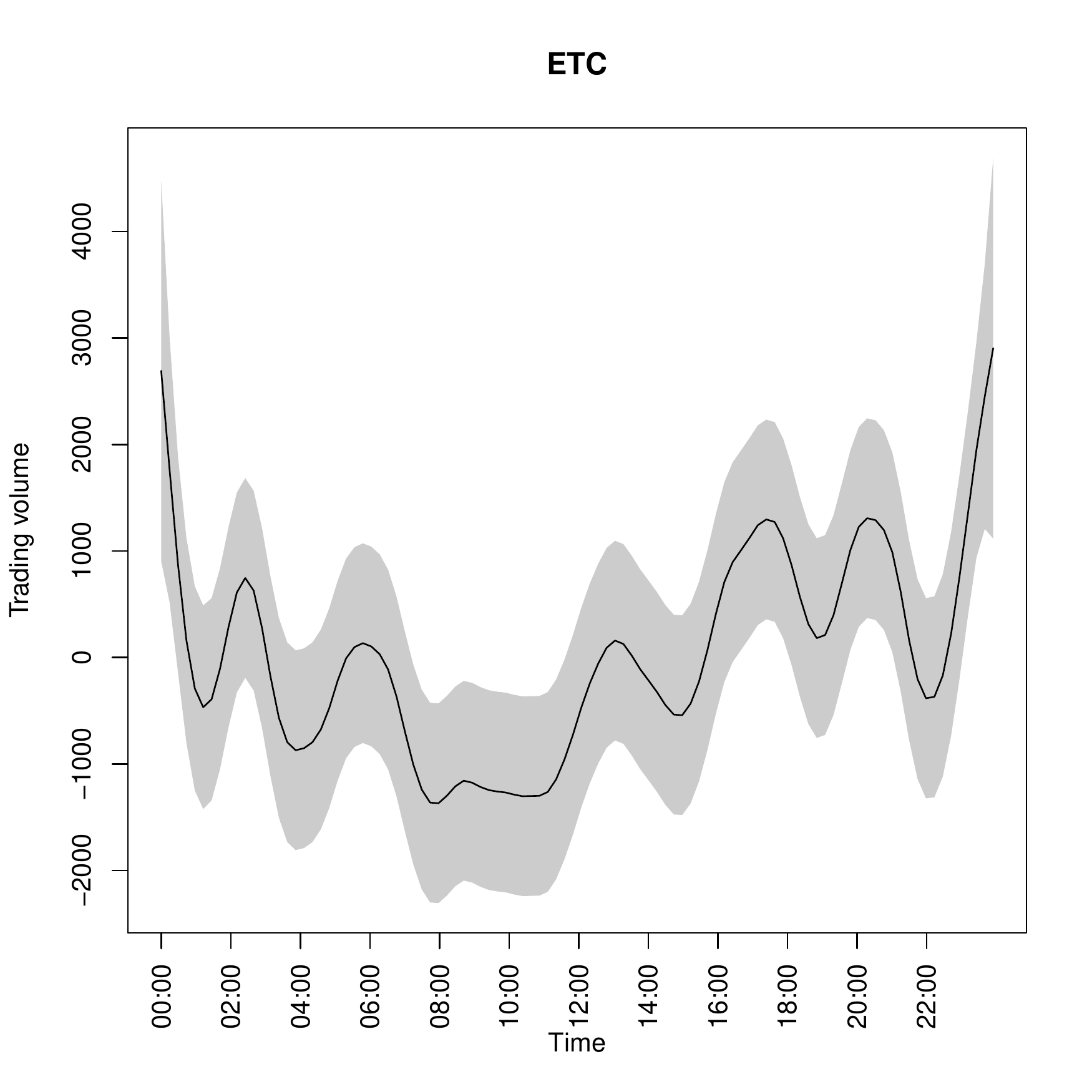}}
\hfill
\subfigure[LTC]{\includegraphics[width=5cm]{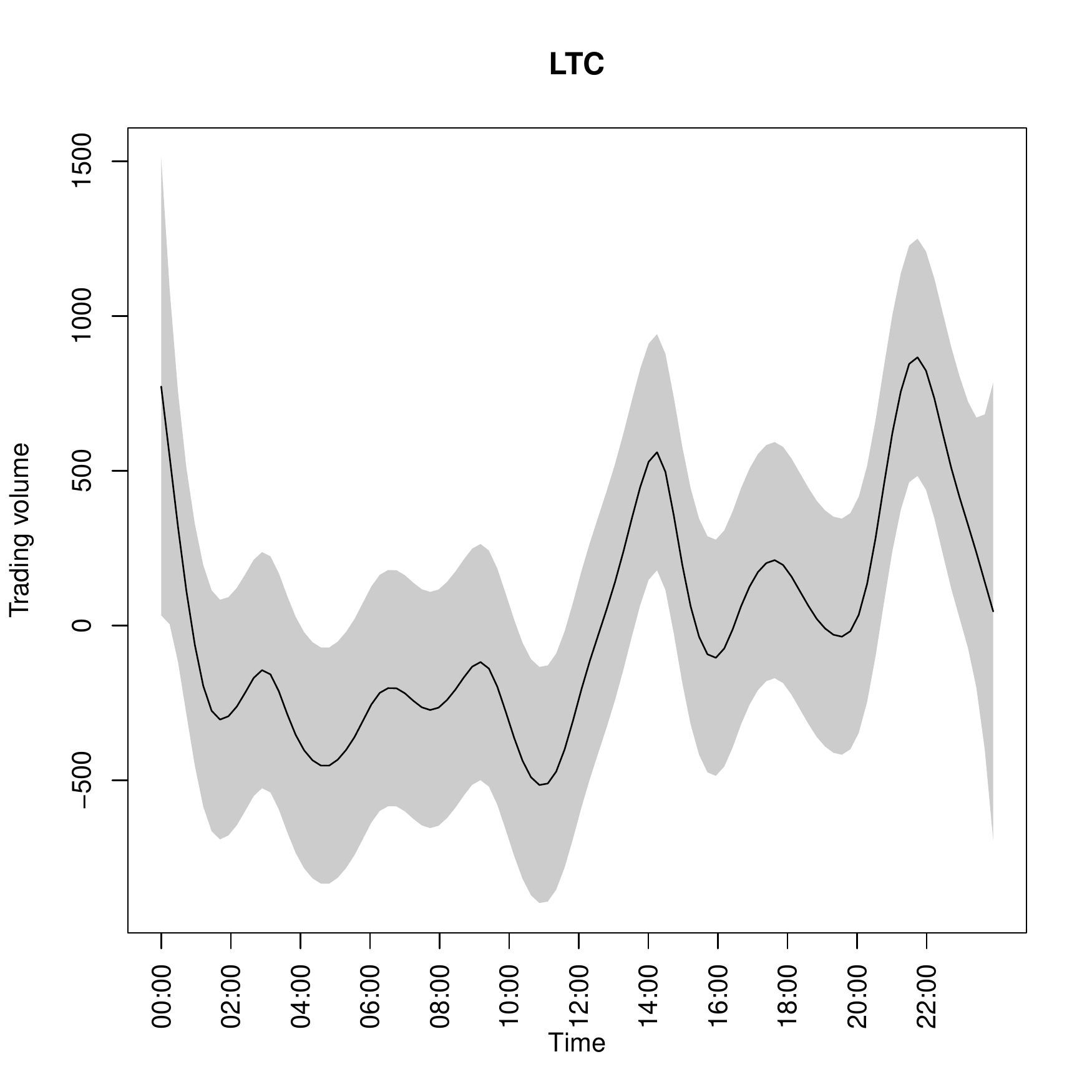}}
\hfill
\href{https://github.com/QuantLet/CCID/tree/master/CCIDvolumeGAM}{\includegraphics[keepaspectratio,width=0.4cm]{media/qletlogo_tr.png}}
\caption{Generalized Additive Model of trading volume of cryptocurrencies. 01. July 2018 - 31. August 2018.}
\end{figure}

\begin{figure}[H]
\hfill
\subfigure[BCH]{\includegraphics[width=5cm]{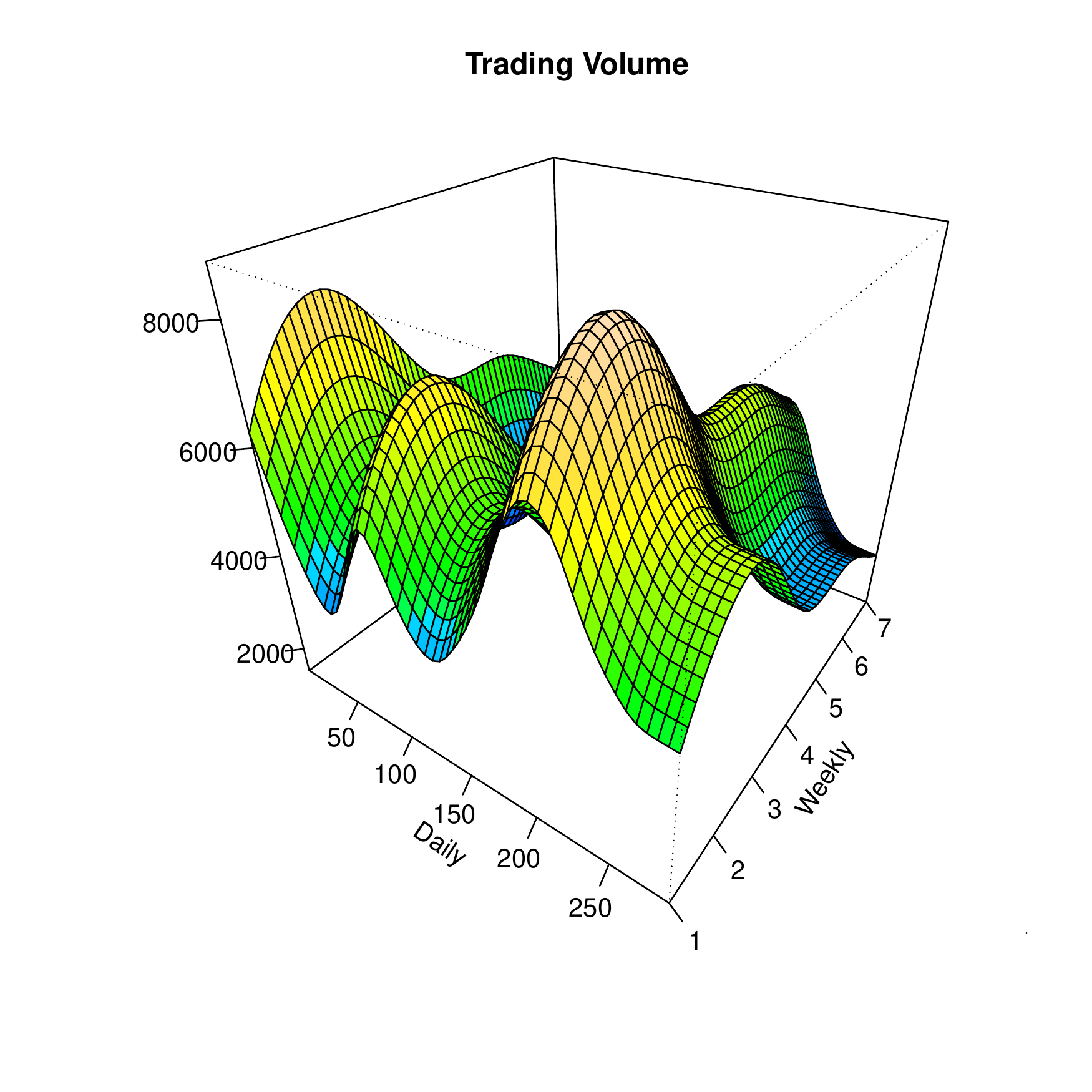}}
\hfill
\subfigure[ETC]{\includegraphics[width=5cm]{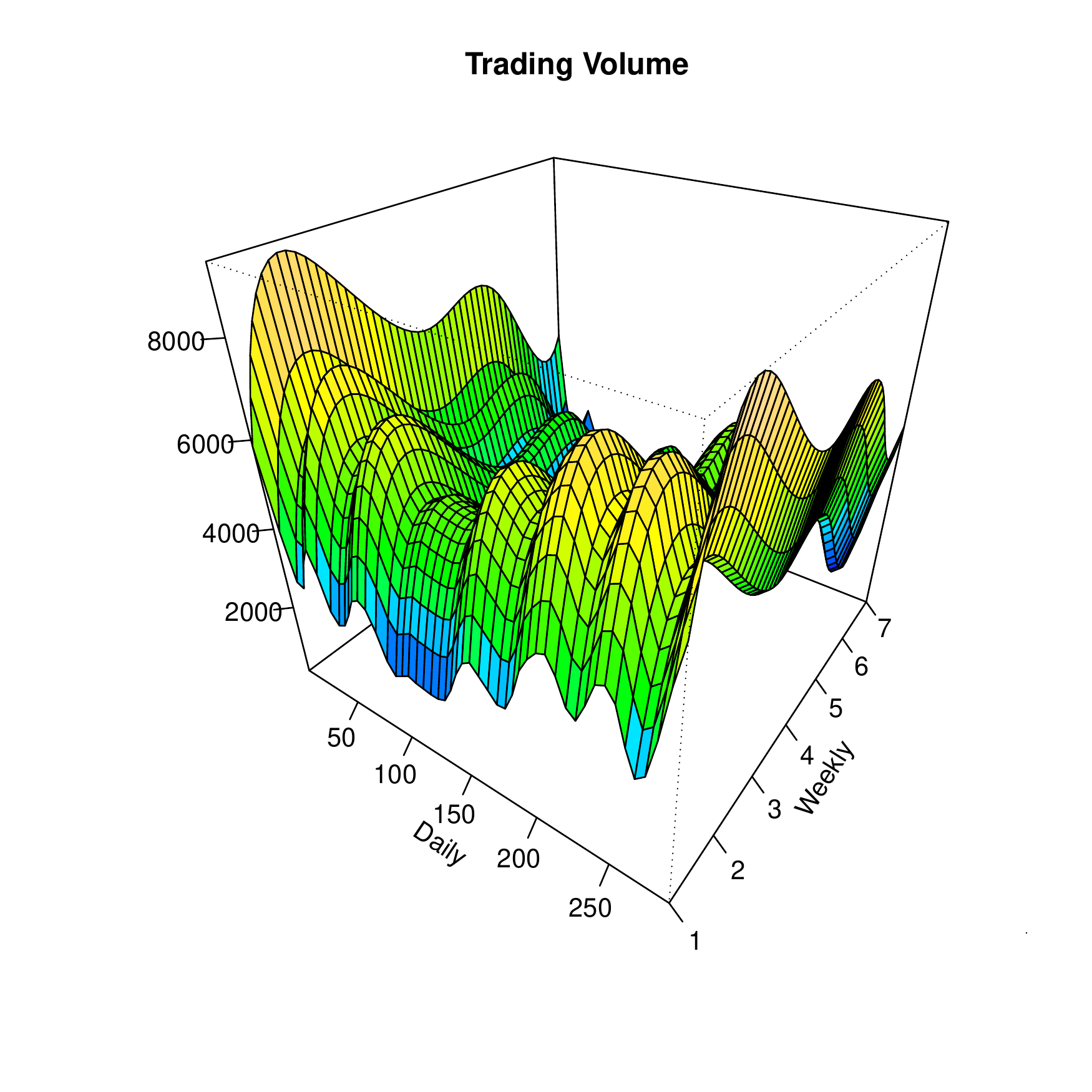}}
\hfill
\subfigure[LTC]{\includegraphics[width=5cm]{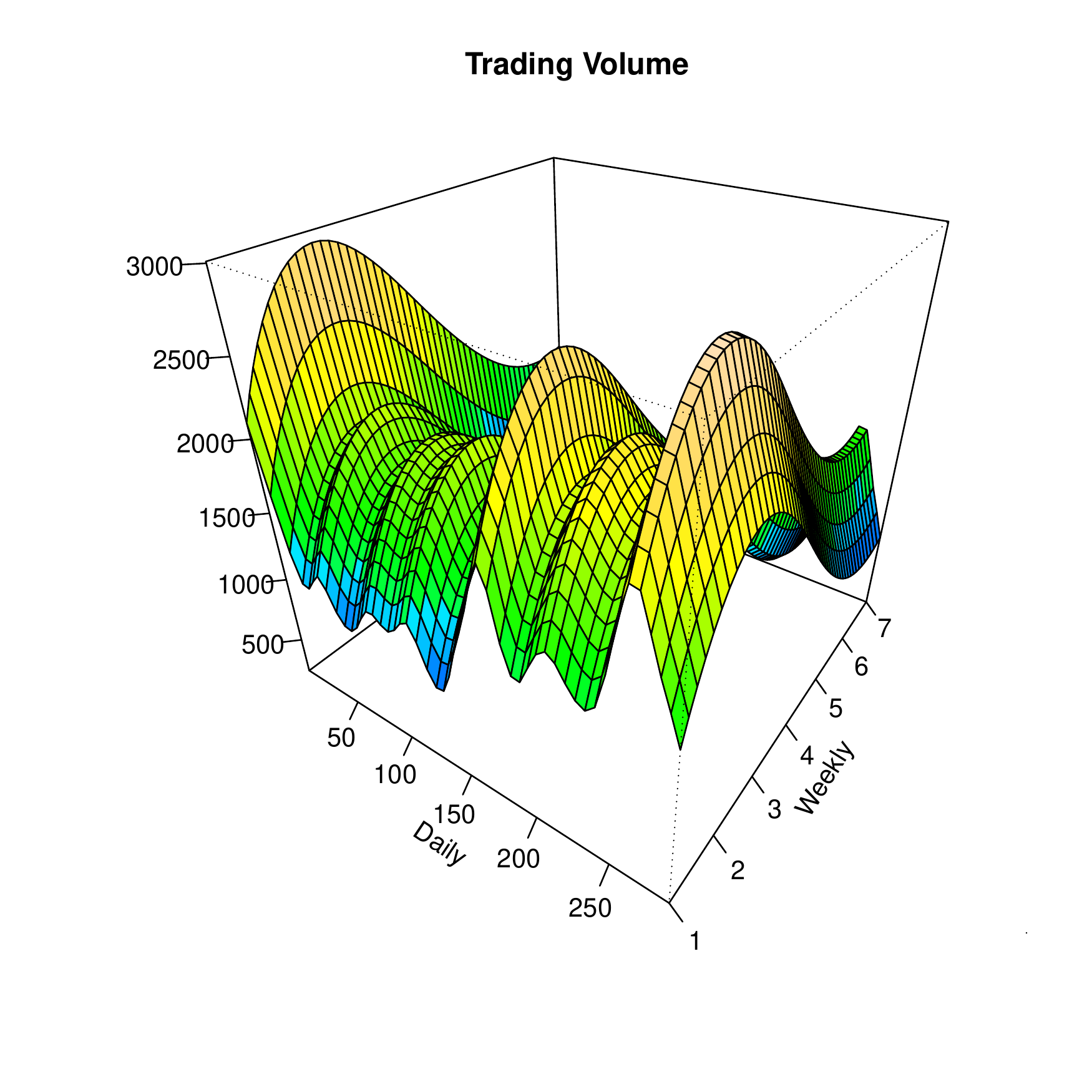}}
\hfill
\href{https://github.com/QuantLet/CCID/tree/master/CCIDvolumeGAM}{\includegraphics[keepaspectratio,width=0.4cm]{media/qletlogo_tr.png}}
\caption{Daily and weekly seasonality: fit of Generalized Additive Model   with cubic and p-splines for trading volume of cryptocurrencies (5 min nodes), 01. July 2018 - 31. August 2018. 01. July 2018 - 31. August 2018.}
\end{figure}

\begin{figure}[H]
\hfill
\subfigure[BCH]{\includegraphics[width=5cm]{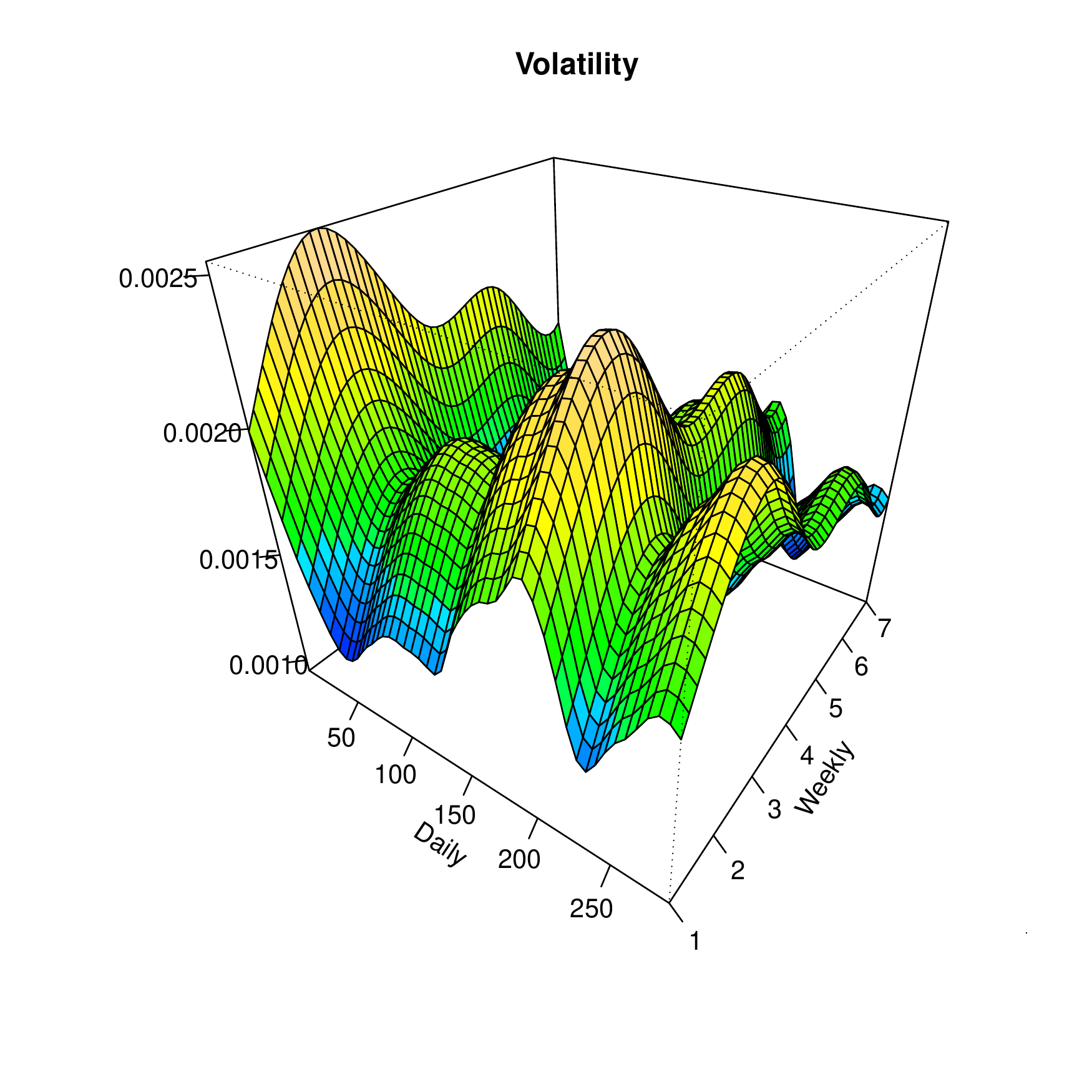}}
\hfill
\subfigure[ETC]{\includegraphics[width=5cm]{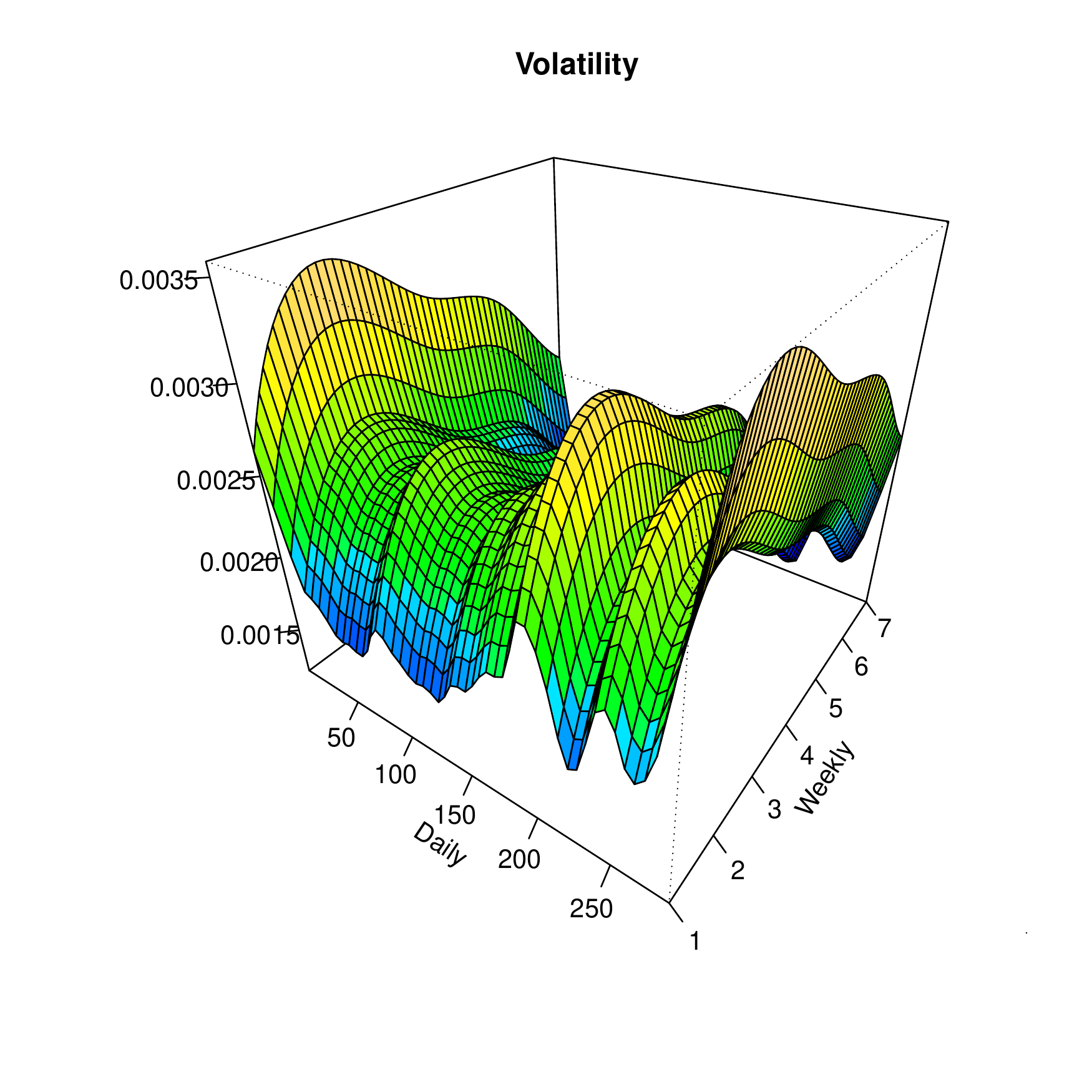}}
\hfill
\subfigure[LTC]{\includegraphics[width=5cm]{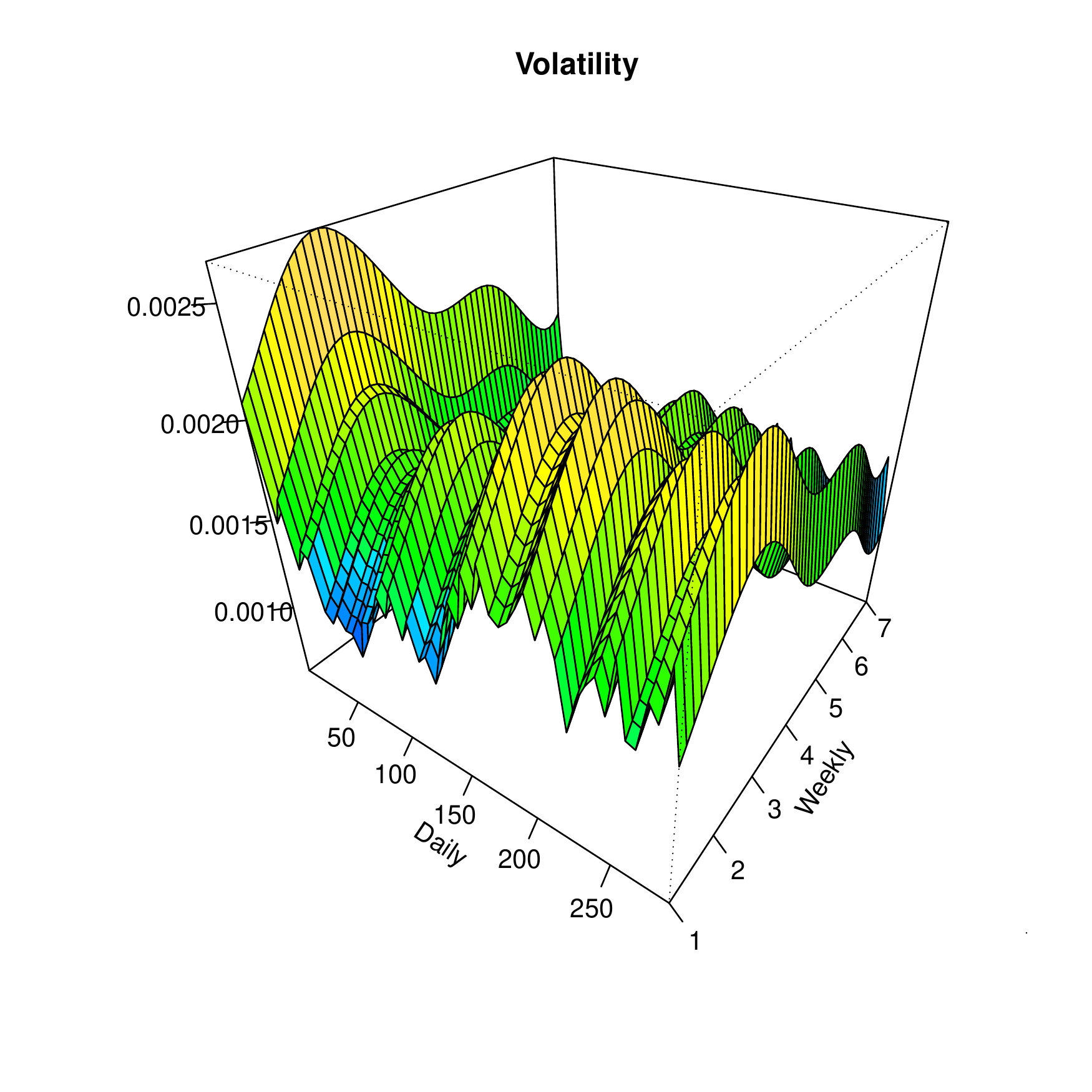}}
\hfill
\href{https://github.com/QuantLet/CCID/tree/master/CCIDvolaGAM}{\includegraphics[keepaspectratio,width=0.4cm]{media/qletlogo_tr.png}}
\caption{Daily and weekly seasonality: fit of Generalized Additive Model   with cubic and p-splines for volatility of cryptocurrencies (5 min nodes), 01. July 2018 - 31. August 2018. 01. July 2018 - 31. August 2018.}
\end{figure}


\subsection{Appendix-Statistics for DASH, REP and STR}

\begin{figure}[H]
\hfill
\subfigure[DASH]{\includegraphics[width=5cm]{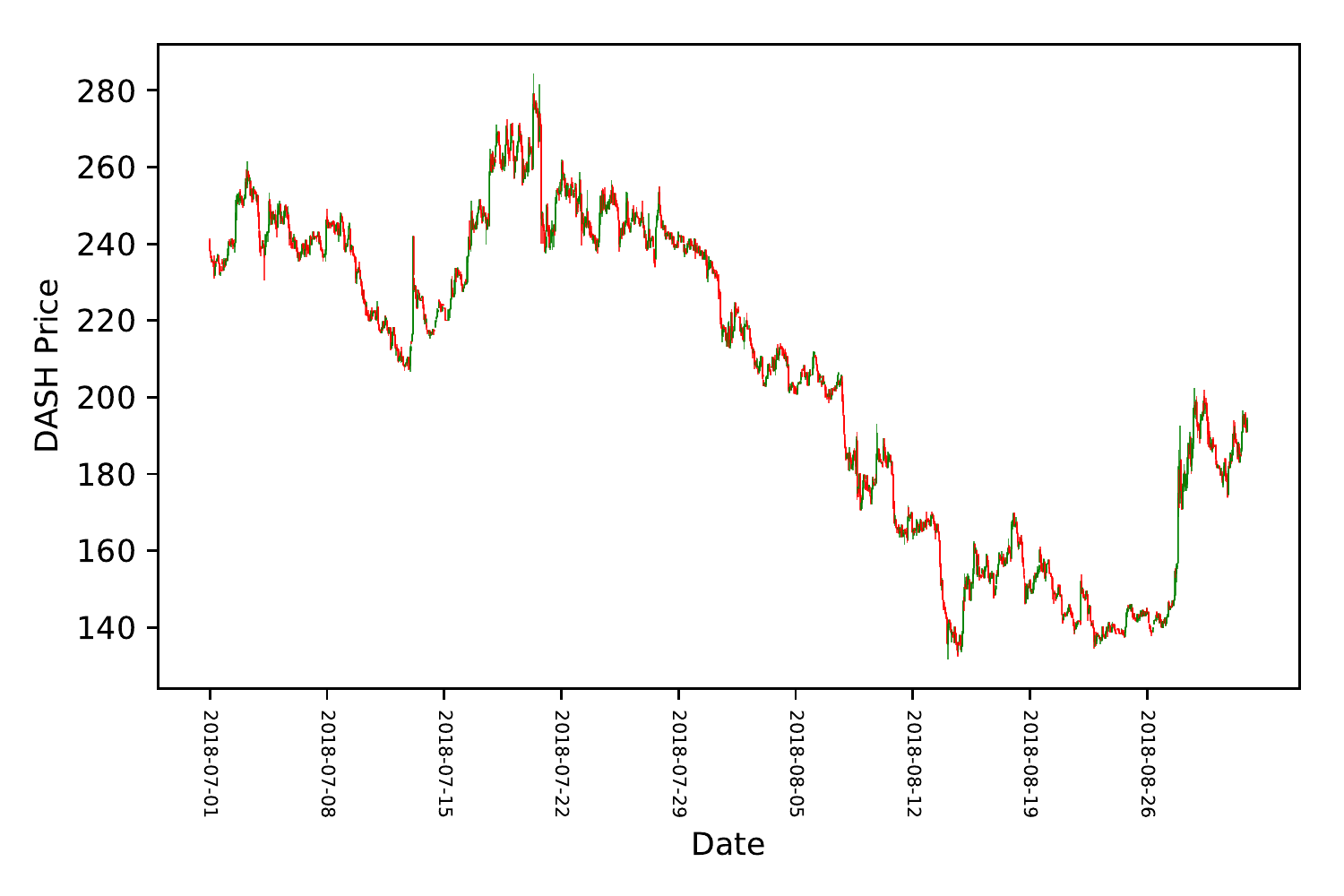}}
\hfill
\subfigure[REP]{\includegraphics[width=5cm]{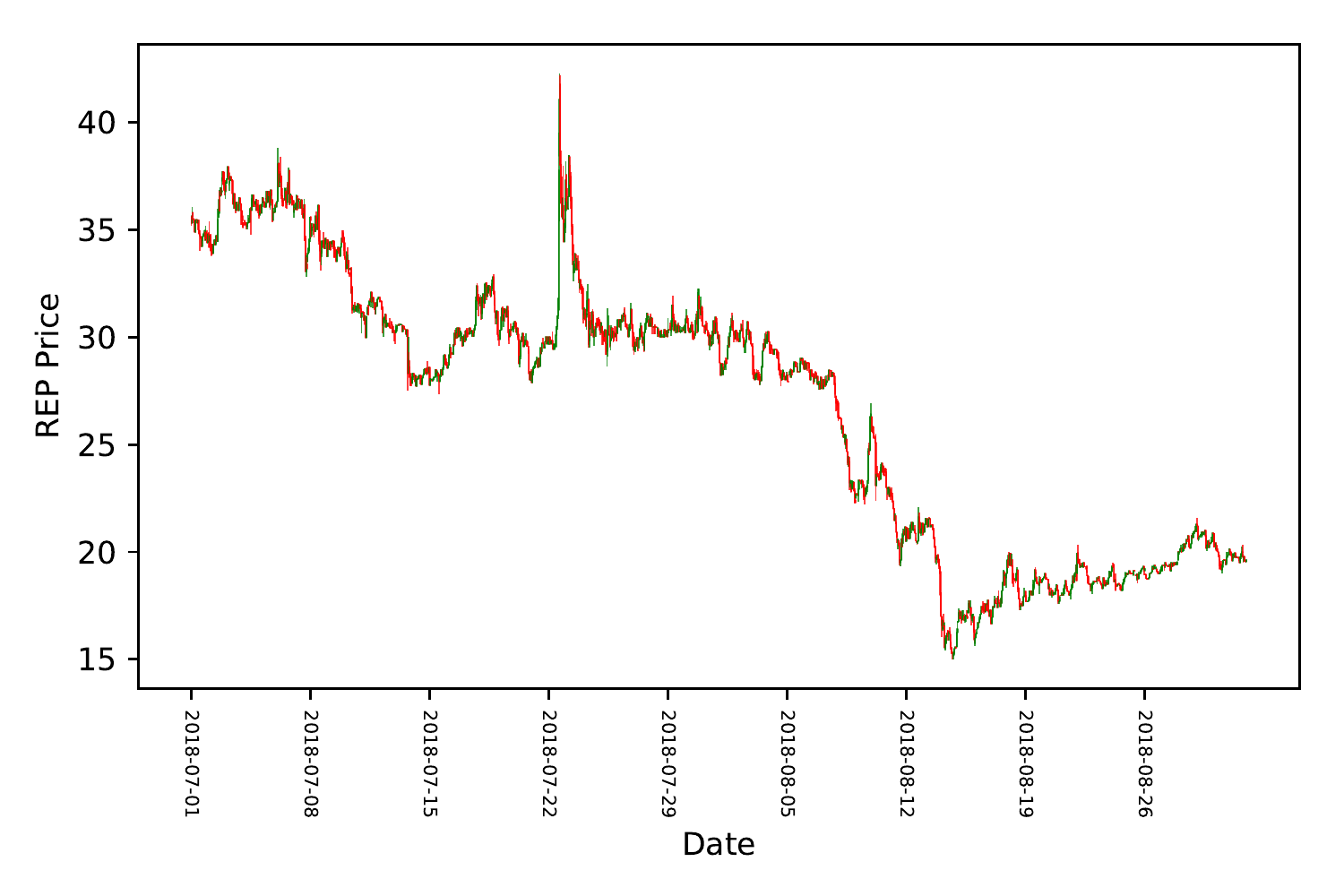}}
\hfill
\subfigure[STR]{\includegraphics[width=5cm]{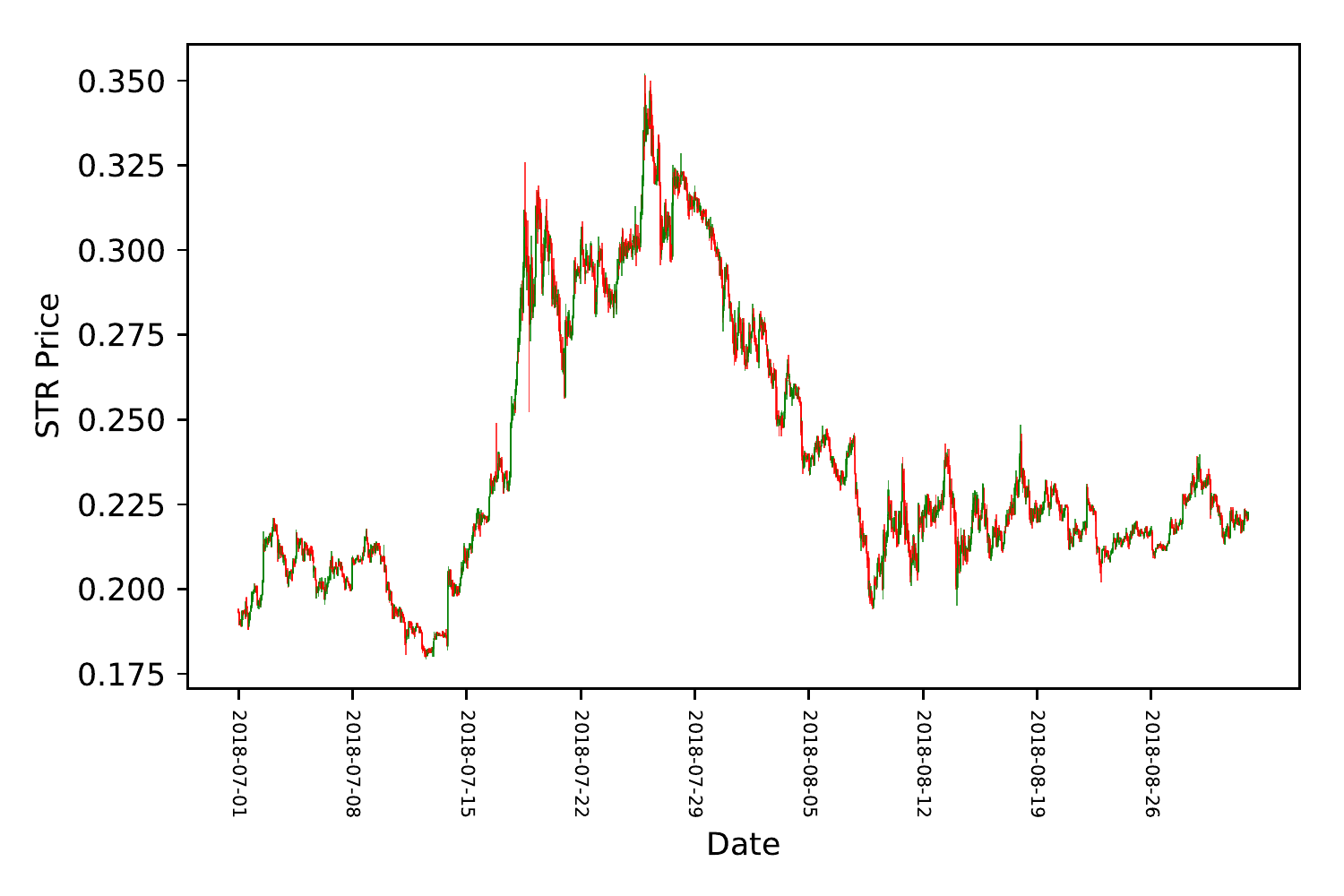}}
\hfill
\href{https://github.com/QuantLet/CCID/tree/master/CCIDCandles}{\includegraphics[keepaspectratio,width=0.4cm]{media/qletlogo_tr.png}}
\caption{Candlestick charts for individual price movements (60-minutes intervals). 01. July 2018 - 31. August 2018.}
\end{figure}

\begin{figure}[H]
\hfill
\subfigure[DASH]{\includegraphics[width=5cm]{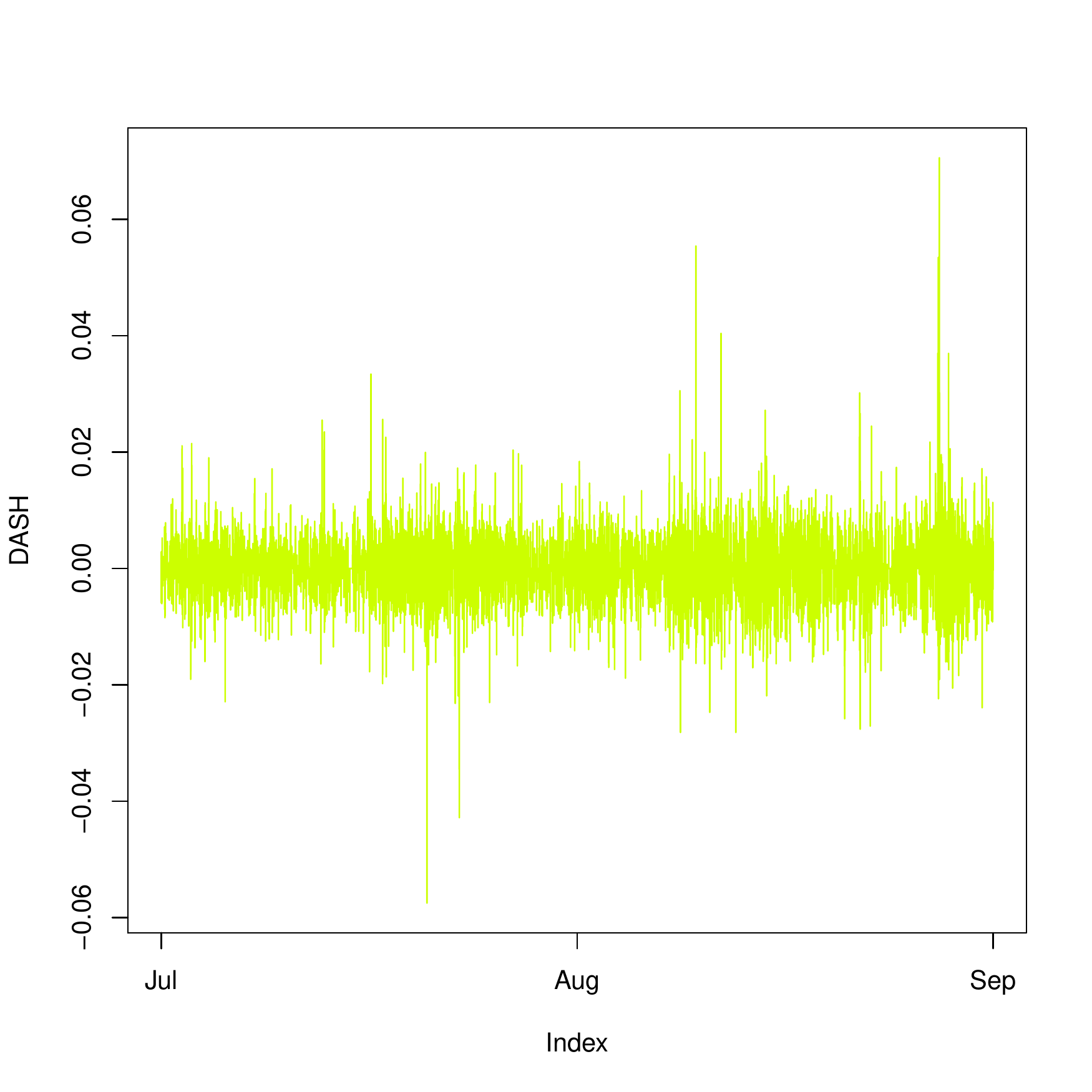}}
\hfill
\subfigure[REP]{\includegraphics[width=5cm]{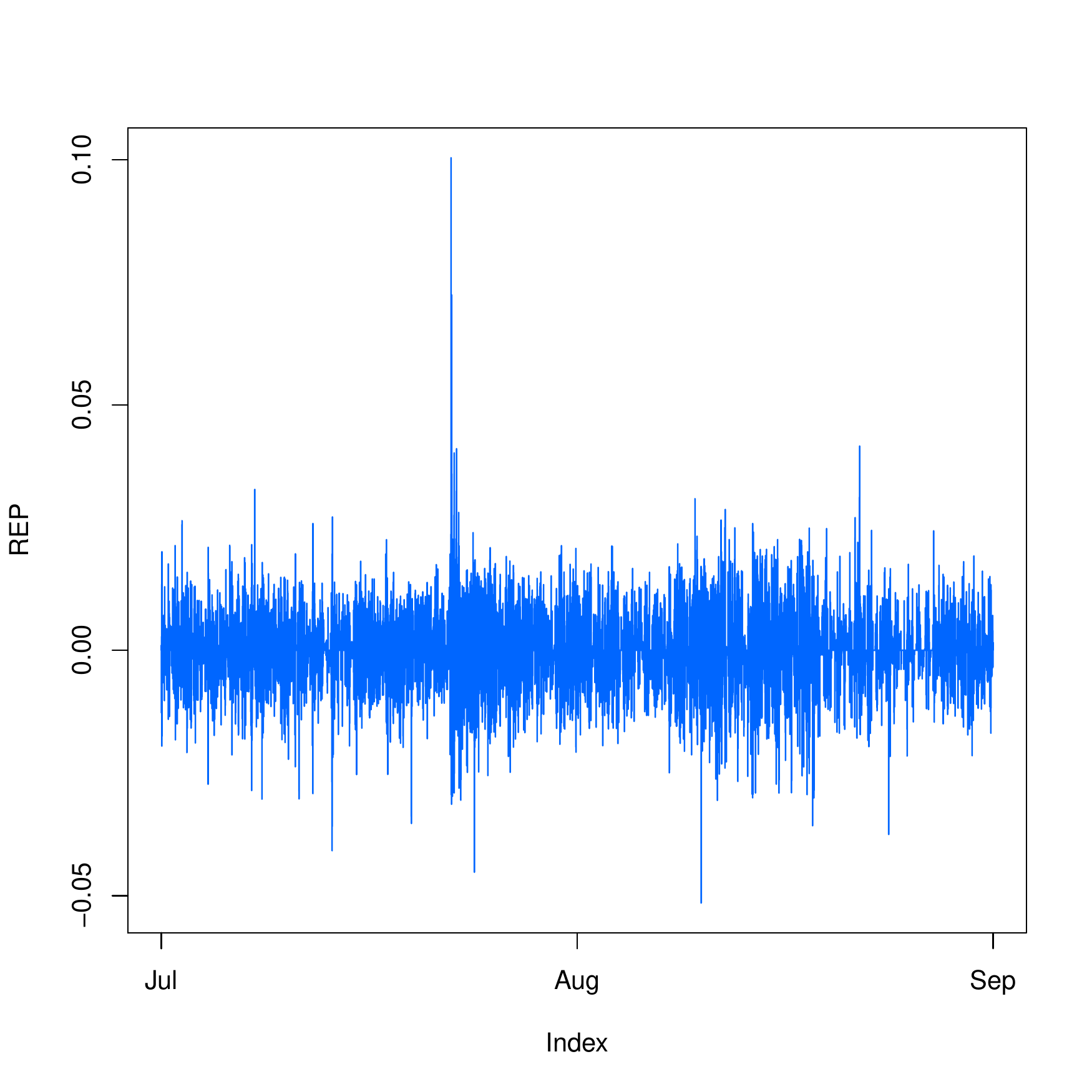}}
\hfill
\subfigure[STR]{\includegraphics[width=5cm]{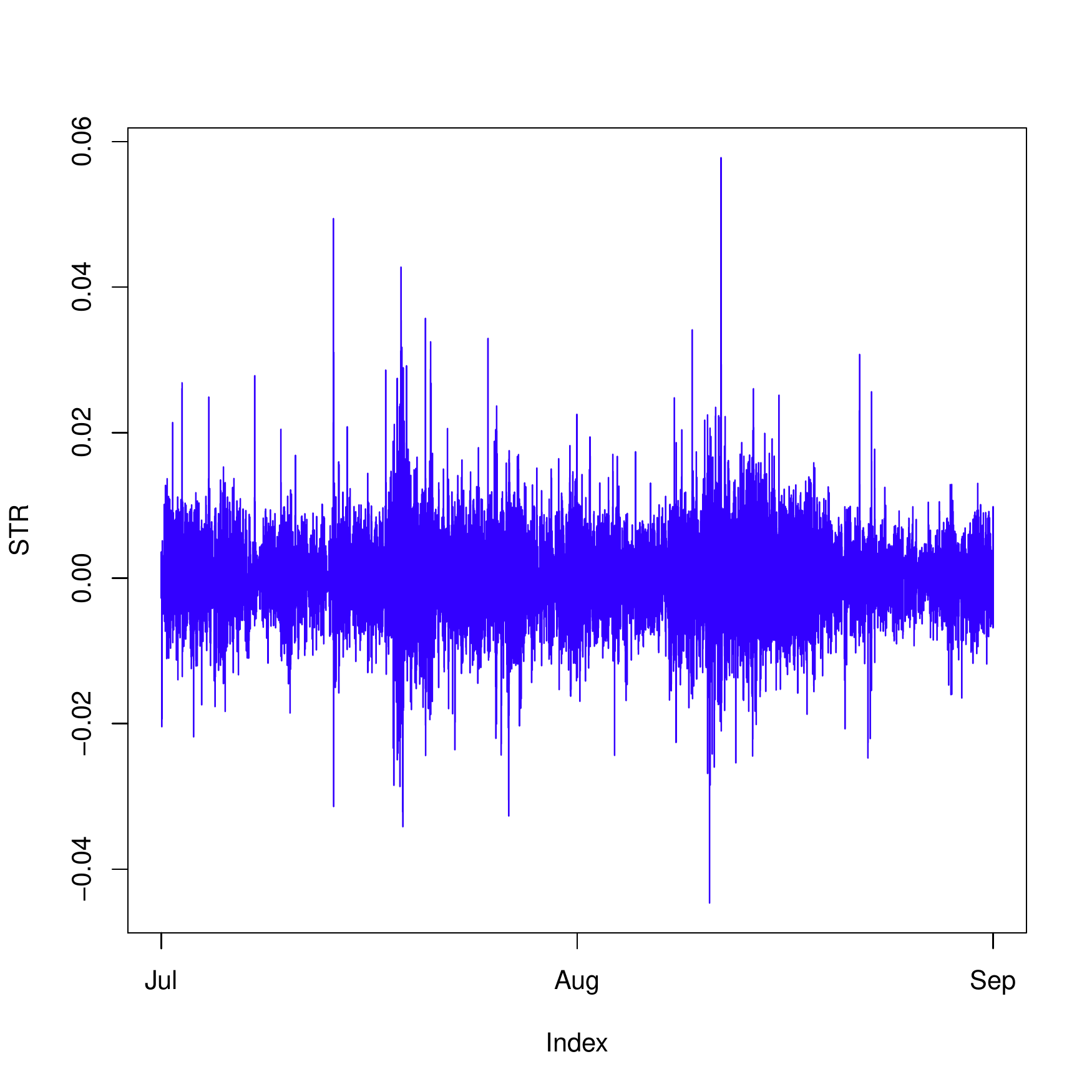}}
\hfill
\href{https://github.com/QuantLet/CCID/tree/master/CCIDHistRet}{\includegraphics[keepaspectratio,width=0.4cm]{media/qletlogo_tr.png}}
\caption{Intraday log-returns (5-minutes). 01. July 2018 - 31. August 2018.}
\end{figure}

\begin{figure}[H]
\hfill
\subfigure[DASH]{\includegraphics[width=5cm]{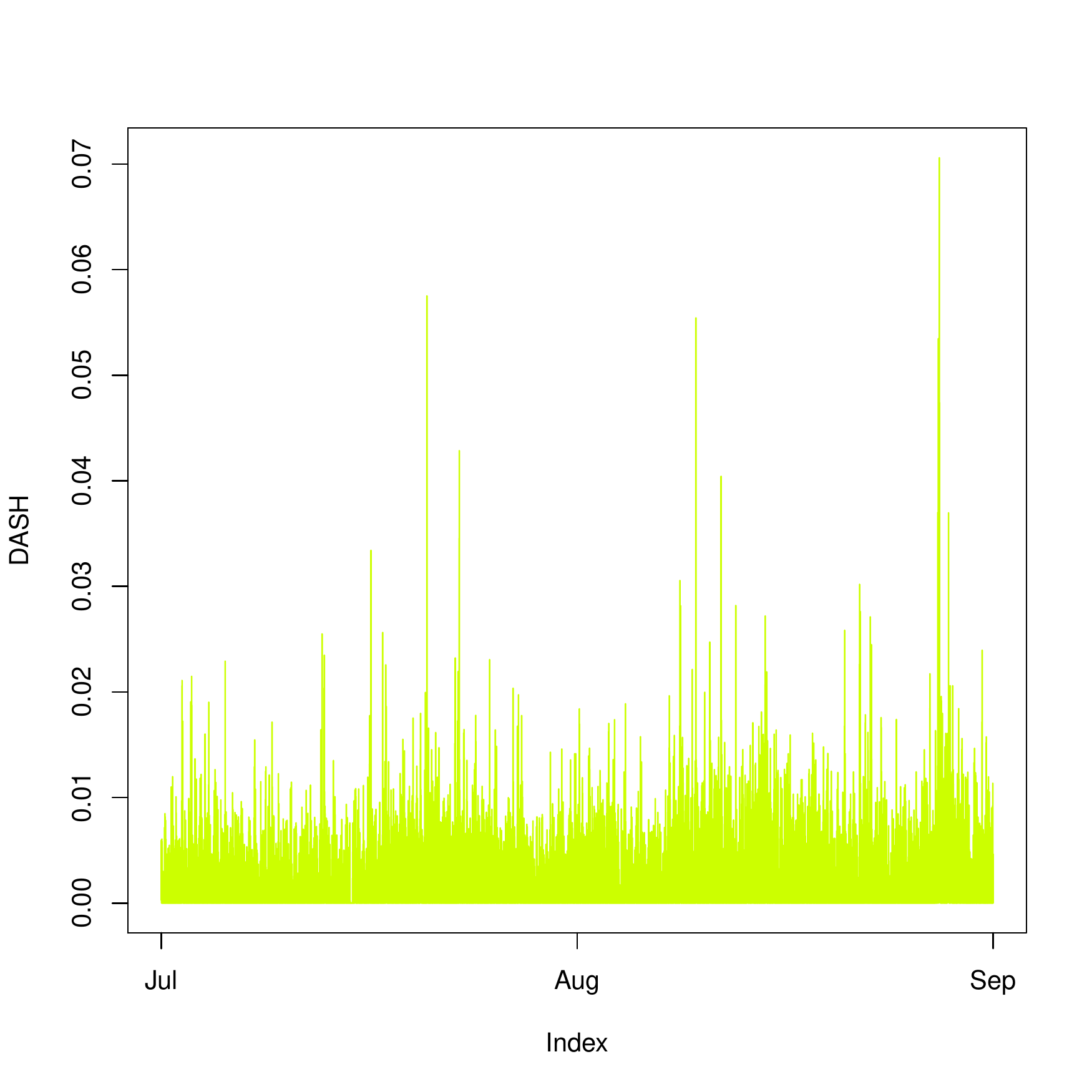}}
\hfill
\subfigure[REP]{\includegraphics[width=5cm]{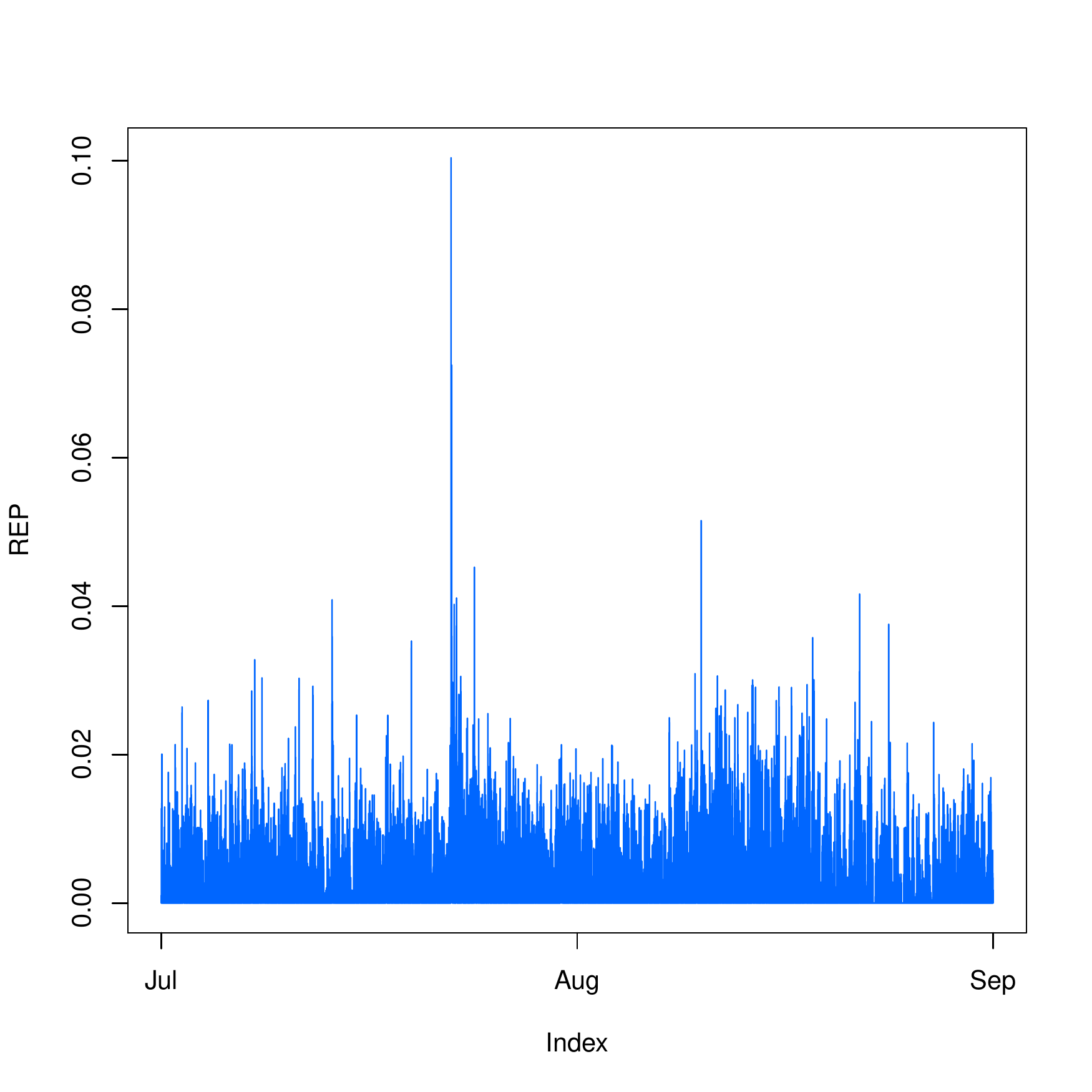}}
\hfill
\subfigure[STR]{\includegraphics[width=5cm]{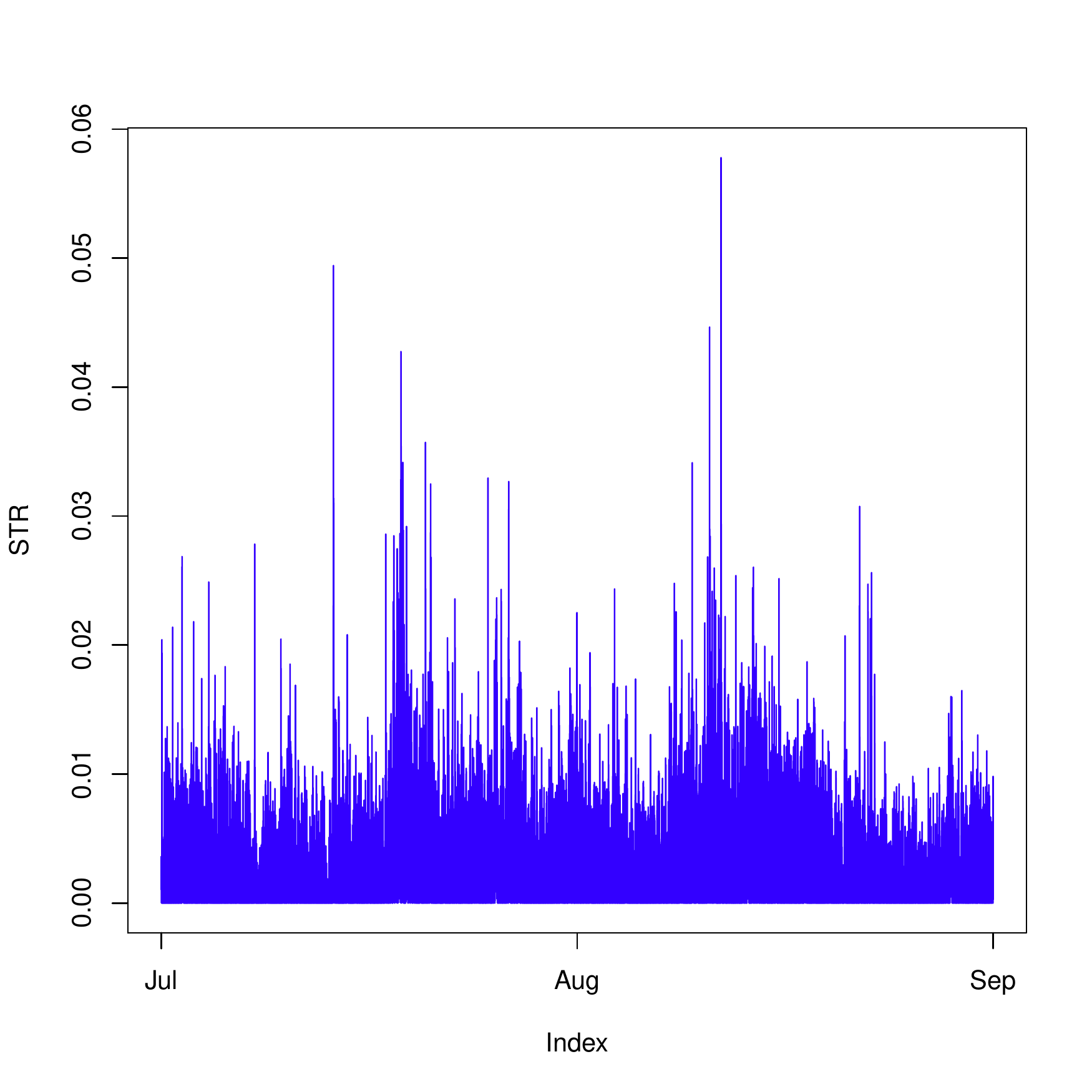}}
\hfill
\href{https://github.com/QuantLet/CCID/tree/master/CCIDHistVola}{\includegraphics[keepaspectratio,width=0.4cm]{media/qletlogo_tr.png}}
\caption{Intraday volatility (absolute 5-minutes log-returns). 01. July 2018 - 31. August 2018.}
\end{figure}

\begin{figure}[H]
\hfill
\subfigure[DASH]{\includegraphics[width=5cm]{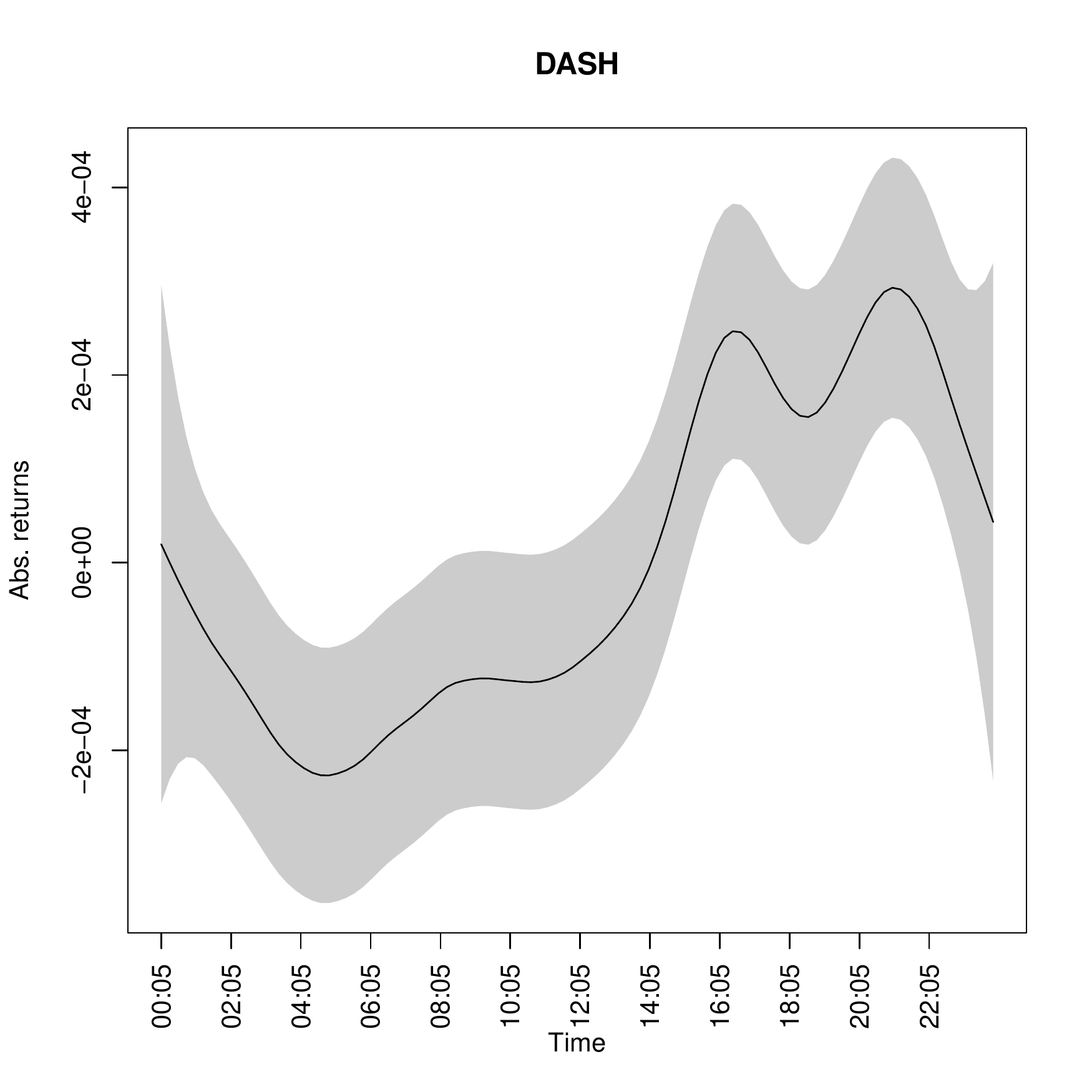}}
\hfill
\subfigure[REP]{\includegraphics[width=5cm]{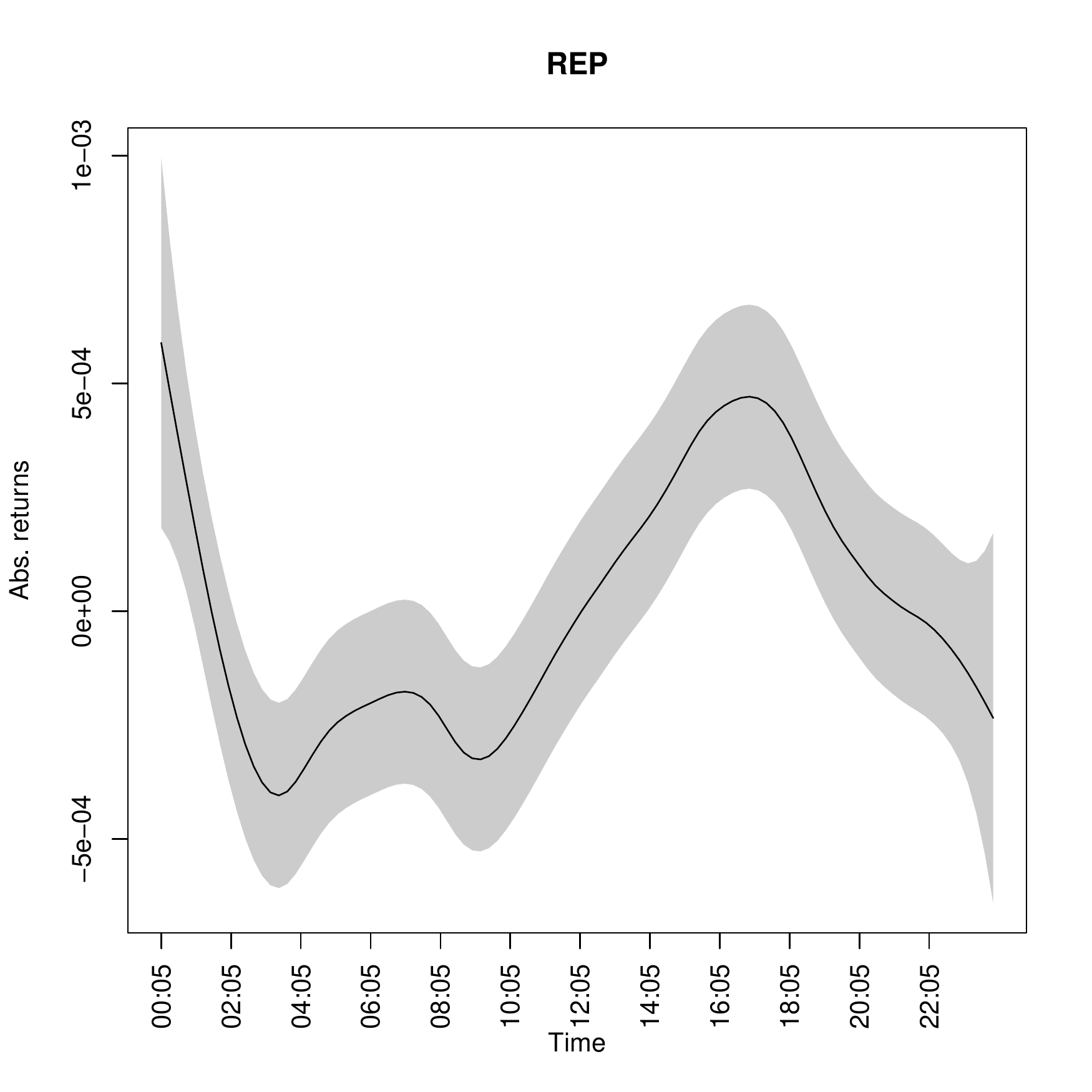}}
\hfill
\subfigure[STR]{\includegraphics[width=5cm]{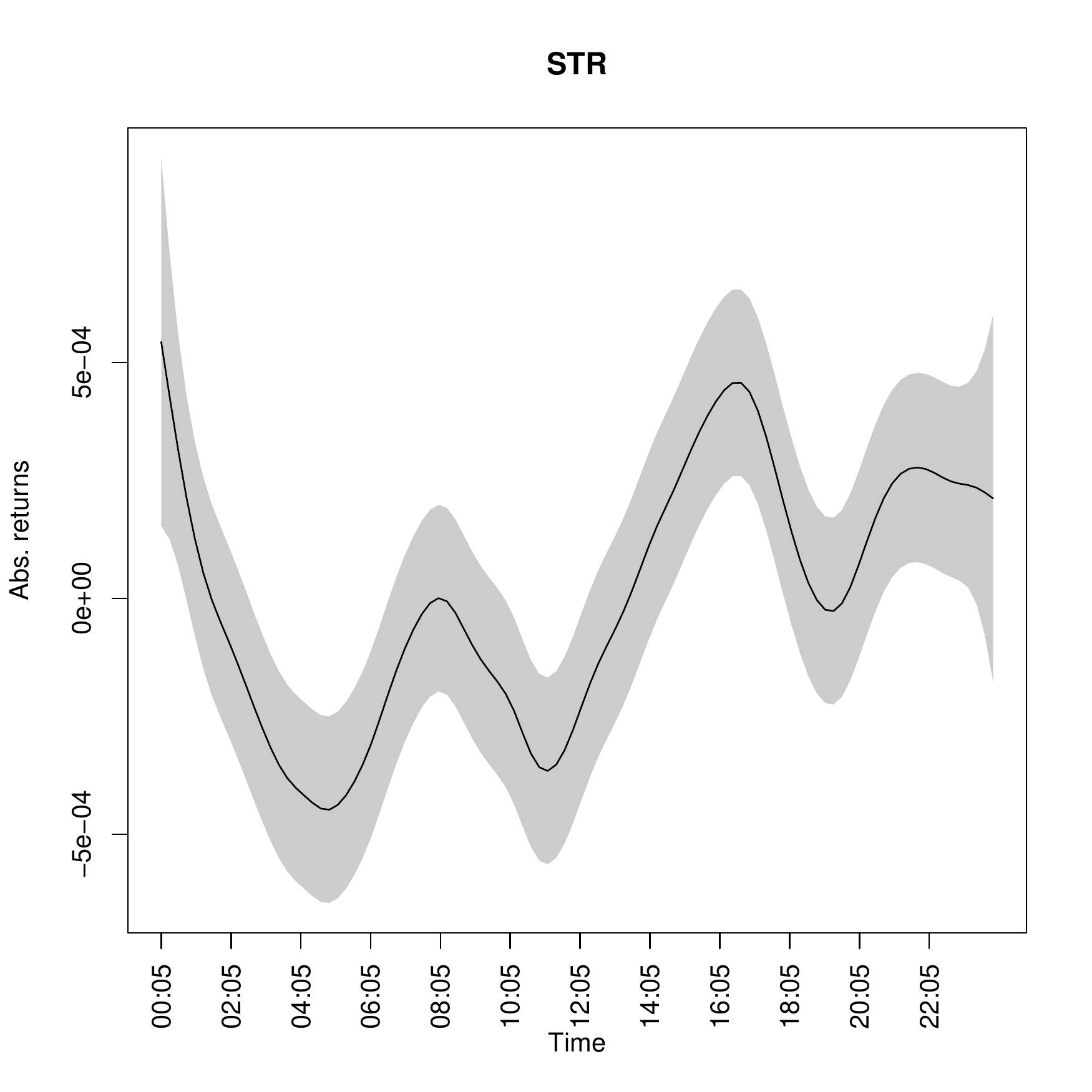}}
\hfill
\href{https://github.com/QuantLet/CCID/tree/master/CCIDvolaGAM}{\includegraphics[keepaspectratio,width=0.4cm]{media/qletlogo_tr.png}}
\caption{Generalized Additive Model of volatility of cryptocurrencies. 01. July 2018 - 31. August 2018.}
\end{figure}

\begin{figure}[H]
\hfill
\subfigure[DASH]{\includegraphics[width=5cm]{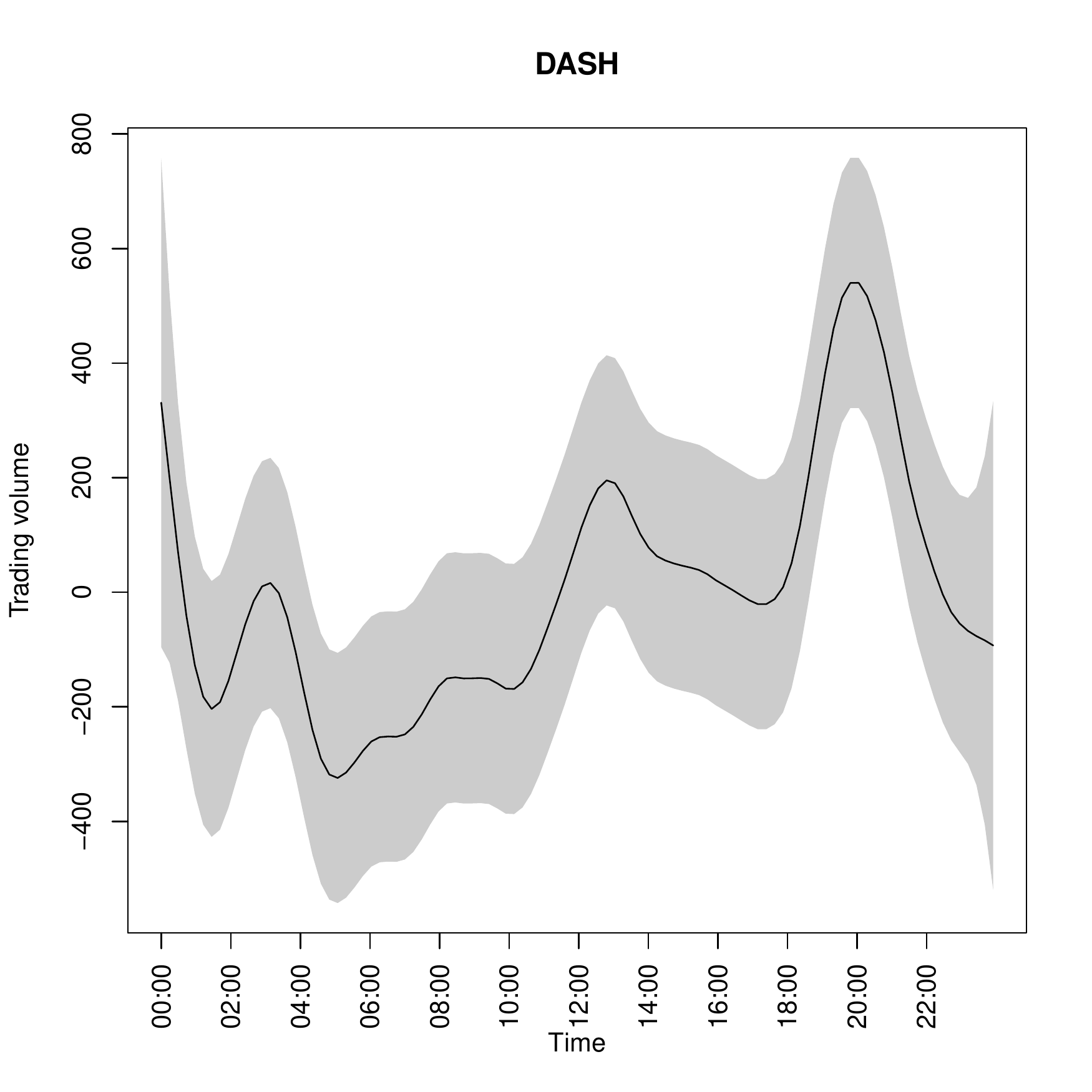}}
\hfill
\subfigure[REP]{\includegraphics[width=5cm]{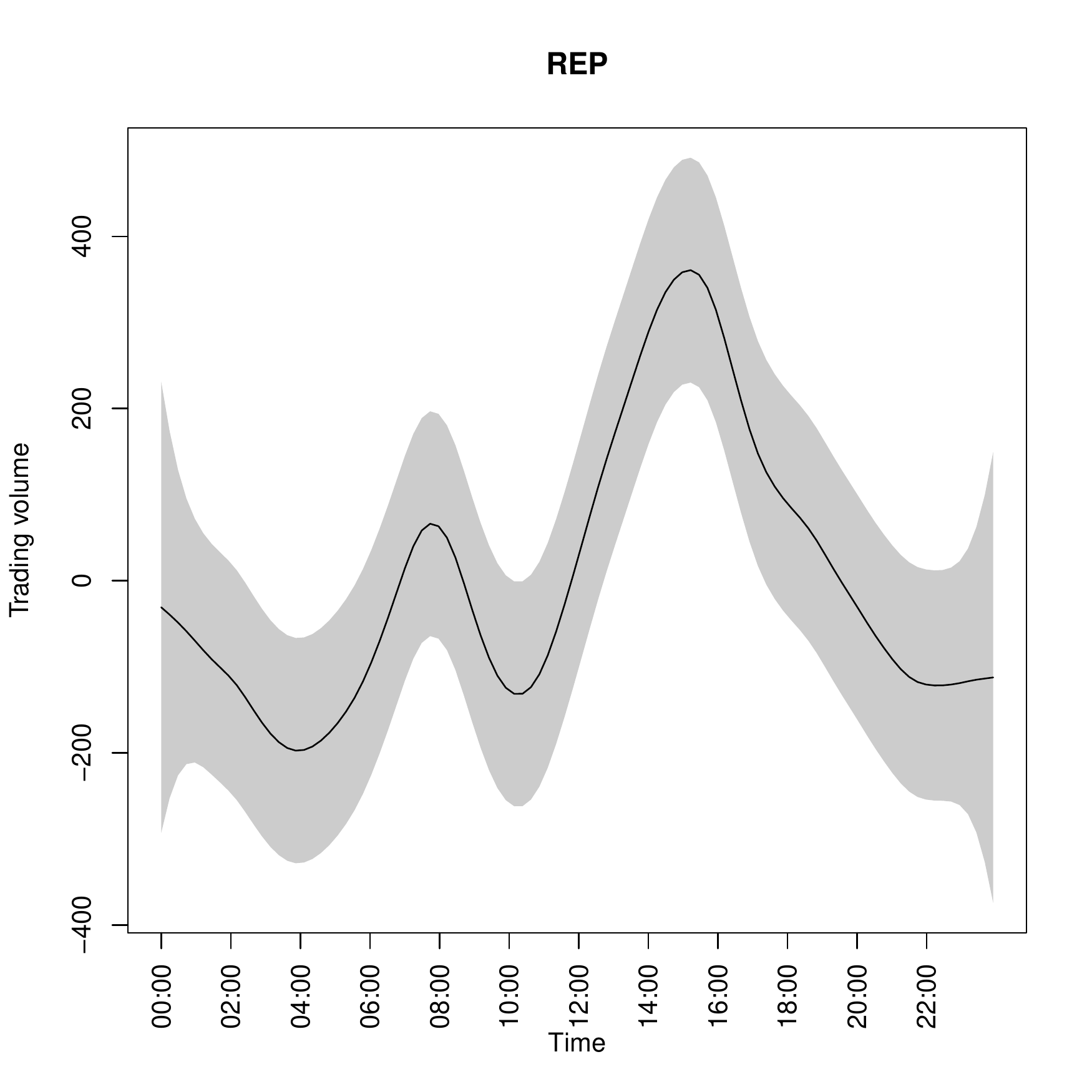}}
\hfill
\subfigure[STR]{\includegraphics[width=5cm]{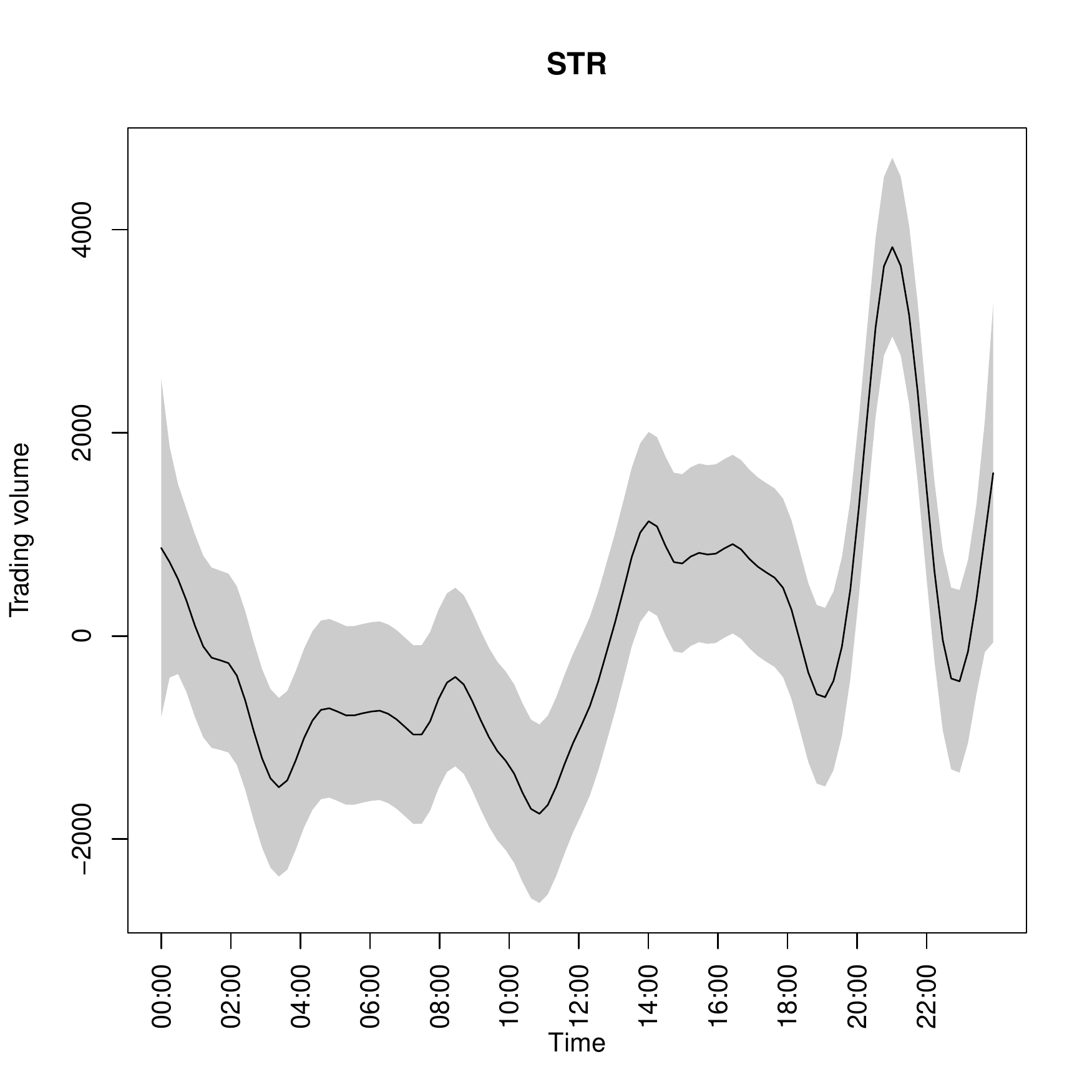}}
\hfill
\href{https://github.com/QuantLet/CCID/tree/master/CCIDvolumeGAM}{\includegraphics[keepaspectratio,width=0.4cm]{media/qletlogo_tr.png}}
\caption{Generalized Additive Model of intraday trading volume of cryptocurrencies. 01. July 2018 - 31. August 2018.}
\end{figure}

\begin{figure}[H]
\hfill
\subfigure[DASH]{\includegraphics[width=5cm]{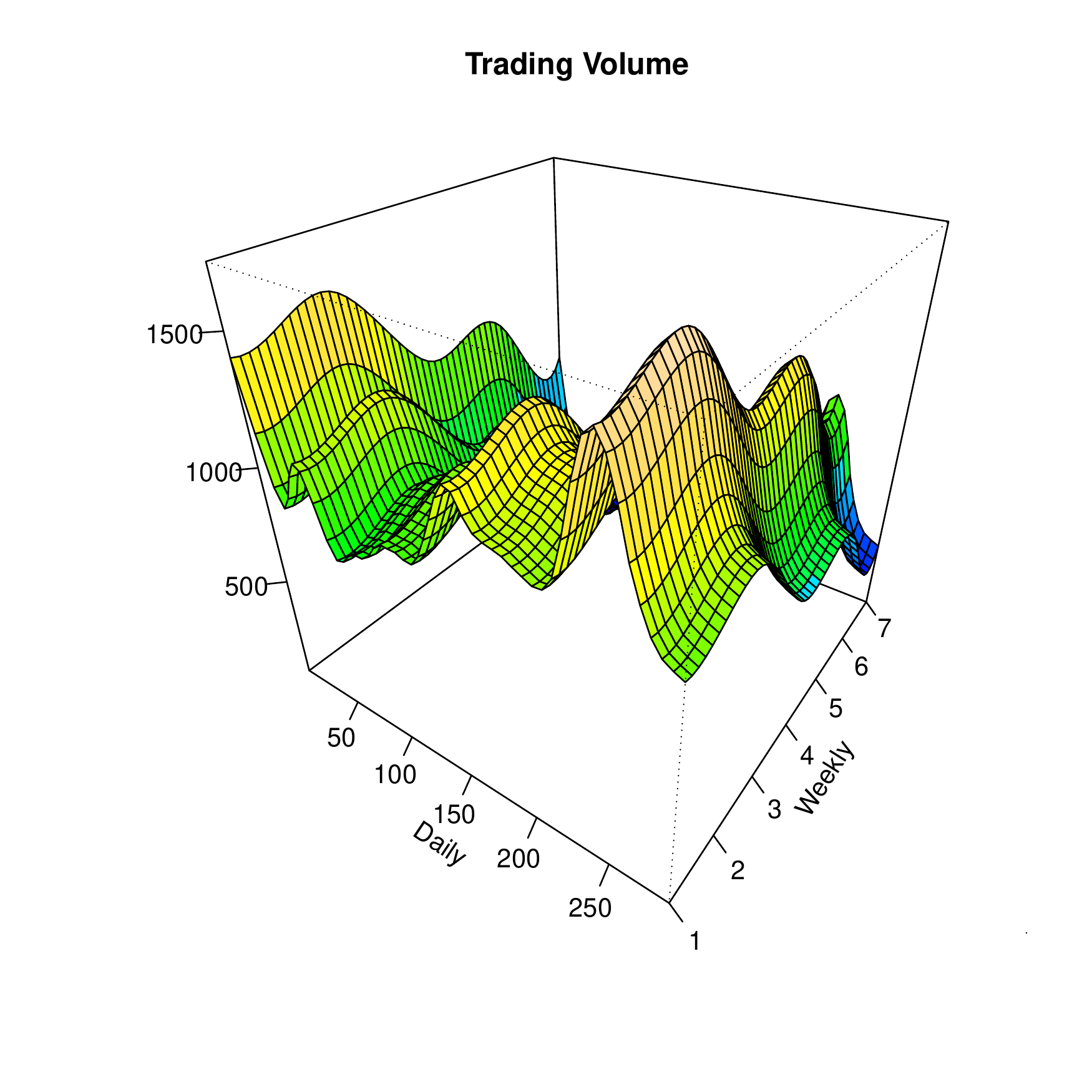}}
\hfill
\subfigure[REP]{\includegraphics[width=5cm]{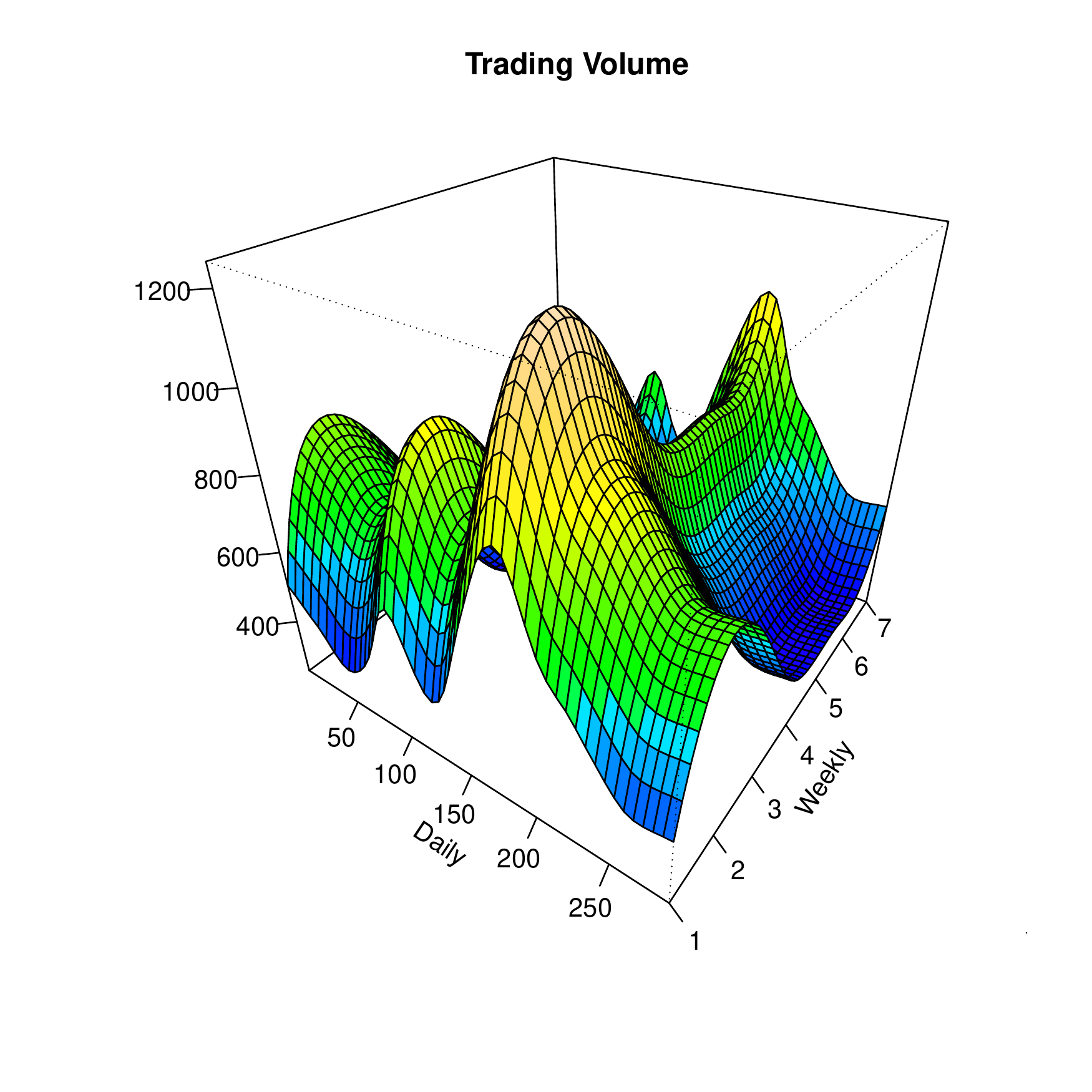}}
\hfill
\subfigure[STR]{\includegraphics[width=5cm]{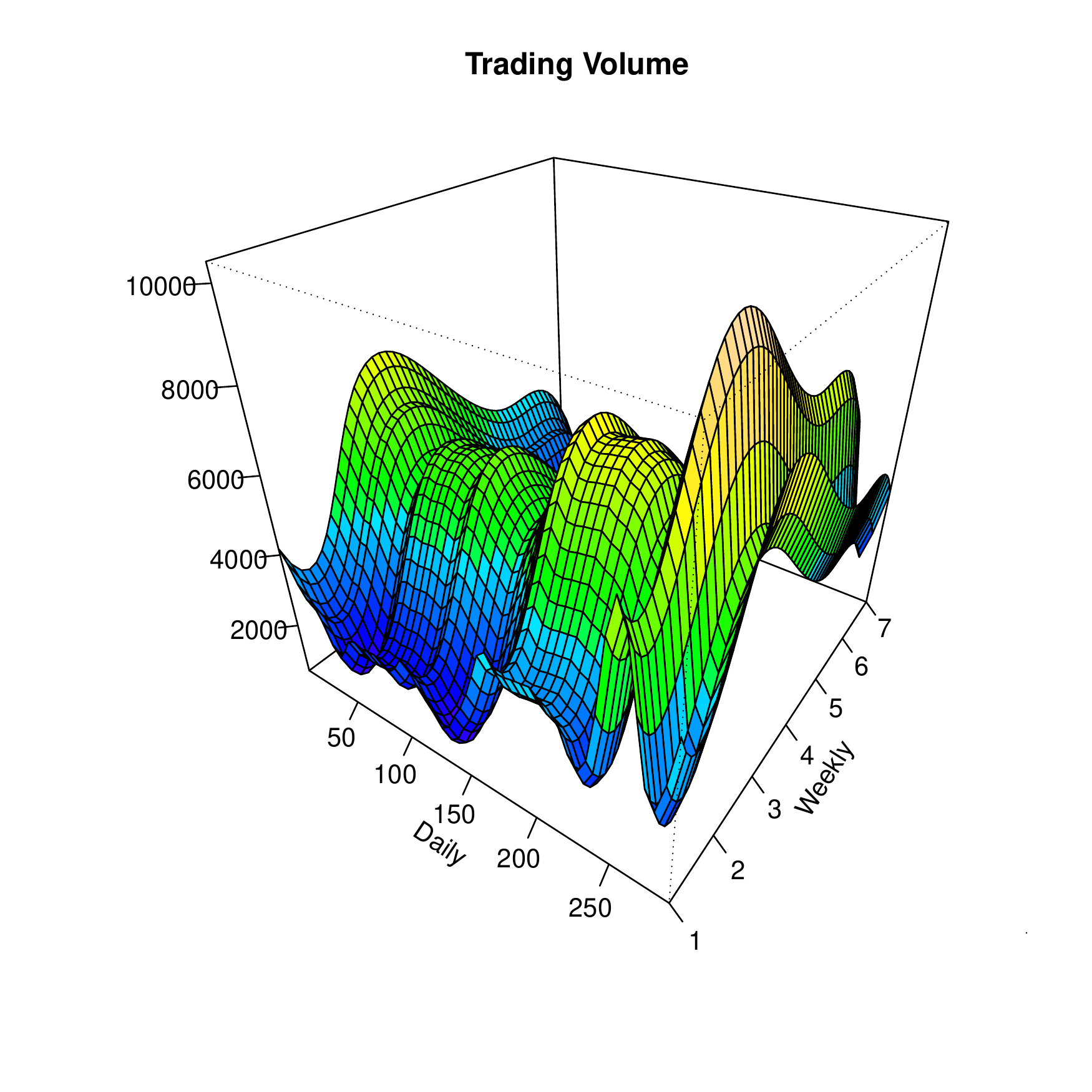}}
\hfill
\href{https://github.com/QuantLet/CCID/tree/master/CCIDvolumeGAM}{\includegraphics[keepaspectratio,width=0.4cm]{media/qletlogo_tr.png}}
\caption{Daily and weekly seasonality: fit of Generalized Additive Model   with cubic and p-splines for trading volume of cryptocurrencies (5 min nodes), 01. July 2018 - 31. August 2018. 01. July 2018 - 31. August 2018.}
\end{figure}

\begin{figure}[H]
\hfill
\subfigure[DASH]{\includegraphics[width=5cm]{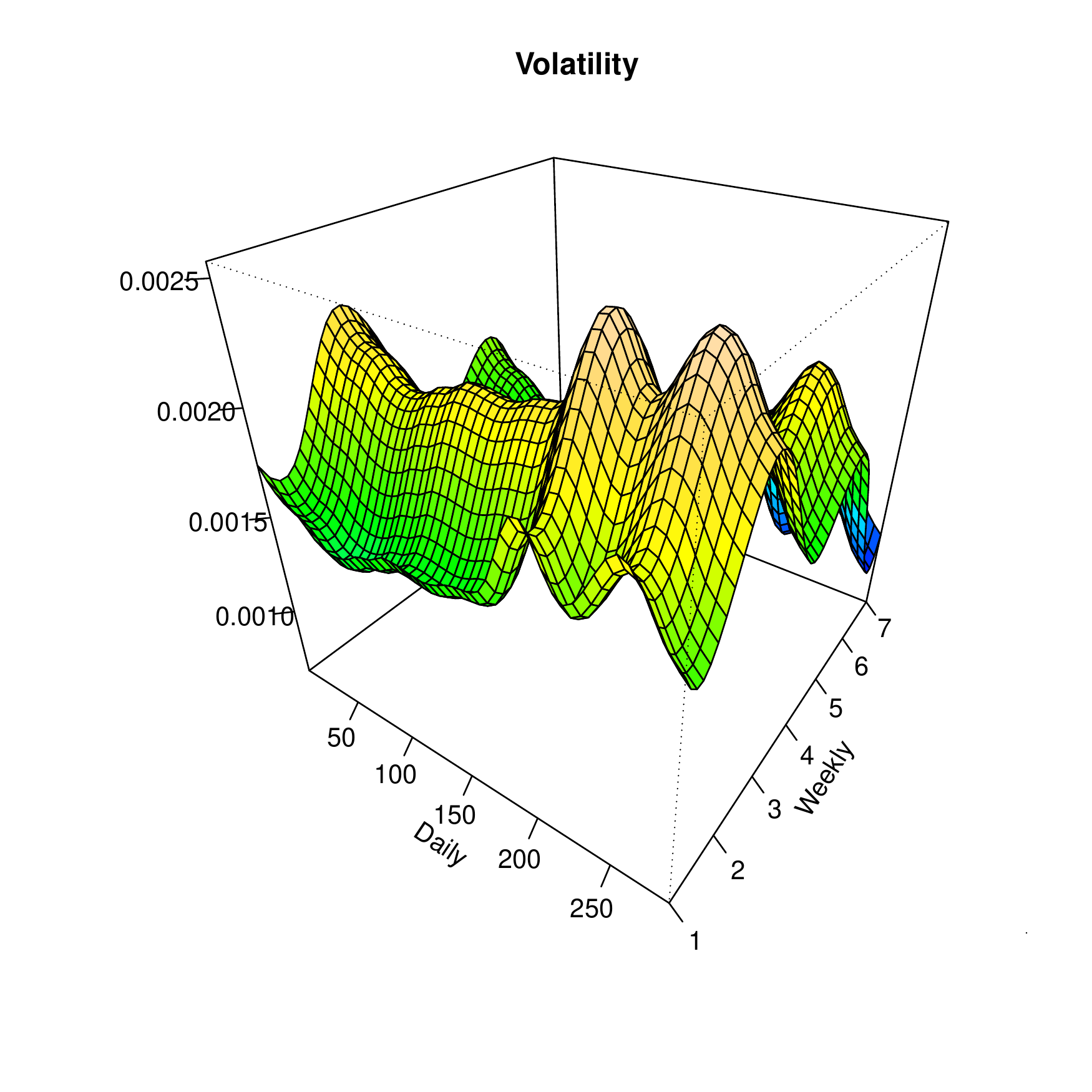}}
\hfill
\subfigure[REP]{\includegraphics[width=5cm]{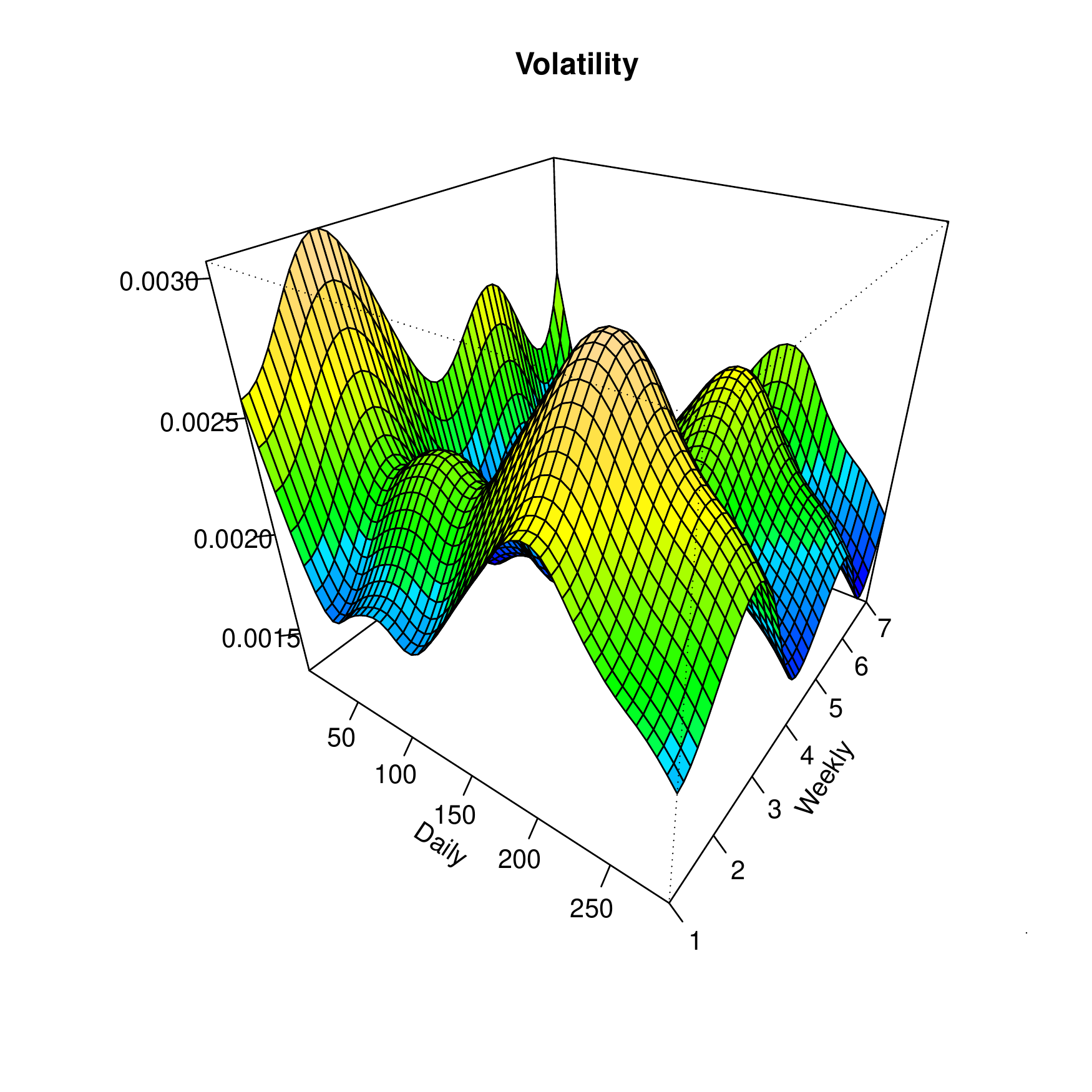}}
\hfill
\subfigure[STR]{\includegraphics[width=5cm]{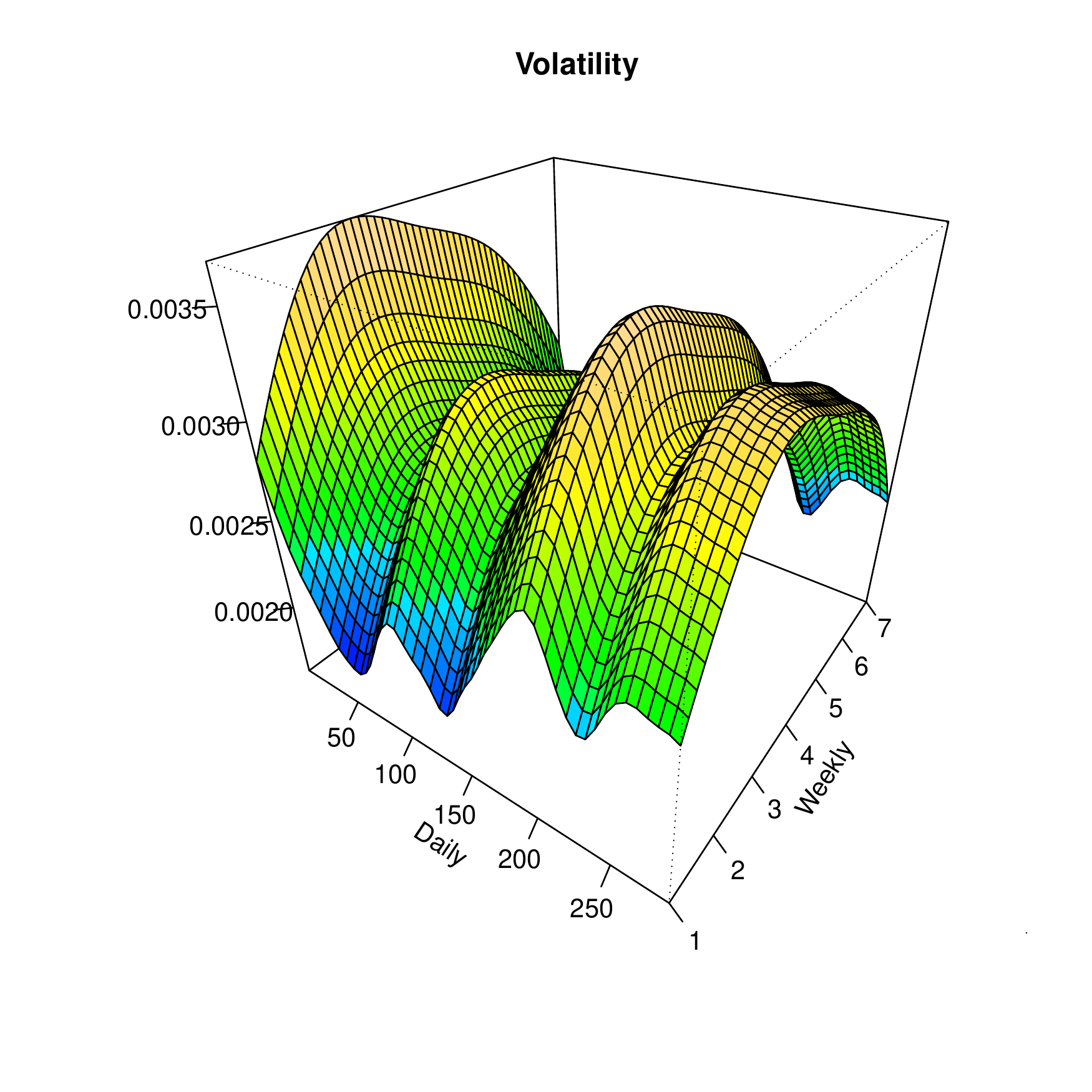}}
\hfill
\href{https://github.com/QuantLet/CCID/tree/master/CCIDvolaGAM}{\includegraphics[keepaspectratio,width=0.4cm]{media/qletlogo_tr.png}}
\caption{Daily and weekly seasonality: fit of Generalized Additive Model   with cubic and p-splines for volatility of cryptocurrencies (5 min nodes), 01. July 2018 - 31. August 2018. 01. July 2018 - 31. August 2018.}
\end{figure}


\subsection{Appendix-Statistics for XMR and ZEC}

\begin{figure}[H]
\hfill
\subfigure[XMR]{\includegraphics[width=5cm]{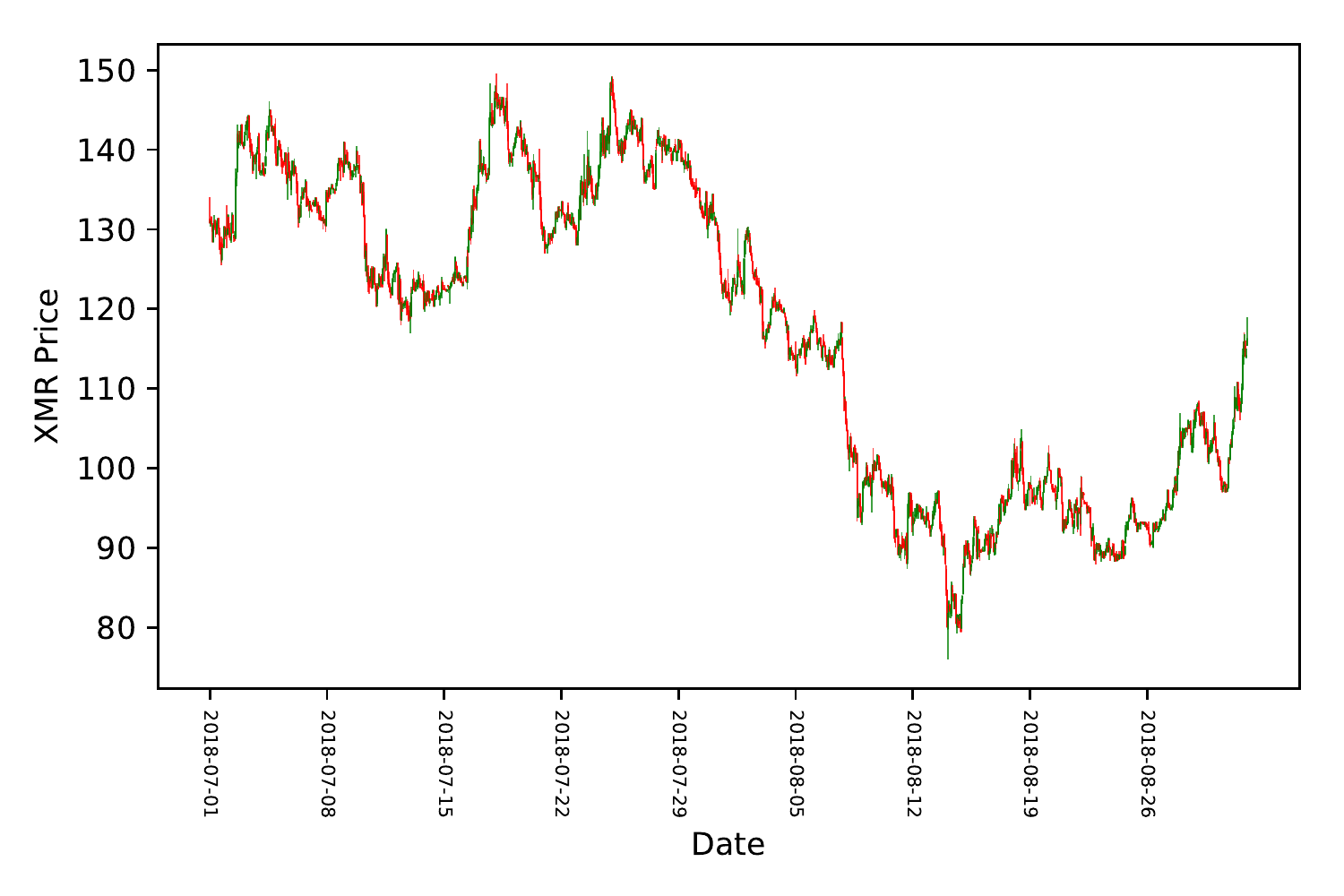}}
\hfill
\subfigure[ZEC]{\includegraphics[width=5cm]{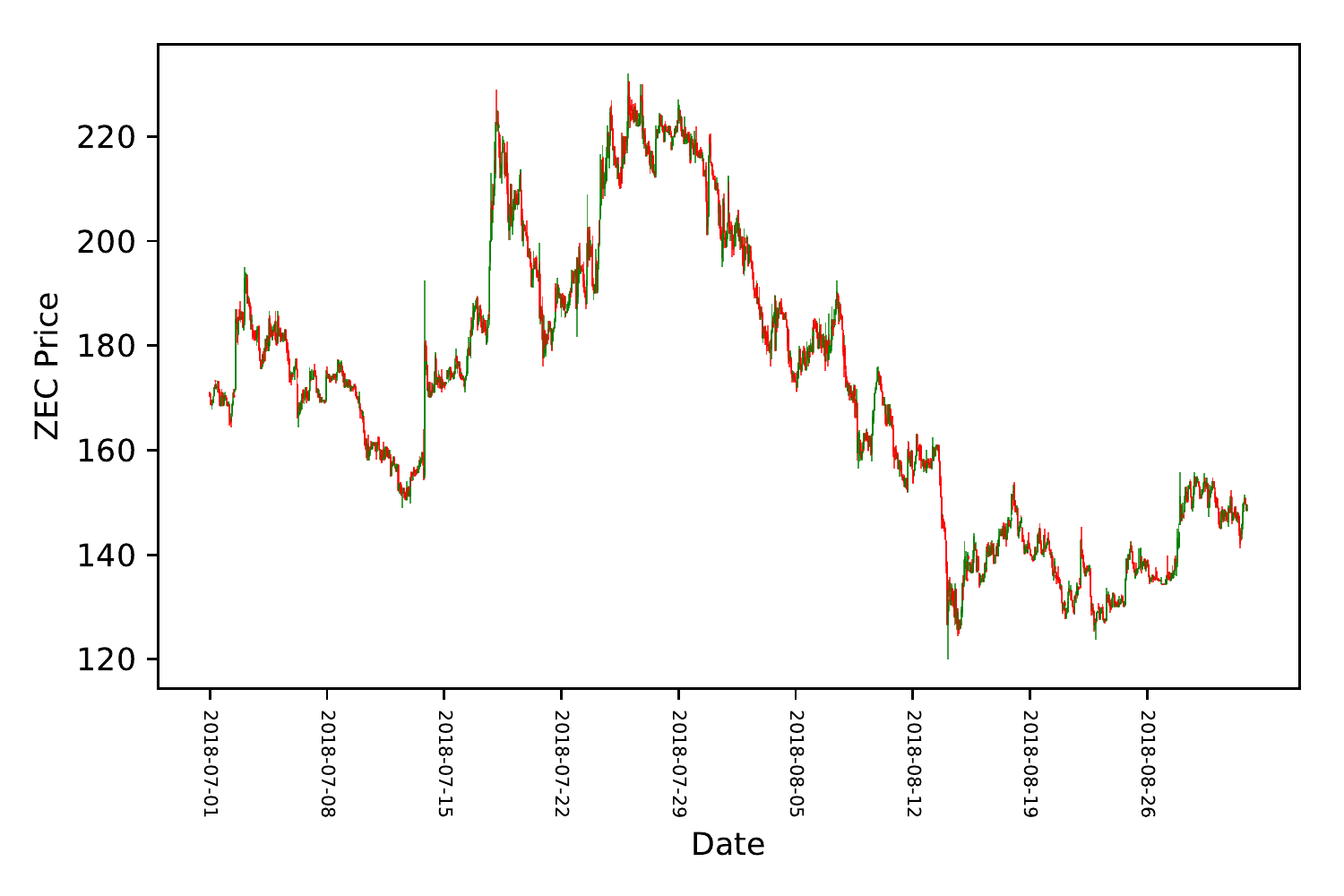}}
\hfill
\href{https://github.com/QuantLet/CCID/tree/master/CCIDCandles}{\includegraphics[keepaspectratio,width=0.4cm]{media/qletlogo_tr.png}}
\caption{Chandlestick charts for individual price movements. 01. July 2018 - 31. August 2018.}
\end{figure}

\begin{figure}[H]
\hfill
\subfigure[XMR]{\includegraphics[width=5cm]{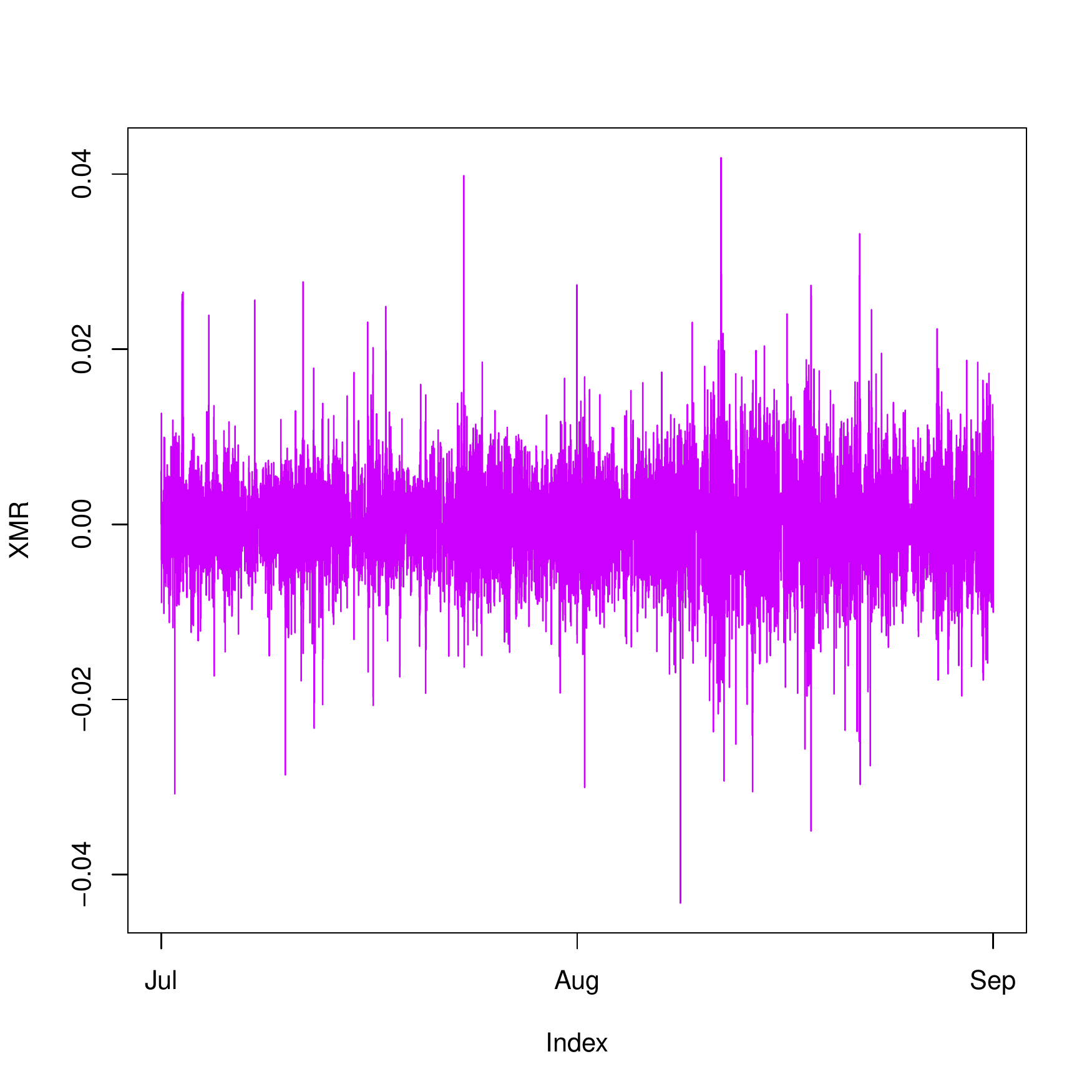}}
\hfill
\subfigure[ZEC]{\includegraphics[width=5cm]{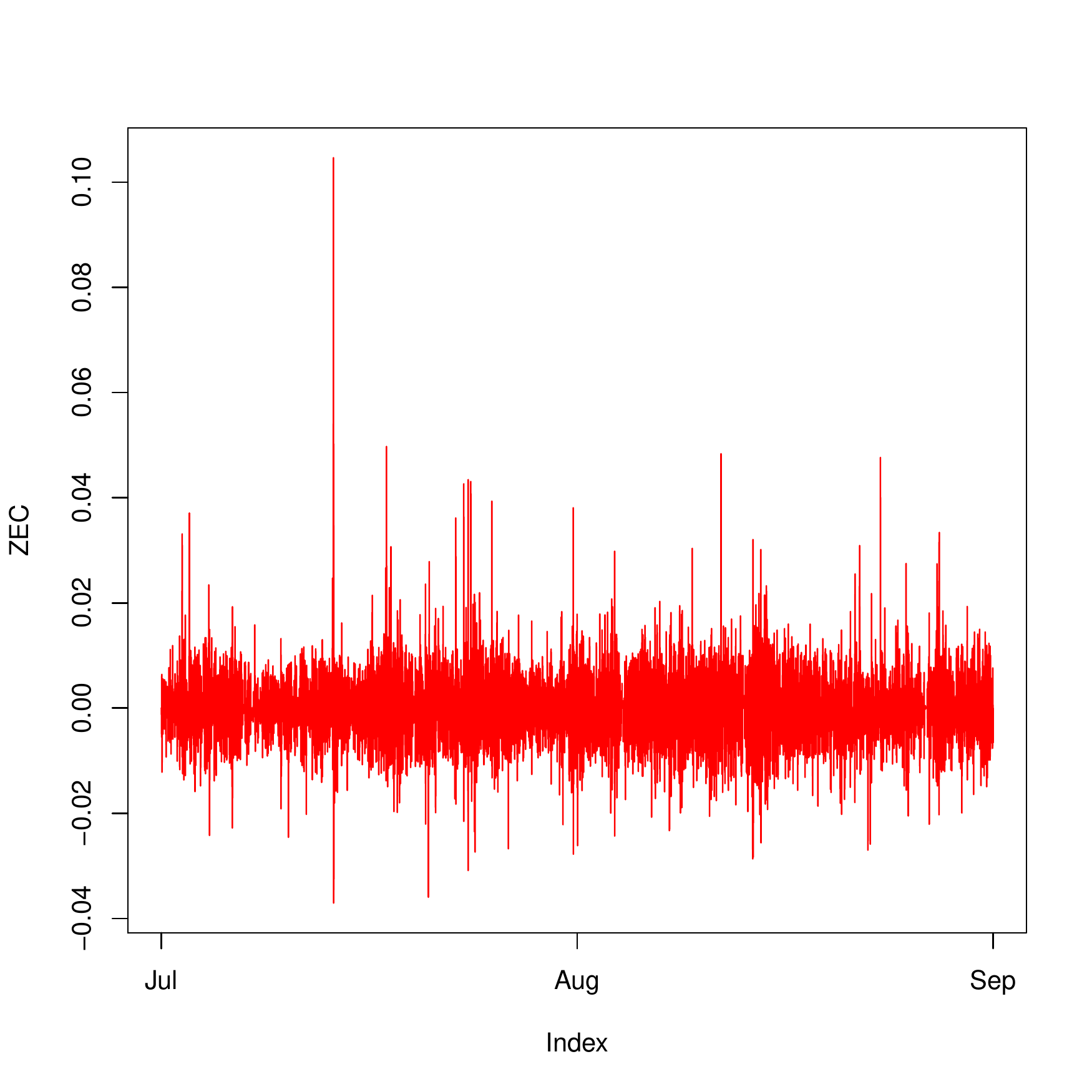}}
\hfill
\href{https://github.com/QuantLet/CCID/tree/master/CCIDHistRet}{\includegraphics[keepaspectratio,width=0.4cm]{media/qletlogo_tr.png}}
\caption{Intraday 5-minutes log-returns. 01. July 2018 - 31. August 2018.}
\end{figure}

\begin{figure}[H]
\hfill
\subfigure[XMR]{\includegraphics[width=5cm]{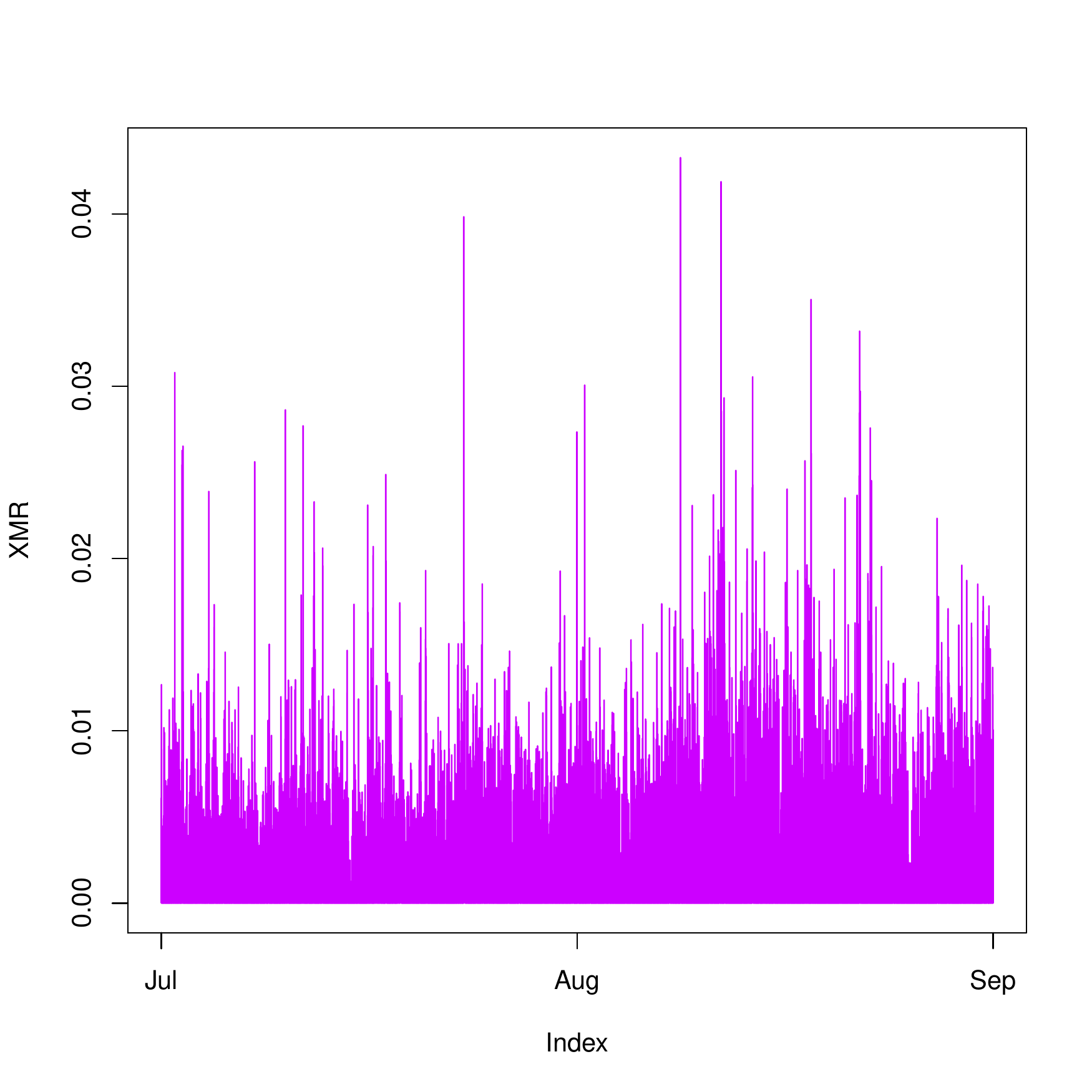}}
\hfill
\subfigure[ZEC]{\includegraphics[width=5cm]{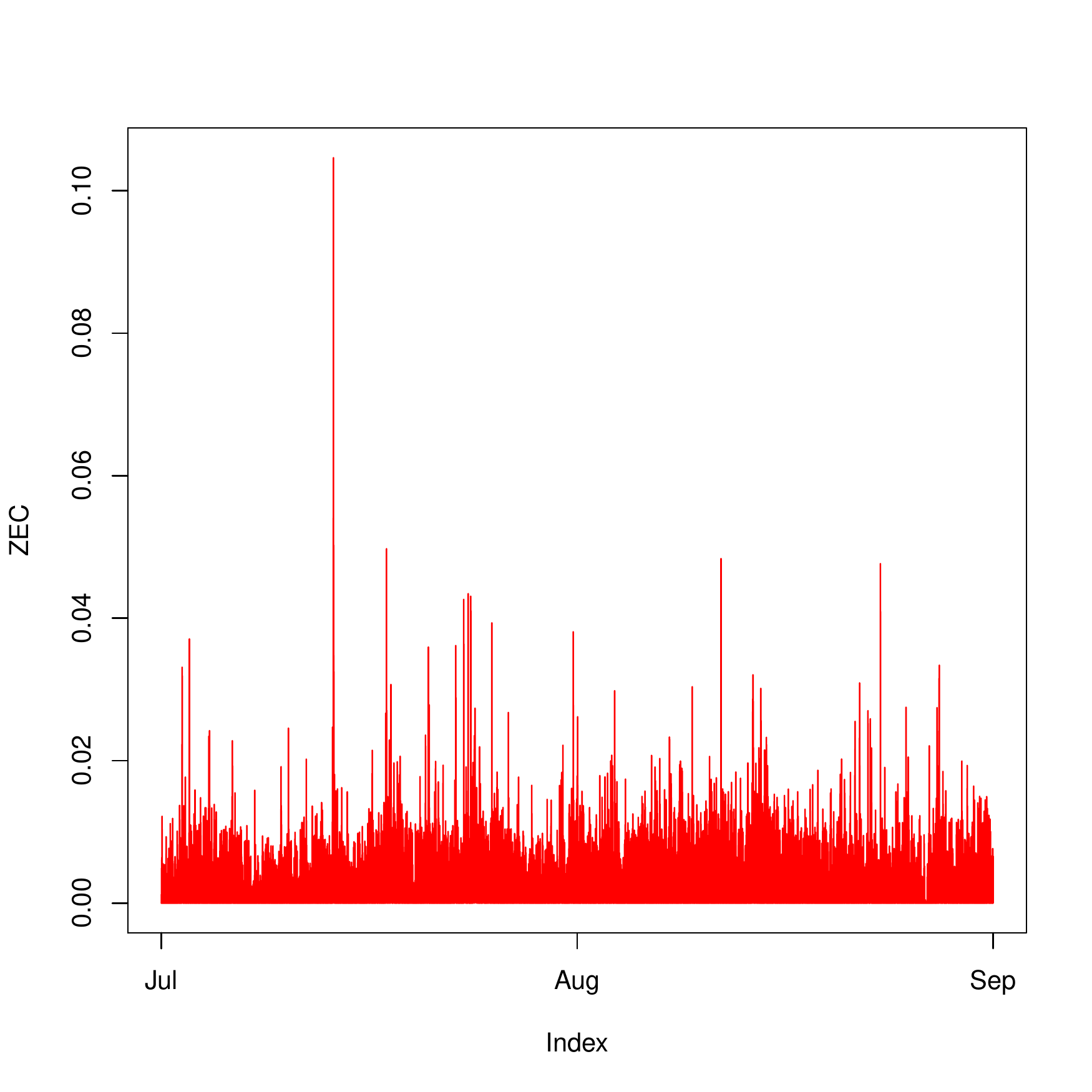}}
\hfill
\href{https://github.com/QuantLet/CCID/tree/master/CCIDHistRet}{\includegraphics[keepaspectratio,width=0.4cm]{media/qletlogo_tr.png}}
\caption{Intraday Volatility. 01. July 2018 - 31. August 2018.}
\end{figure}

\begin{figure}[H]
\hfill
\subfigure[XMR]{\includegraphics[width=5cm]{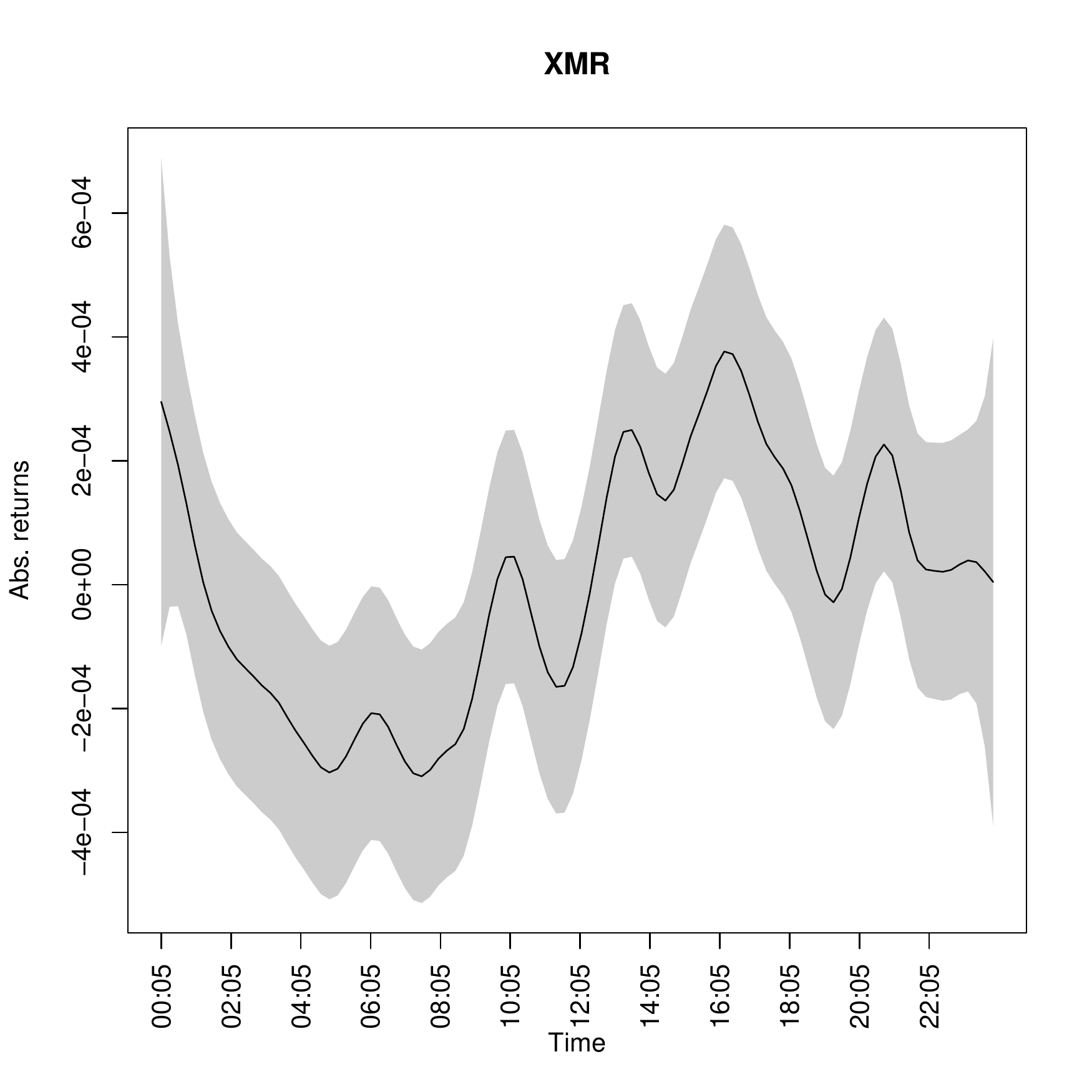}}
\hfill
\subfigure[ZEC]{\includegraphics[width=5cm]{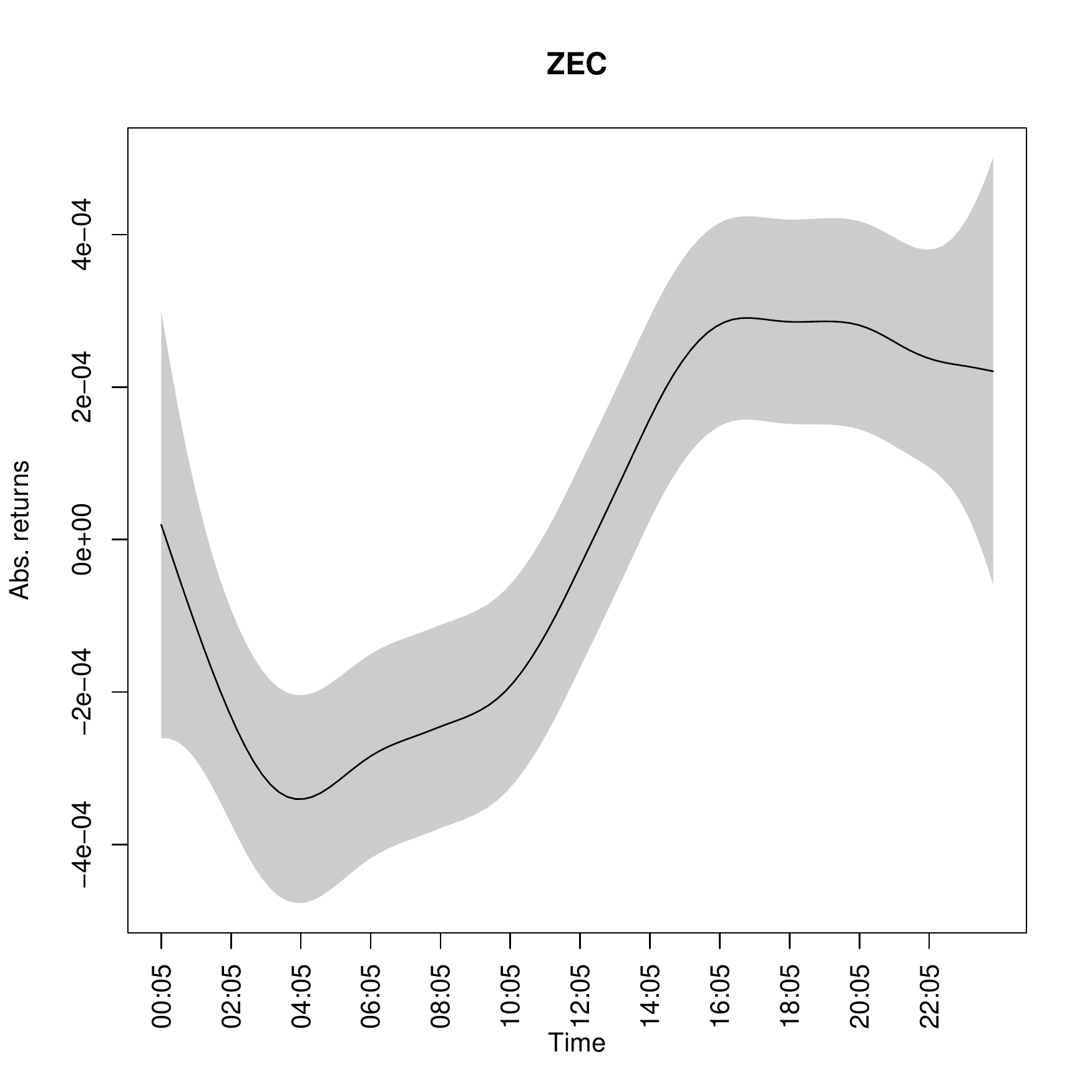}}
\hfill
\href{https://github.com/QuantLet/CCID/tree/master/CCIDvolaGAM}{\includegraphics[keepaspectratio,width=0.4cm]{media/qletlogo_tr.png}}
\caption{Generalized Additive Model of volatility of cryptocurrencies. 01. July 2018 - 31. August 2018.}
\end{figure}

\begin{figure}[H]
\hfill
\subfigure[XMR]{\includegraphics[width=5cm]{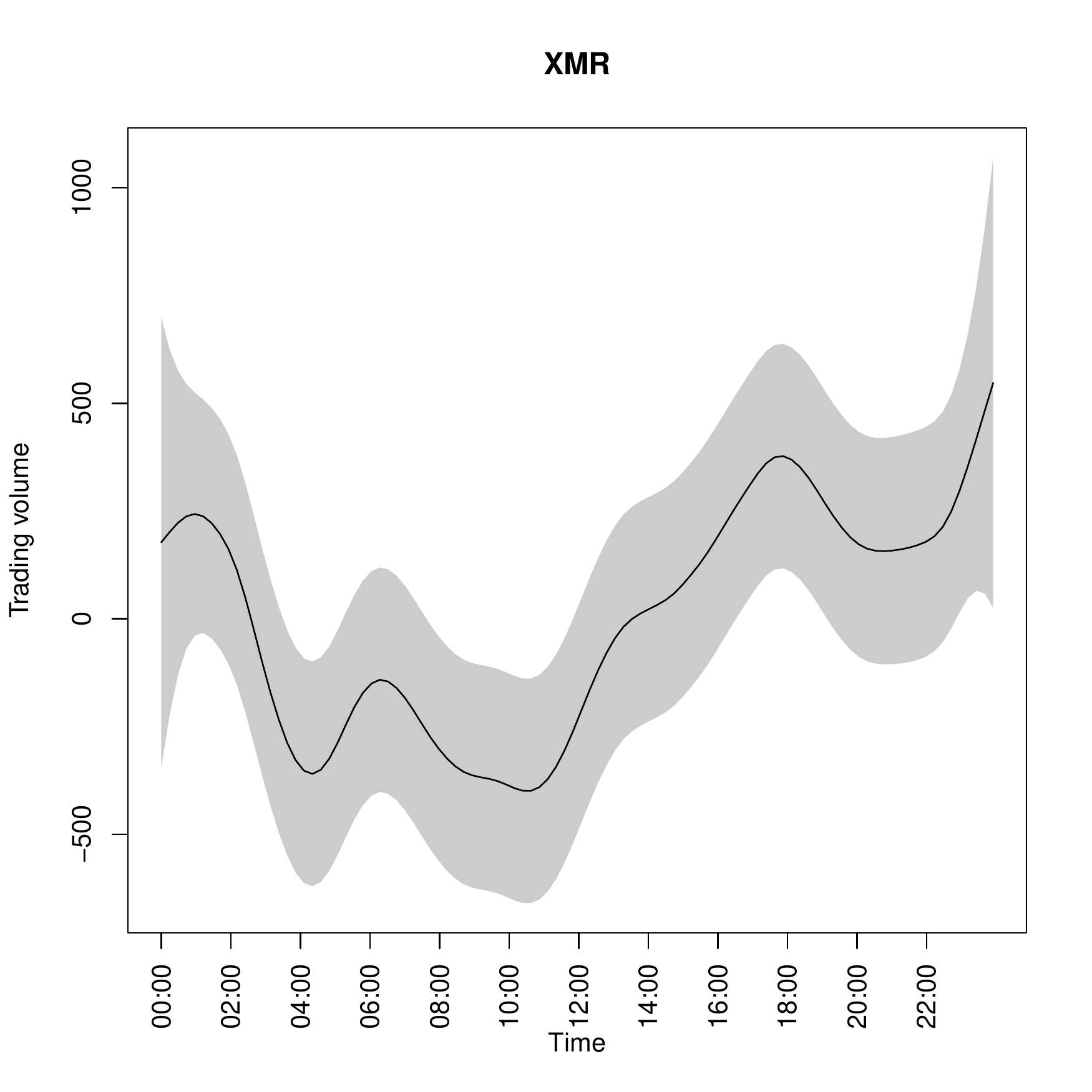}}
\hfill
\subfigure[ZEC]{\includegraphics[width=5cm]{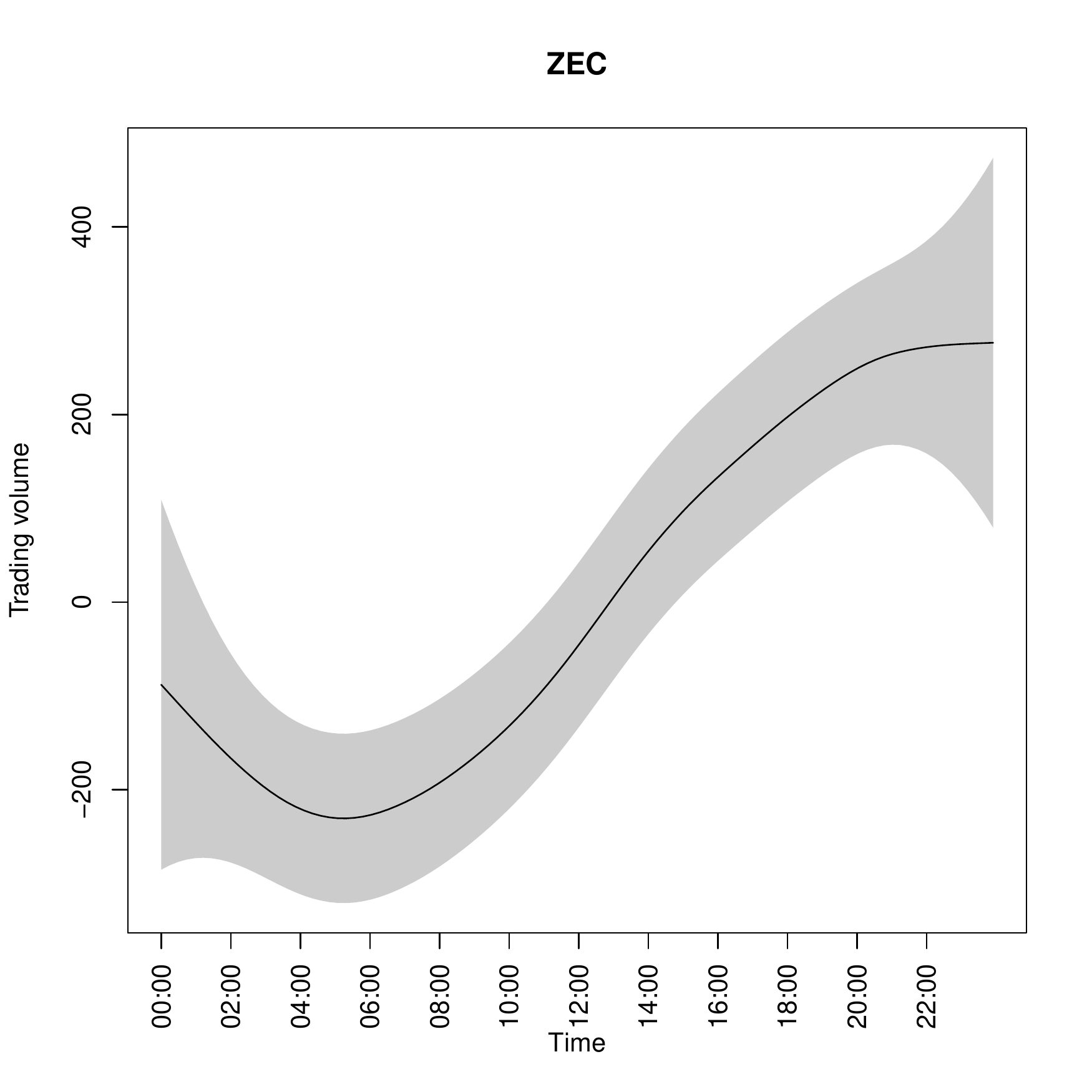}}
\hfill
\href{https://github.com/QuantLet/CCID/tree/master/CCIDvolumeGAM}{\includegraphics[keepaspectratio,width=0.4cm]{media/qletlogo_tr.png}}
\caption{Generalized Additive Model of the 62 intraday trading volume of cryptocurrencies. 01. July 2018 - 31. August 2018.}
\end{figure}

\begin{figure}[H]
\hfill
\subfigure[XMR]{\includegraphics[width=5cm]{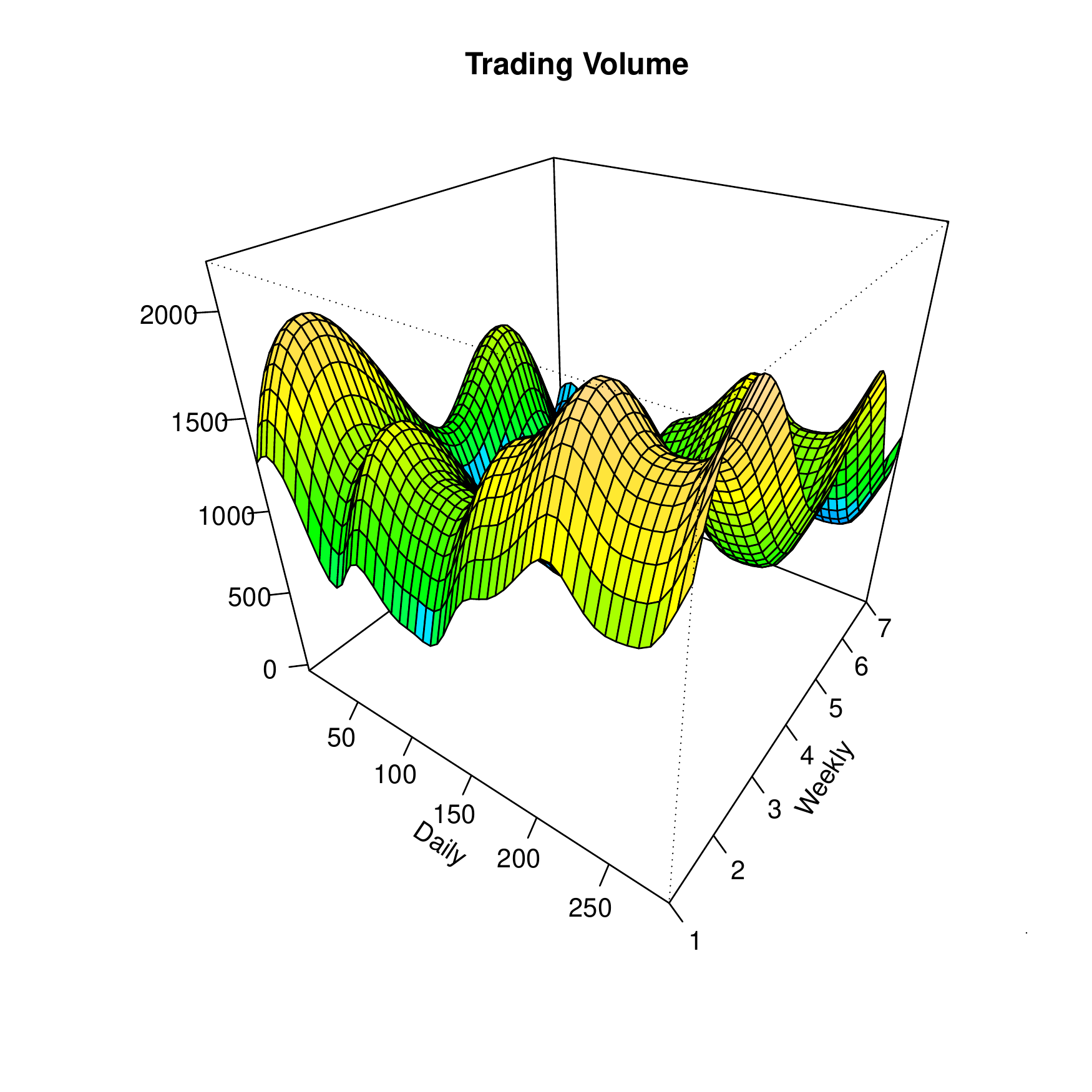}}
\hfill
\subfigure[ZEC]{\includegraphics[width=5cm]{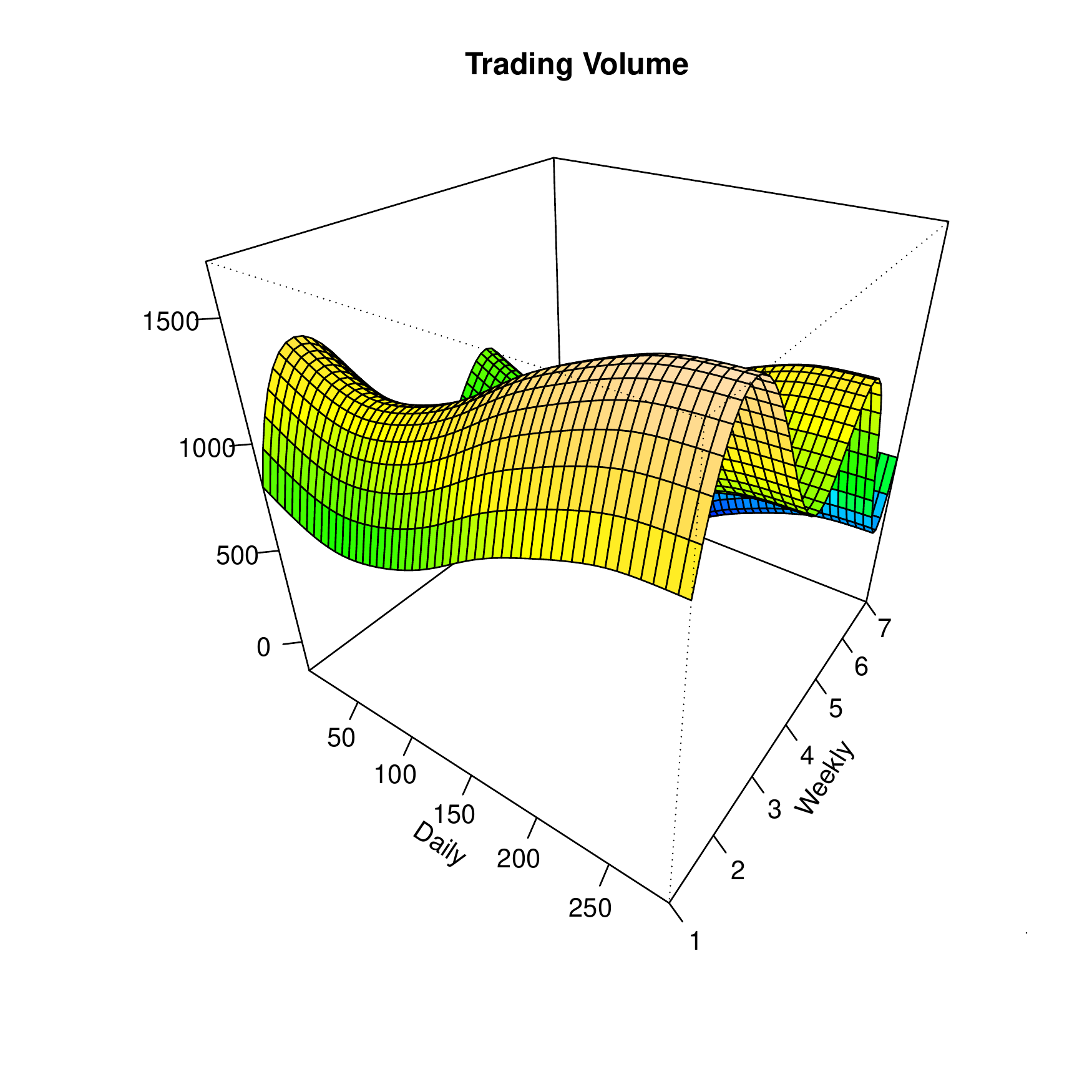}}
\hfill
\href{https://github.com/QuantLet/CCID/tree/master/CCIDvolumeGAM}{\includegraphics[keepaspectratio,width=0.4cm]{media/qletlogo_tr.png}}
\caption{Daily and weekly seasonality: fit of Generalized Additive Model   with cubic and p-splines for trading volume of cryptocurrencies (5 min nodes), 01. July 2018 - 31. August 2018. 01. July 2018 - 31. August 2018.}
\end{figure}

\begin{figure}[H]
\hfill
\subfigure[XMR]{\includegraphics[width=5cm]{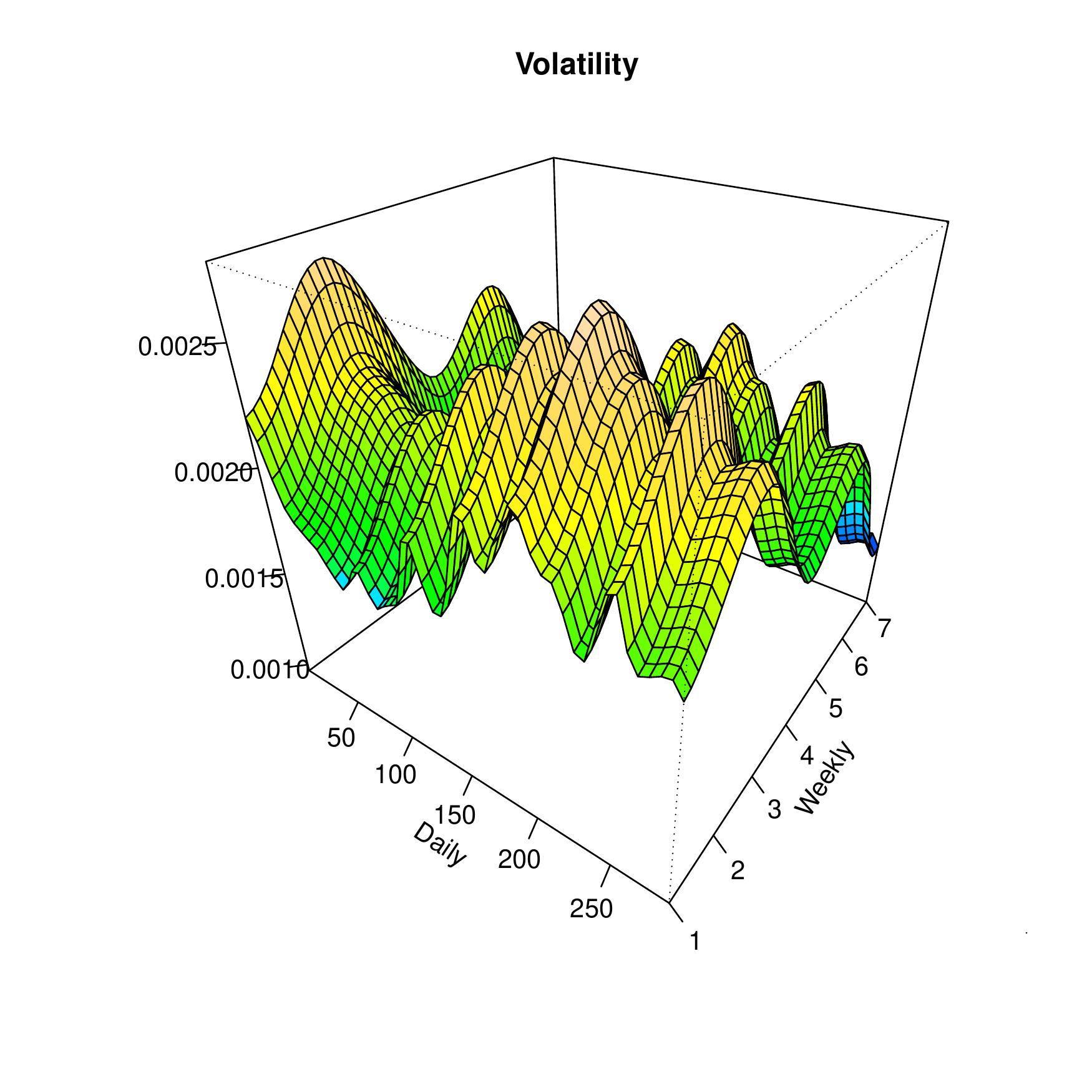}}
\hfill
\subfigure[ZEC]{\includegraphics[width=5cm]{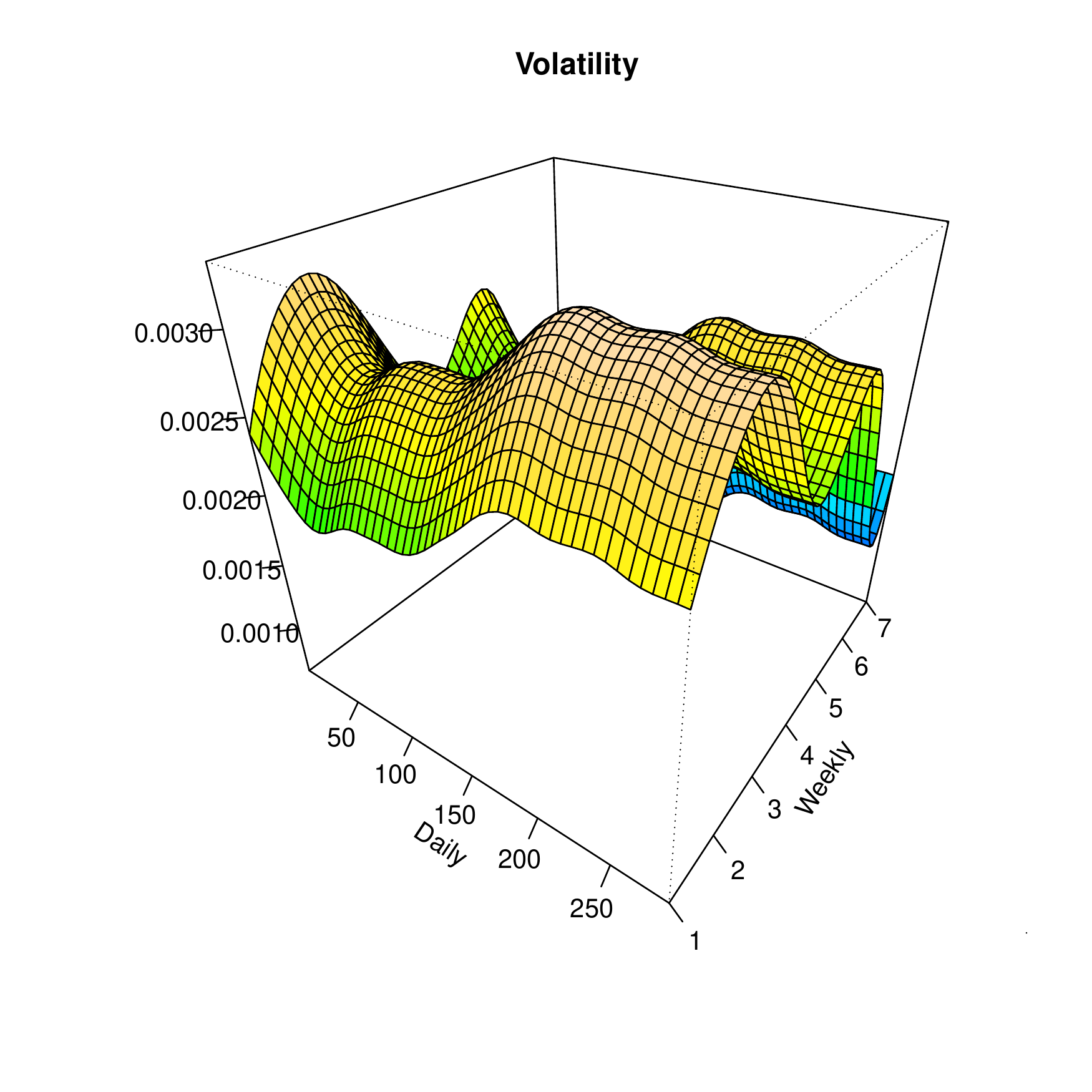}}
\hfill
\href{https://github.com/QuantLet/CCID/tree/master/CCIDvolaGAM}{\includegraphics[keepaspectratio,width=0.4cm]{media/qletlogo_tr.png}}
\caption{Daily and weekly seasonality: fit of Generalized Additive Model   with cubic and p-splines for volatility of cryptocurrencies (5 min nodes), 01. July 2018 - 31. August 2018. 01. July 2018 - 31. August 2018.}
\end{figure}



\end{document}